\documentclass[twocolumn]{aastex63}

\usepackage{amsmath,amssymb}
\usepackage{xspace}
\usepackage{CJK}

\usepackage{hyperref}
\hypersetup{
    colorlinks=true,
    linkcolor=blue,
    filecolor=magenta,      
    urlcolor=cyan,
}

\usepackage[normalem]{ulem} 
\usepackage{cancel} 

\usepackage{color}

\newcommand{\um}{\ensuremath{\mu\rm{m}}\xspace}
\newcommand{\kms}{\ensuremath{\rm{km\,s}^{-1}}\xspace}
\newcommand{\alphaco}{\ensuremath{\alpha_{\rm{CO}}}\xspace}

\newcommand{\lfir}{\ensuremath{L_{\rm{FIR}}}\xspace}
\newcommand{\lir}{\ensuremath{L_{\rm{IR}}}\xspace}
\newcommand{\fagn}{\ensuremath{f_\mathrm{AGN}}\xspace}

\newcommand{\Mstar}{\ensuremath{M_{\rm{star}}}\xspace}

\newcommand{\MHt}{\ensuremath{M_{\rm{H}_2}}\xspace}

\newcommand{\Msol}{\ensuremath{\rm{M}_\odot}\xspace}
\newcommand{\Lsol}{\ensuremath{\rm{L}_\odot}\xspace}

\newcommand{\percc}{\ensuremath{\rm{cm}^{-3}}\xspace}
\newcommand{\nht}{\ensuremath{n_\mathrm{H_2}}\xspace}

\newcommand{\vmax}{\ensuremath{v_\mathrm{max}}\xspace}
\newcommand{\vfifty}{\ensuremath{v_{50}}\xspace}
\newcommand{\vef}{\ensuremath{v_{84}}\xspace}
\newcommand{\cii}{[C{\scriptsize II}]\xspace}

\newcommand{\etal}{et~al.\xspace}
\newcommand{\hst}{\textit{HST}\xspace}
\newcommand{\arc}{\ensuremath{''}\xspace}

\shortauthors{J.~S.~Spilker, et~al.}
\shorttitle{Ubiquitous $z$$>$4 Molecular Outflows I. Outflow Structure}

\begin{document}
\begin{CJK*}{UTF8}{gbsn}

\defcitealias{spilker18a}{S18}
\defcitealias{spilker20}{Paper~I}
\defcitealias{spilker20a}{Paper~II}

\title{Ubiquitous Molecular Outflows in $z$ $>$ 4 Massive, Dusty Galaxies\\
I. Sample Overview and Clumpy Structure in Molecular Outflows on 500pc Scales}

\correspondingauthor{Justin S. Spilker}
\email{spilkerj@gmail.com}

\author[0000-0003-3256-5615]{Justin~S.~Spilker}
\altaffiliation{NHFP Hubble Fellow}
\affiliation{Department of Astronomy, University of Texas at Austin, 2515 Speedway, Stop C1400, Austin, TX 78712, USA}

\author[0000-0001-7946-557X]{Kedar~A.~Phadke}
\affiliation{Department of Astronomy, University of Illinois, 1002 West Green St., Urbana, IL 61801, USA}

\author[0000-0002-6290-3198]{Manuel~Aravena}
\affiliation{N\'{u}cleo de Astronom{\'i}a de la Facultad de Ingenier\'{i}a y Ciencias, Universidad Diego Portales, Av. Ej\'{e}rcito Libertador 441, Santiago, Chile}

\author[0000-0002-3915-2015]{Matthieu~B{\'e}thermin}
\affiliation{Aix Marseille Univ., CNRS, CNES, LAM, Marseille, France}

\author{Scott~C.~Chapman}
\affiliation{Department of Physics and Astronomy, University of British Columbia, 6225 Agricultural Rd., Vancouver, V6T 1Z1, Canada}
\affiliation{National Research Council, Herzberg Astronomy and Astrophysics, 5071 West Saanich Rd., Victoria, V9E 2E7, Canada}
\affiliation{Department of Physics and Atmospheric Science, Dalhousie University, Halifax, Nova Scotia, Canada}

\author[0000-0002-5823-0349]{Chenxing~Dong~(董辰兴)}
\affiliation{Department of Astronomy, University of Florida, 211 Bryant Space Sciences Center, Gainesville, FL 32611, USA}

\author[0000-0002-0933-8601]{Anthony~H.~Gonzalez}
\affiliation{Department of Astronomy, University of Florida, 211 Bryant Space Sciences Center, Gainesville, FL 32611, USA}

\author[0000-0003-4073-3236]{Christopher~C.~Hayward}
\affiliation{Center for Computational Astrophysics, Flatiron Institute, 162 Fifth Avenue, New York, NY, 10010, USA}

\author[0000-0002-8669-5733]{Yashar~D.~Hezaveh}
\affiliation{D\'{e}partement de Physique, Universit\'{e} de Montr\'{e}al, Montreal, Quebec, H3T 1J4, Canada}
\affiliation{Center for Computational Astrophysics, Flatiron Institute, 162 Fifth Avenue, New York, NY, 10010, USA}

\author[0000-0002-5386-7076]{Sreevani~Jarugula}
\affiliation{Department of Astronomy, University of Illinois, 1002 West Green St., Urbana, IL 61801, USA}

\author[0000-0002-4208-3532]{Katrina~C.~Litke}
\affiliation{Steward Observatory, University of Arizona, 933 North Cherry Avenue, Tucson, AZ 85721, USA}

\author[0000-0001-6919-1237]{Matthew~A.~Malkan}
\affiliation{Department of Physics and Astronomy, University of California, Los Angeles, CA 90095-1547, USA}

\author[0000-0002-2367-1080]{Daniel~P.~Marrone}
\affiliation{Steward Observatory, University of Arizona, 933 North Cherry Avenue, Tucson, AZ 85721, USA}

\author[0000-0002-7064-4309]{Desika~Narayanan}
\affiliation{Department of Astronomy, University of Florida, 211 Bryant Space Sciences Center, Gainesville, FL 32611, USA}
\affiliation{University of Florida Informatics Institute, 432 Newell Drive, CISE Bldg E251, Gainesville, FL 32611, USA}
\affiliation{Cosmic Dawn Center (DAWN), DTU-Space, Technical University of Denmark, Elektrovej 327, DK-2800 Kgs. Lyngby, Denmark}

\author[0000-0001-7477-1586]{Cassie~Reuter}
\affiliation{Department of Astronomy, University of Illinois, 1002 West Green St., Urbana, IL 61801, USA}

\author[0000-0001-7192-3871]{Joaquin~D.~Vieira}
\affiliation{Department of Astronomy, University of Illinois, 1002 West Green St., Urbana, IL 61801, USA}
\affiliation{Department of Physics, University of Illinois, 1110 West Green St., Urbana, IL 61801, USA}
\affiliation{National Center for Supercomputing Applications, University of Illinois, 1205 West Clark St., Urbana, IL 61801, USA}

\author[0000-0003-4678-3939]{Axel~Wei{\ss}}
\affiliation{Max-Planck-Institut f\"{u}r Radioastronomie, Auf dem H\"{u}gel 69, D-53121 Bonn, Germany}

\begin{abstract}

Massive galaxy-scale outflows of gas are one of the most commonly-invoked mechanisms to regulate the growth and evolution of galaxies throughout the universe. 
While the gas in outflows spans a large range of temperatures and densities, the cold molecular phase is of particular interest because molecular outflows may be capable of suppressing star formation in galaxies by removing the star-forming gas. 
We have conducted the first survey of molecular outflows at $z>4$, targeting 11 strongly-lensed dusty, star-forming galaxies (DSFGs) with high-resolution Atacama Large Millimeter Array (ALMA) observations of OH 119\,\um absorption as an outflow tracer. In this first paper, we give an overview of the survey, focusing on the detection rate and structure of molecular outflows. We find unambiguous evidence for outflows in 8/11 (73\%) galaxies, more than tripling the number known at $z>4$. This implies that molecular winds in $z>4$ DSFGs must have both a near-unity occurrence rate and large opening angles to be detectable in absorption.  Lensing reconstructions reveal that 500pc-scale clumpy structures in the outflows are common. The individual clumps are not directly resolved, but from optical depth arguments we expect that future observations will require 50--200pc spatial resolution to do so. We do not detect high-velocity \cii wings in any of the sources with clear OH outflows, indicating that \cii is not a reliable tracer of molecular outflows. Our results represent a first step toward characterizing molecular outflows at $z>4$ at the population level, demonstrating that large-scale outflows are ubiquitous among early massive, dusty galaxies.

\end{abstract}

\section{Introduction} \label{intro}

Galactic feedback is now widely recognized as a key component in our modern understanding of galaxy formation and evolution. `Feedback' is an umbrella term for a wide range of physical processes enabling self-regulated galaxy growth, setting the efficiency of star formation and shaping fundamental correlations between galaxy properties such as stellar mass, metallicity, star formation rate (SFR) and supermassive black hole mass. One of the most striking observational windows into galactic feedback is the ubiquitous detection of massive outflows of gas and dust being launched from galaxies, generally thought to be powered by supernovae and/or supermassive black hole accretion. Outflows of ionized and neutral atomic gas have been detected in galaxies over a wide range of mass and redshift for decades \citep[e.g.][]{heckman90,rupke05,weiner09,chisholm15}, including in massive dusty galaxies at high redshifts \citep[e.g.][]{banerji11,casey17,schechter18}. More recently, observations that probe the cold molecular gas in outflows have been a focus of recent interest because molecular gas is the direct fuel for future star formation and is often the dominant phase in the outflow mass budget (see \citealt{veilleux20} for a recent review of cold galactic winds).

In the high-redshift universe, spatially-resolved studies of massive quenching galaxies at $z \gtrsim 2$ typically find evidence for an inside-out suppression of star formation \citep[e.g.][]{tacchella15,tacchella18,nelson16,spilker19} accompanying a sharp overall decrease in the molecular gas fraction compared to equally-massive star-forming galaxies at the same epoch \citep[e.g.][]{spilker16a,popping17,tadaki17,talia18}. Indeed, massive ($\Mstar \sim 10^{11}$\,\Msol) quiescent galaxies have been identified in sizable numbers as early as $z\sim4$ \citep[e.g.][]{straatman14,guarnieri19,carnall20,valentino20}, implying a very rapid formation history with star formation rates (SFRs) of hundreds of \Msol/yr and subsequent rapid quenching of star formation \citep[e.g.][]{glazebrook17,estradacarpenter20,forrest20}. The required high SFRs are generally only found in very infrared-luminous systems, in which the UV radiation from young stars is absorbed and reprocessed by dust. These observations paint an appealing picture in which initially gas-rich, dusty, star-forming galaxies (DSFGs) at least temporarily suppress star formation via the consumption, heating, and/or ejection of the molecular gas fuel through a self-regulating feedback process or processes in order to create the early passive galaxy population \citep[e.g.][]{narayanan15}.

There are two primary tracers of the cold molecular phase of galactic winds in both the nearby and distant universe, neither of which is easily detectable, carbon monoxide (CO) and the hydroxyl molecule OH \citep[e.g.][]{veilleux20}. Low-order transitions of CO can be used to detect cold molecular outflows just as they are often used to probe the overall molecular contents of galaxies more generally \citep[e.g.][]{walter02,alatalo11,barcosmunoz18}. For unresolved observations, the outflow signature is an excess of CO emission at high velocities relative to systemic that is not plausibly related to rotational or non-rotational motions within the galaxies. CO observations have the benefit of sensitivity to gas at all distances and lines of sight to the host galaxy, but the line wings are very faint and the geometry of the emitting gas is difficult to constrain (for example, it is hard to distinguish outflowing from inflowing gas because the line-of-sight location of the emission is unknown, or to rule out that the emission is from a separate galaxy in a merger that may not be apparent even in deep imaging data). An alternative is far-infrared transitions of OH, demonstrated to be a very good tracer of outflowing and inflowing gas in dozens of nearby galaxies over the lifetime of the \textit{Herschel Space Observatory} \citep[e.g.][]{sturm11,spoon13,veilleux13,stone16,gonzalezalfonso17}. In this case the outflow (or inflow) signature takes the form of broad blueshifted (redshifted) absorption profiles against the continuum emission of the host galaxies. Because the gas flows are seen in absorption, the geometrical interpretation of the line profiles is simplified, but OH studies consequently require galaxies with bright continuum emission, and are not sensitive to outflowing material that does not intersect the line of sight to the hosts.

In the best-studied examples in the local universe, the geometry and structure of the winds can be spatially resolved, allowing for detailed pictures of the location, kinematics, and conditions within the molecular gas contained in the outflows. The prototypical starburst-driven outflows in M82 and NGC253, for example, are both seen nearly edge-on, and both show clumpy streamers of molecular gas extending a kpc or more out of the disks \citep[e.g.][]{walter02,leroy15,walter17,krieger19}. Meanwhile, very high-resolution observations of the outflow from the nuclear region of the nearby Seyfert galaxy NGC1068 in multiple molecular tracers reveal the impact of the AGN wind on the surrounding torus, also likely extending to larger spatial scales in the host galaxy \citep[e.g.][]{garciaburillo19}.

Such detailed views of molecular outflows have thus far been confined to the local universe. Despite a handful of past successes, merely detecting molecular outflows in the early universe at all continues to be extremely challenging: detecting the faint CO line wings associated with outflows requires substantial observational investments even with ALMA and is generally only possible out to $z\sim2$ except in extreme cases, and the detection of OH absorption requires bright continuum fluxes that limit the plausible target galaxies to very IR-luminous QSOs and DSFGs. Thus far only three objects at $z>4$ have reported molecular outflows \citep{spilker18a,jones19,herreracamus20}, and even in these cases the interpretation of the observations is not necessarily clear-cut given limitations in signal-to-noise and the complex galactic dynamics at play in the early universe. All told, at best a handful of galaxies at $z \gtrsim 2$ have detected molecular outflows, and while selected in very heterogeneous ways, all are limited to luminous dusty galaxies and/or AGN hosts \citep{weiss12,george14,feruglio17,fan18,herreracamus19}.

The structure of molecular outflows is of special interest because the emergence of molecular winds with properties similar to those observed in real galaxies has proven to be especially challenging for hydrodynamical simulations. In particular, accelerating initially at-rest molecular gas up to velocities of hundreds of \kms through direct ram pressure from a hot, fast wind or through entrainment in such a wind has proven extremely difficult. The cold and dense gas is shredded by hydrodynamical instabilities long before it reaches speeds like those observed in real galaxies \citep[e.g.][]{klein94,scannapieco13,schneider17}. One possible alternative is that molecules in outflows re-form at large galactocentric distances, cooling out of a hotter wind fluid having already reached velocities like those observed \citep[e.g.][]{zubovas14,mccourt18,richings18,schneider18}. Depending on the details of the simulation and the outflow energetics, this cold gas can show either kpc-scale clumpy structures or a fine mist-like morphology on very small scales.

This is the first in a series of papers in which we present the first constraints on the occurrence, structure, and physical properties of molecular outflows in a sample of DSFGs at $z>4$ targeting the OH 119\,\um doublet. This sample expands on our work in \citet{spilker18a} (hereafter \citetalias{spilker18a}), in which we reported the highest-redshift detection of a molecular outflow towards a $z=5.3$ galaxy. All targets are gravitationally lensed by foreground galaxies, which allows us to spatially resolve both the rest-frame 120\,\um dust continuum emission as well as the OH absorption at systemic and blueshifted velocities. While still not representative of the general population of high-redshift galaxies, our goal with this survey is to take a first step towards constraining the occurrence rate and properties of molecular outflows in the early universe in a statistical sense at the population level.

In this work we give an overview of the sample objects and present our new ALMA data. We focus here on the broad sample properties, outflow detection rates, and the resolved structure of the molecular outflows we determine from gravitational lensing reconstructions of the sources. In a companion paper (\citet{spilker20a}, hereafter \citetalias{spilker20a}), we characterize the physical properties of the molecular outflows we detect, focusing on the outflow rates, energetics, and wind driving mechanisms. Section~\ref{data} gives an overview of the sample objects, ALMA observations, and ancillary data for our objects and literature comparison samples. Section~\ref{analysis} describes our analysis methods for the OH spectra, how we classify whether or not objects show signs of outflow, and our lens modeling methodology and tests. Section~\ref{results} gives our main observational results, with additional discussion in Section~\ref{discussion}. We summarize our principle findings and conclude in Section~\ref{conclusions}. We assume a flat $\Lambda$CDM cosmology with $\Omega_m=0.307$ and $H_0=67.7$\,\kms\,Mpc$^{-1}$ \citep{planck15}, and we take the total infrared and far-infrared luminosities \lir and \lfir to be integrated over rest-frame 8--1000 and 40--120\,\um, respectively. Tables of the sample properties from this work, as well as the outflow properties from \citetalias{spilker20a}, are available in electronic form at \url{https://github.com/spt-smg/publicdata}.

\section{Sample and Observations} \label{data}

\subsection{Parent Sample and Source Selection} \label{selection}

We designed an observing campaign targeting $^{16}$OH $^2\Pi_{3/2}$ $J = 3/2 \rightarrow 5/2$ absorption. This transition is a $\Lambda$ doublet with components at rest-frame 2509.9 and 2514.3\,GHz (separated by $\sim$520\,\kms) and additional hyperfine structure that remains spectrally unresolved. We selected sources for OH observations from the point-source catalog of the 2500\,deg$^2$ SPT survey at 1.4 and 2\,mm \citep{carlstrom11,vieira10,mocanu13,everett20}. From the survey data and subsequent observations using the APEX/LABOCA camera at 870\,\um, a total of 81 objects were selected with spectral indices consistent with thermal dust emission (namely $S_{\mathrm{1.4mm}}/S_{\mathrm{2mm}} > 1.8$), raw 1.4\,mm flux density greater than 20\,mJy, flux density at 870\,\um greater than 25\,mJy, and that were not detected in various shallow multiwavelength surveys to reject low-redshift interlopers. Given their extreme brightness, the vast majority of these sources were expected to be gravitationally lensed by foreground galaxies. High-resolution ALMA imaging confirmed the lensed nature of these sources, with typical magnifications of 3--30 \citep{vieira13,hezaveh13,spilker16}. Extensive spectroscopic campaigns subsequently measured spectroscopic redshifts for the entire sample, which range from 1.87--6.90 with a median of 3.9 \citep{weiss13,strandet16,marrone18,reuter20}, although not all sources had known redshifts at the time the present outflow survey was designed.

Our primary selection criterion for OH 119\,\um observations was that the source redshift place the OH doublet lines at frequencies of relatively good atmospheric transmission in ALMA Band 8 (385--500\,GHz), requiring $z_\mathrm{source} \gtrsim 4.02$ (to reach ALMA Band 7 requires $z_\mathrm{source} > 5.8$, where we have few available targets). The atmospheric transmission at these frequencies is strongly affected by telluric water, oxygen, and ozone features, so OH observations are not feasible for all sources at $z>4$. In particular, OH observations are not possible for redshift windows $4.07 \lesssim z \lesssim 4.18$, $4.50 \lesssim z \lesssim 4.68$,   $4.86 \lesssim z \lesssim 4.99$, and $5.43 \lesssim z \lesssim 5.53$ due to especially deep atmospheric features. We restricted the sample to sources with redshifts that avoided frequencies of poor transmission, and predicted 119\,\um continuum flux densities bright enough that ALMA would be able to reach sensitivities of 5\% of the continuum level in $\sim200$\,\kms channels in less than an hour of observing time after resolving the source over 5--20 resolution elements. These predicted continuum flux densities were estimated using the available far-IR photometry, which provides very good sampling of the long-wavelength spectral energy distributions (SEDs; Section~\ref{ancillary}). We finally required that all targets have lens models from ALMA 870\,\um observations \citep{spilker16} and chose objects to span a wide range in \lir. The sources selected for OH observations are not obviously biased with respect to the full SPT sample of $z>4$ DSFGs in intrinsic (lensing-corrected) \lir, dust mass, or effective dust temperature \citep{reuter20}, although because we lack lens models for every SPT DSFG this remains somewhat uncertain.  Even after lensing correction these remain extremely luminous objects ($\log \lir/\Lsol = 12.5-13.5$); they are certainly not `typical' galaxies at these redshifts by any conceivable definition.

\begin{figure*}[t]
\begin{centering}
\includegraphics[width=\textwidth]{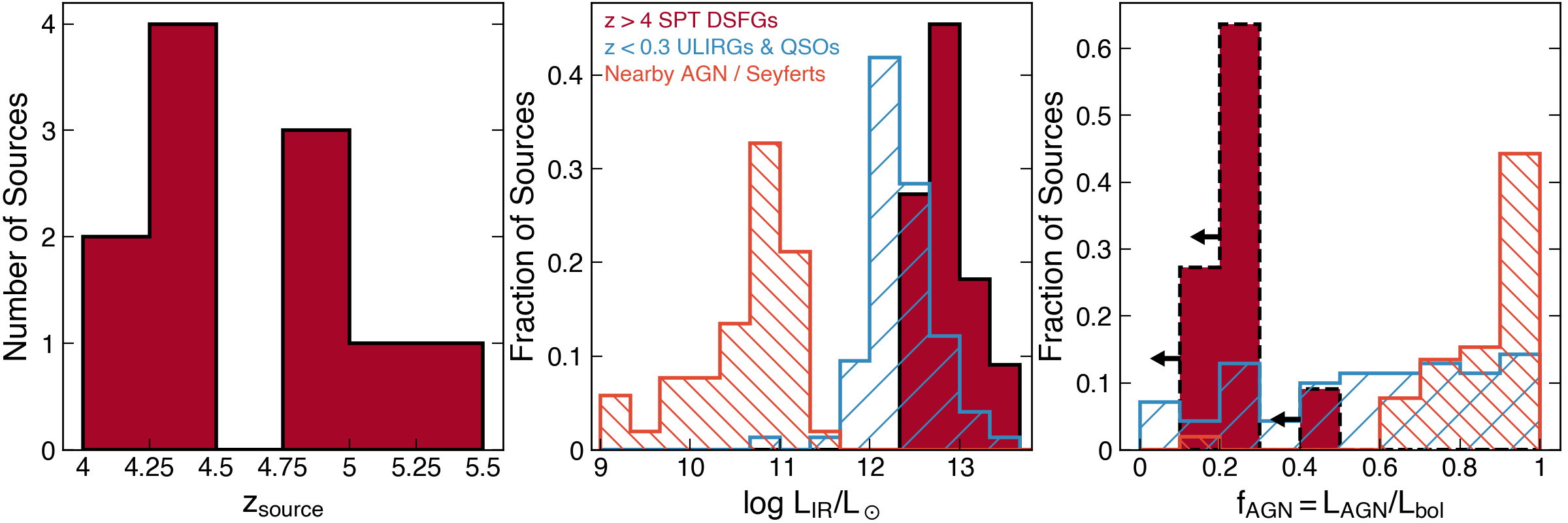}\\
\end{centering}
\caption{
Summary of source properties for our SPT-selected high-redshift DSFG outflows sample and literature comparison samples detailed in Section~\ref{lowzcomp}. All \fagn values for the SPT sample are upper limits, although the rest-frame mid-IR data used to compute \fagn may fail in the event of extremely high column densities hiding highly obscured AGN.
}\label{fig:sample}
\end{figure*}

The final sample consists of 11 objects at $4.09 < z < 5.30$ including SPT2319-55, previously published in \citetalias{spilker18a}. Basic properties of the sample are given in Tables~\ref{tab:almaobs} and \ref{tab:sample}, with a few salient properties shown in Figure~\ref{fig:sample}.

\subsection{ALMA Observations} \label{almaobs}

ALMA observed our sample galaxies across several projects from 2016--2019, summarized in Table~\ref{tab:almaobs}. For each object, we configured the correlator to observe the OH doublet with two slightly overlapping 1.875\,GHz wide basebands and 3.9MHz channels, providing contiguous coverage over 2200--2700\,\kms around the OH lines, depending on the redshift of each source. These basebands were placed such that the lower frequency edge corresponded to $\approx$1200\,\kms redward of the upper OH transition (or $\approx$680\,\kms redward of the lower-frequency transition), leaving $\approx$1000--1700\,\kms to the blueshifted side of the upper-frequency doublet transition. This setup was chosen to maximize the amount of blueshifted velocity coverage while still allowing both doublet transitions to be detected. Unfortunately it does not allow for the detection of strongly redshifted emission (or absorption), as expected for the classical P Cygni profile and sometimes observed in the OH spectra of local ULIRGs and quasar hosts \citep[e.g.,][]{veilleux13}. We also placed an additional two basebands of 1.875\,GHz width each for continuum coverage in the other sideband of the ALMA correlator. Given the fixed 4--8\,GHz intermediate frequency of the ALMA Band 8 receivers, these continuum measurements are centered either 12\,GHz above or below the OH frequencies in Table~\ref{tab:almaobs}, depending on the atmospheric transmission. For SPT0459-59 and SPT2132-58 the atmospheric transmission is poor both above and below the OH observed frequencies, making half (SPT0459-59) or all (SPT2132-58) of the continuum bandwidth unusable.

The observing time and requested spatial resolution were estimated using the available far-IR and sub/mm photometry and the lens models available from high-resolution ALMA imaging for each individual source \citep{spilker16,reuter20}. The array configuration(s) varied for each source, with maximum baseline lengths ranging from 700--1600m and minimum baseline lengths $\sim$15\,m for all observations. The shortest baselines lead to maximum recoverable scales $\gtrsim$2.5\arc depending on observing frequency; we do not expect significant emission on larger spatial scales. The data were reduced with the standard pipelines available for each ALMA cycle, with additional manual calibration and flagging where necessary. Besides the typical bandpass, flux, and complex gain calibrators, all observing blocks also recorded data for a quasar near each DSFG used as a test of the astrometry and calibration quality; this is standard for high-frequency and high-resolution ALMA observing blocks. Images of these test sources showed astrometric shifts of up to $\sim$0.1\arc and atmospheric decorrelation of up to 30\% depending on the target, evidence of residual atmospheric phase noise varying faster than the source-calibrator observing cycle. To mitigate this noise, we attempted one or two rounds of phase-only self-calibration on the test and target sources, using solution intervals of the scan length or half the scan length. This self-calibration was successful for all sources except SPT2103-60, decreasing the image rms by up to a factor of two. We note that self-calibration makes absolute astrometry impossible, so the astrometry of these data should be considered accurate to $\sim$0.1\arc, as measured from the test source observations before self-calibration. Self-calibration has no influence on the relative astrometry within the ALMA data (e.g. across the line profiles or between sidebands).

\begin{figure*}[t]
\begin{centering}
\includegraphics[width=0.47\textwidth]{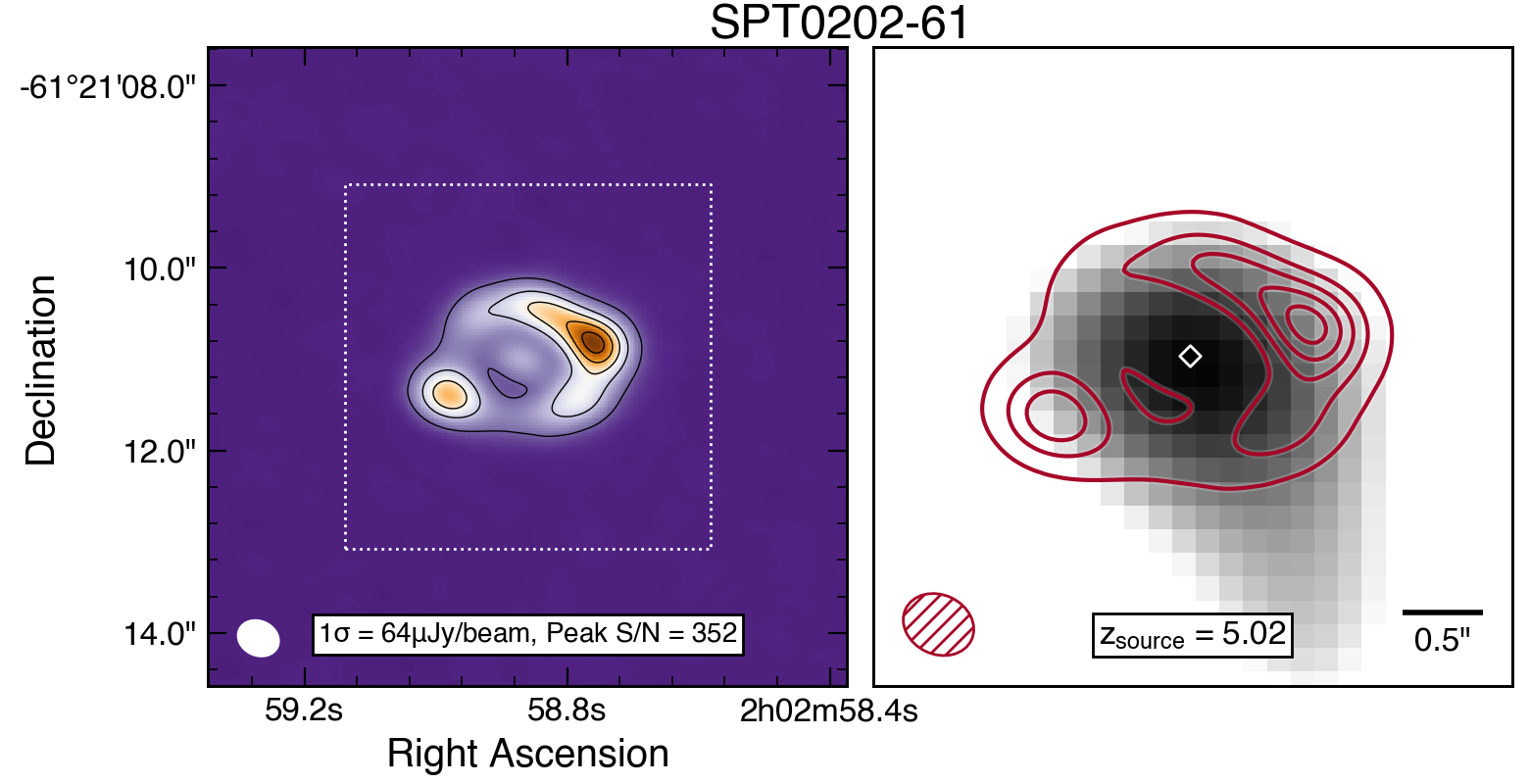}
\includegraphics[width=0.47\textwidth]{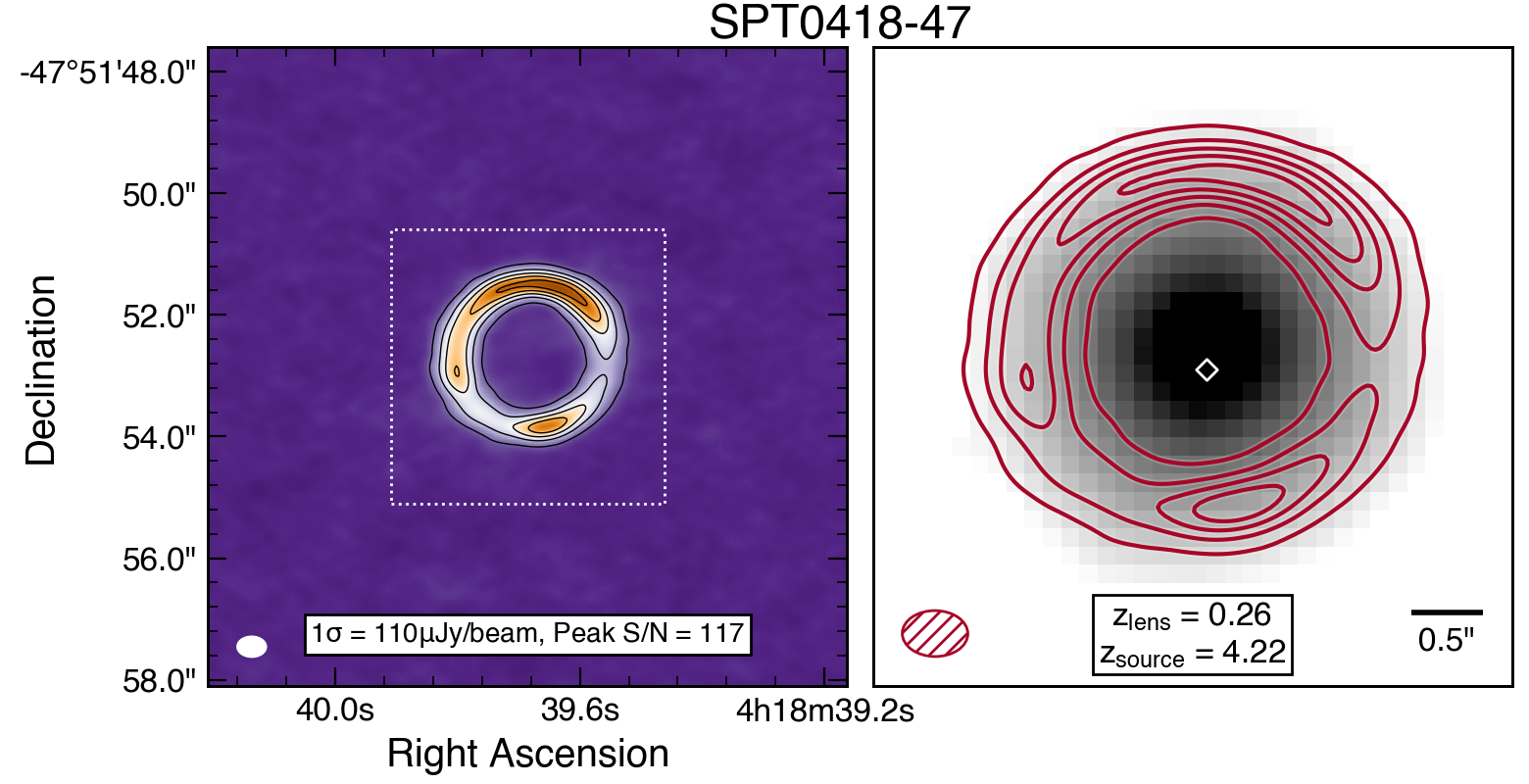}
\includegraphics[width=0.47\textwidth]{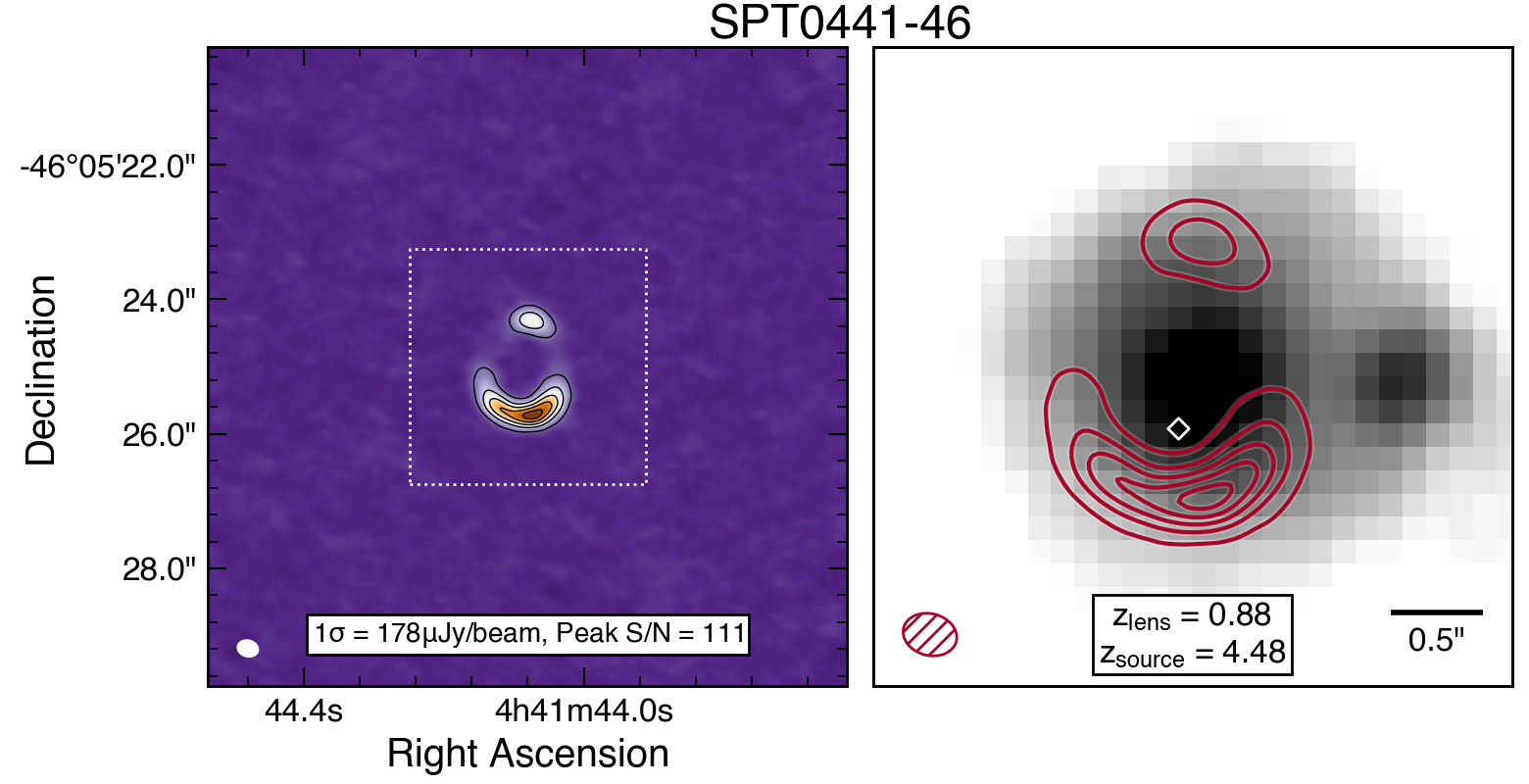}
\includegraphics[width=0.47\textwidth]{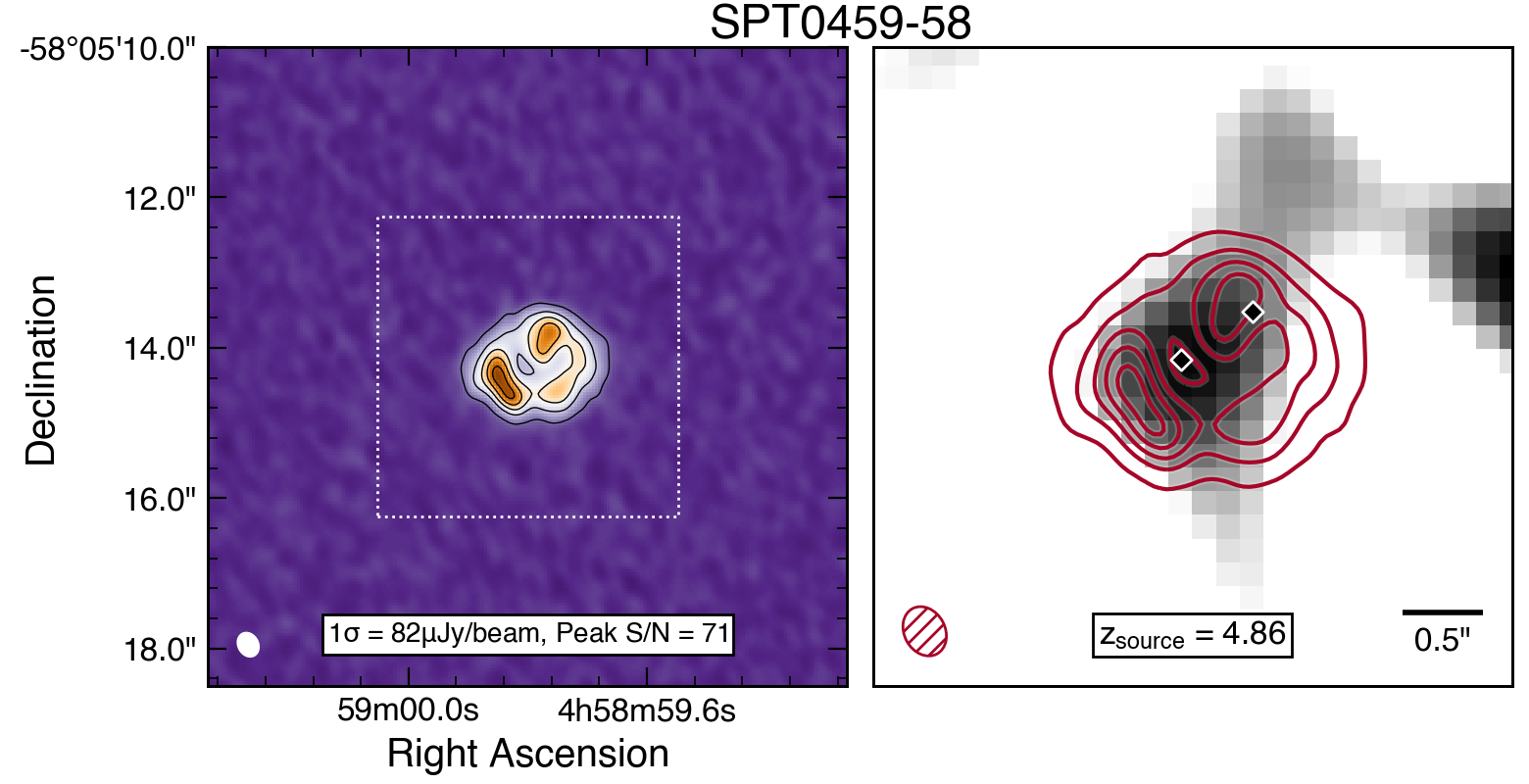}
\includegraphics[width=0.47\textwidth]{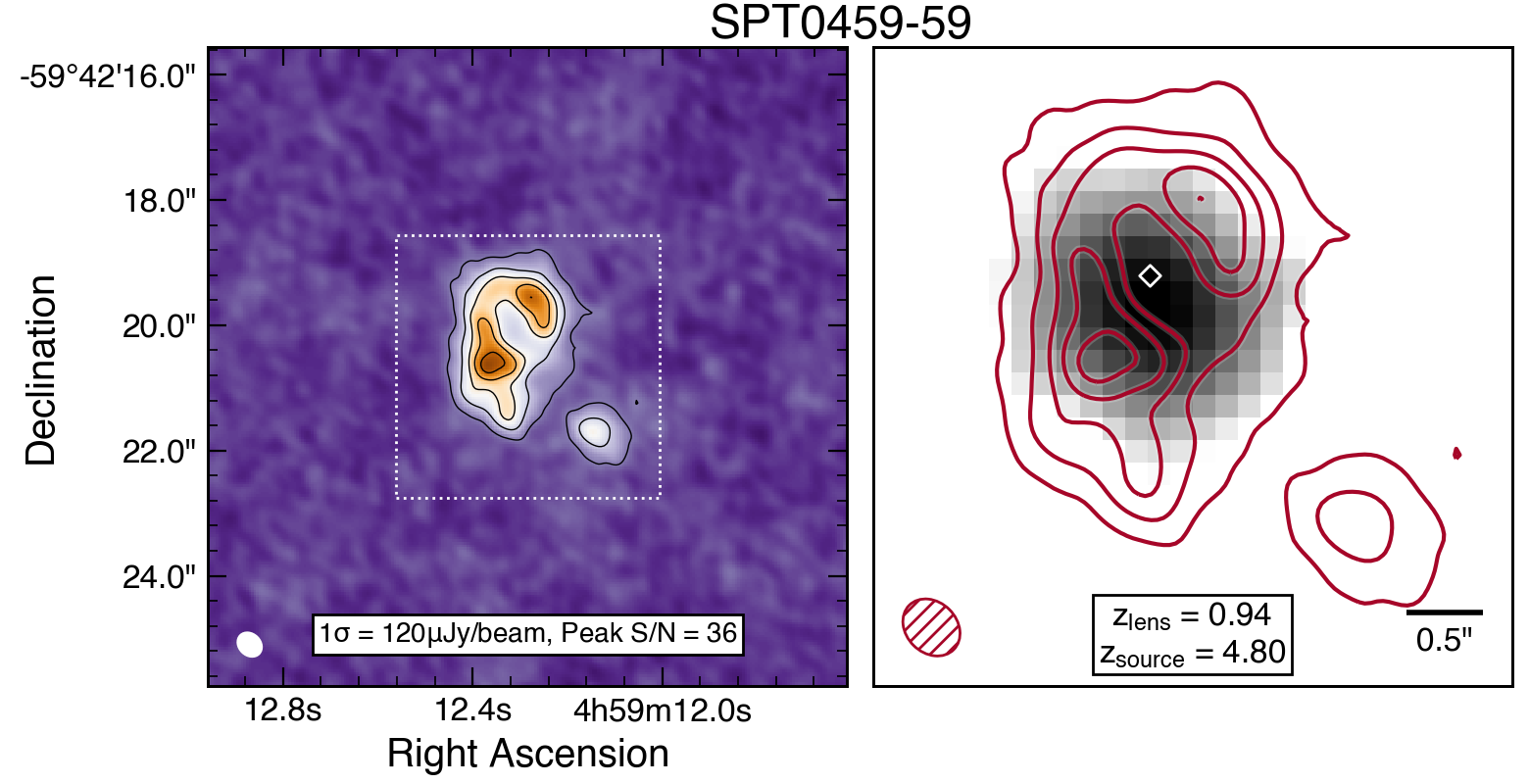}
\includegraphics[width=0.47\textwidth]{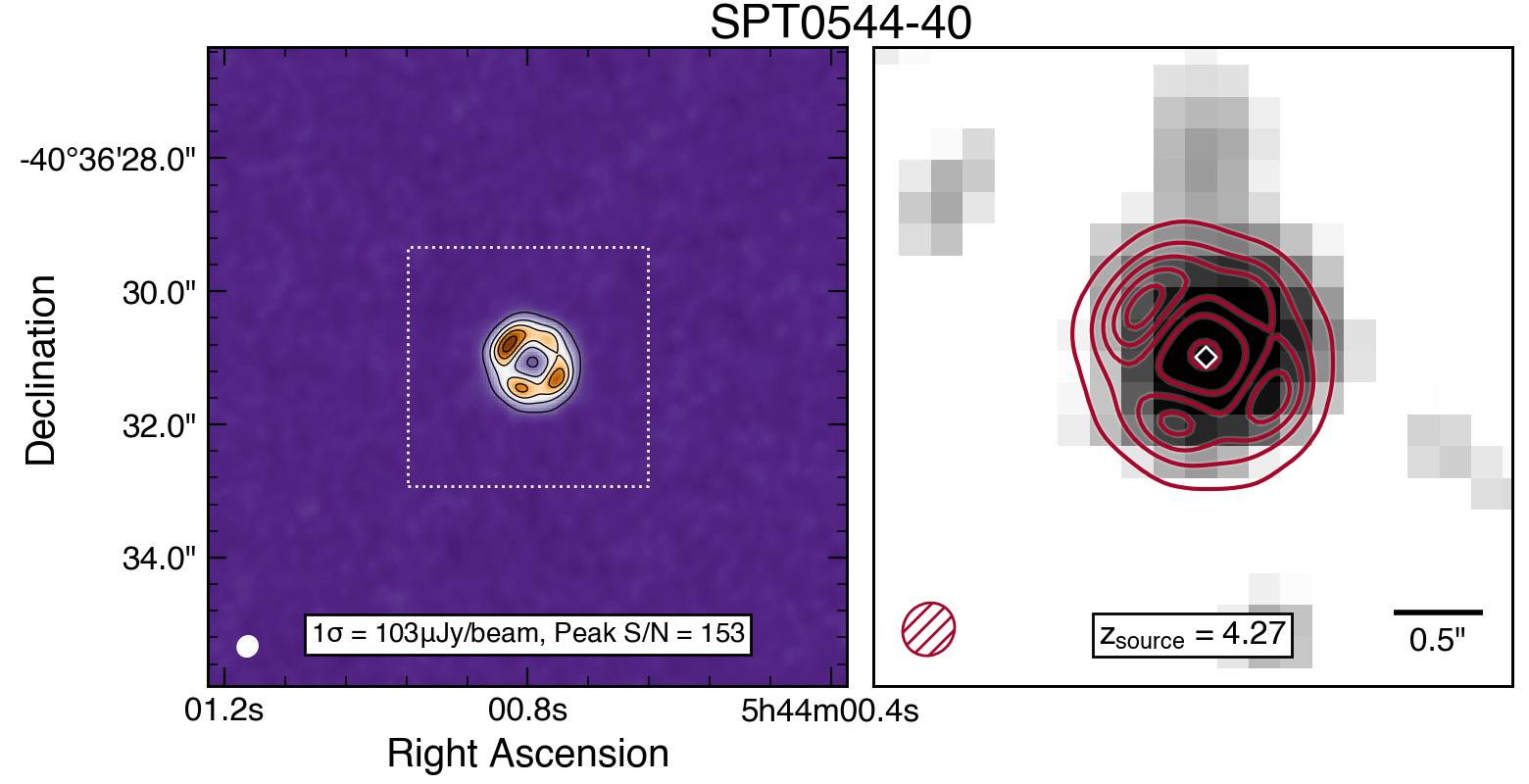}
\includegraphics[width=0.47\textwidth]{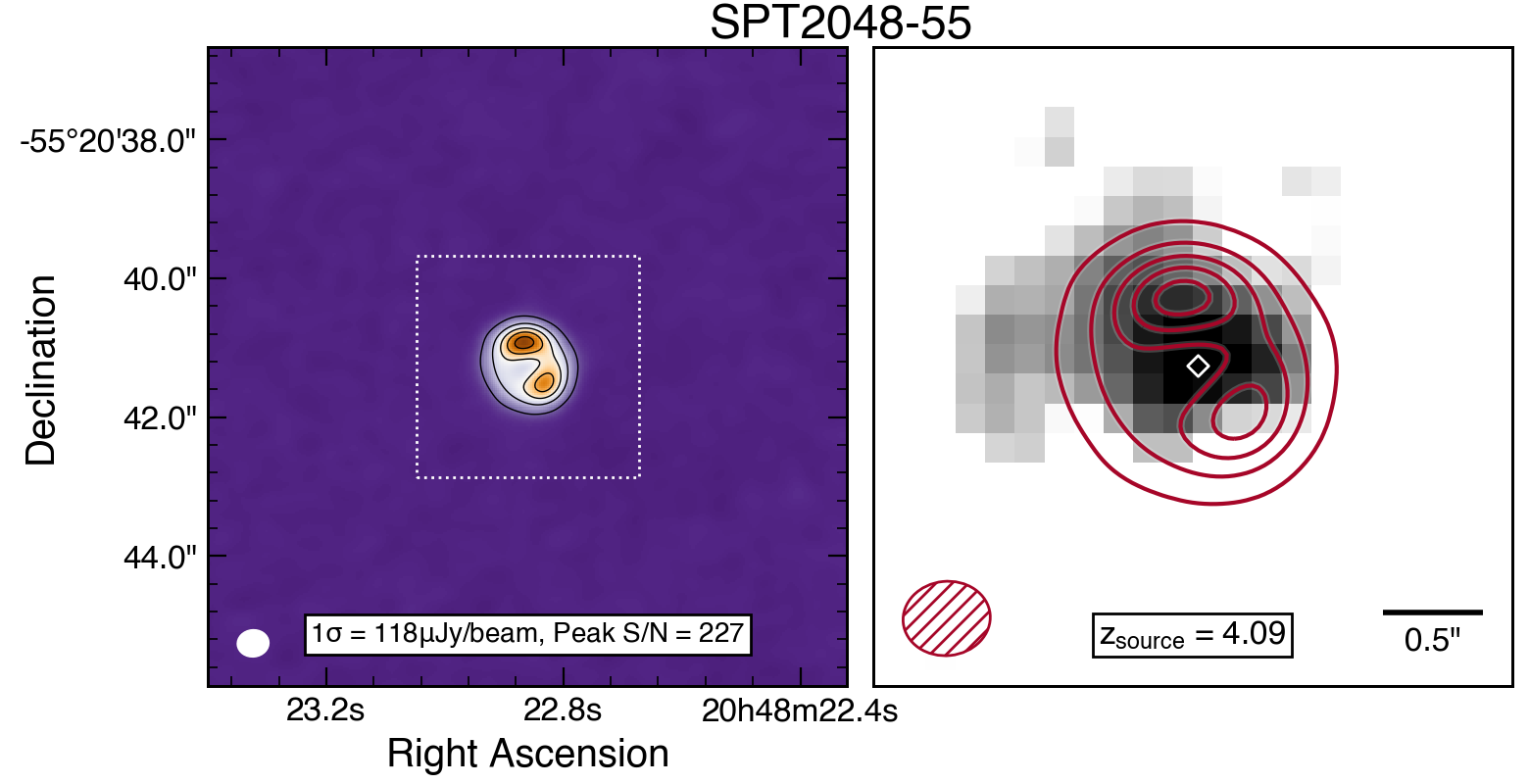}
\includegraphics[width=0.47\textwidth]{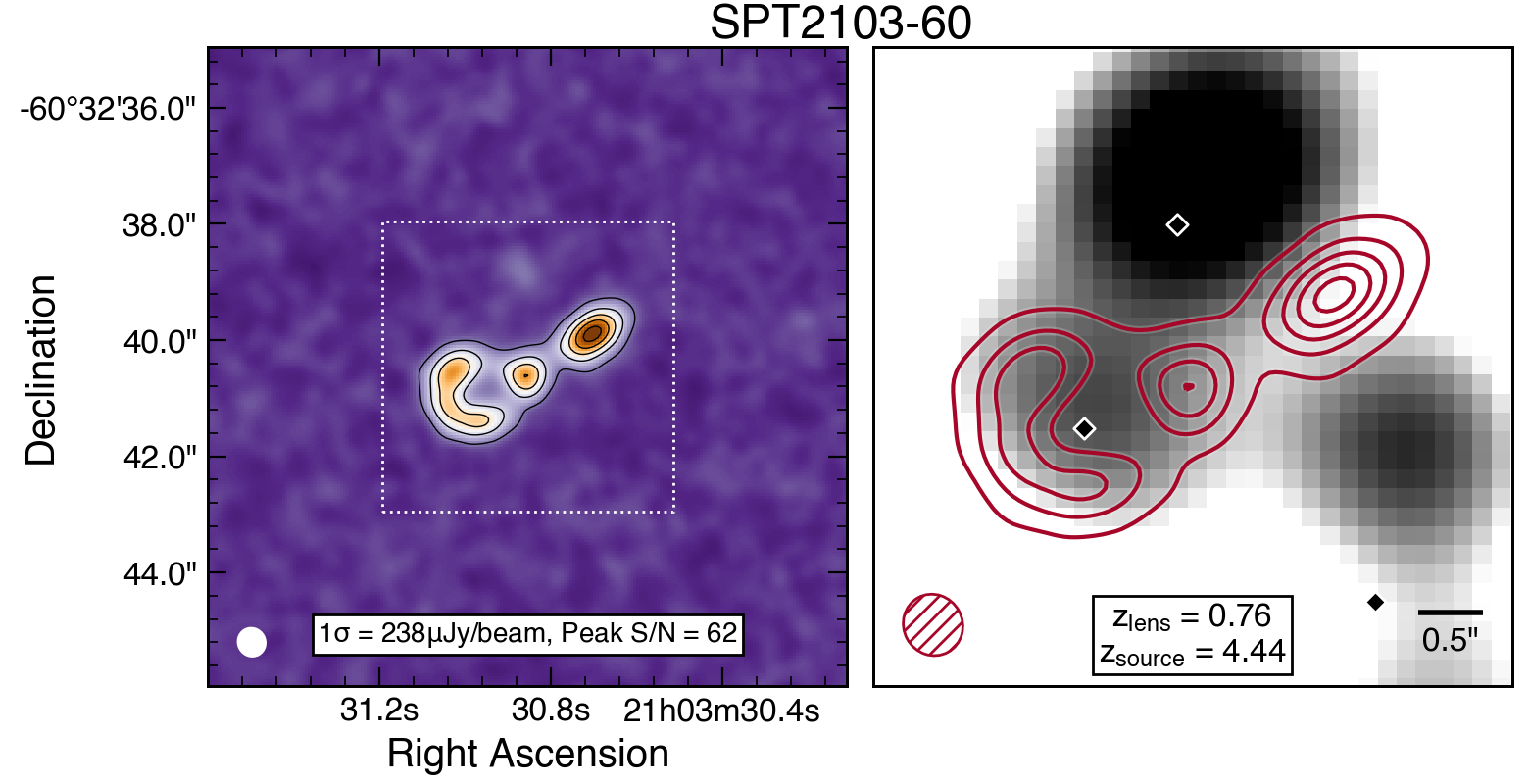}
\includegraphics[width=0.47\textwidth]{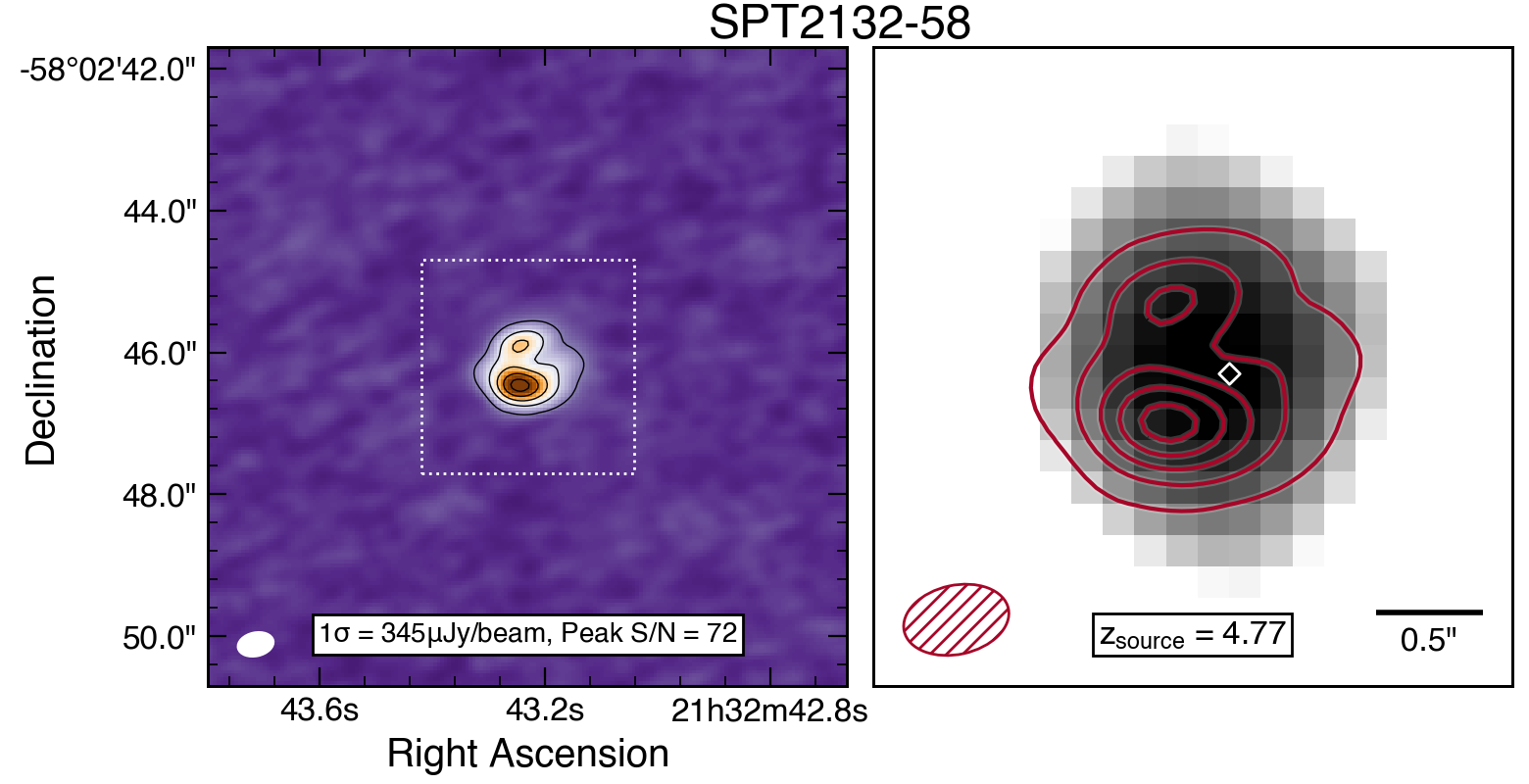}
\includegraphics[width=0.47\textwidth]{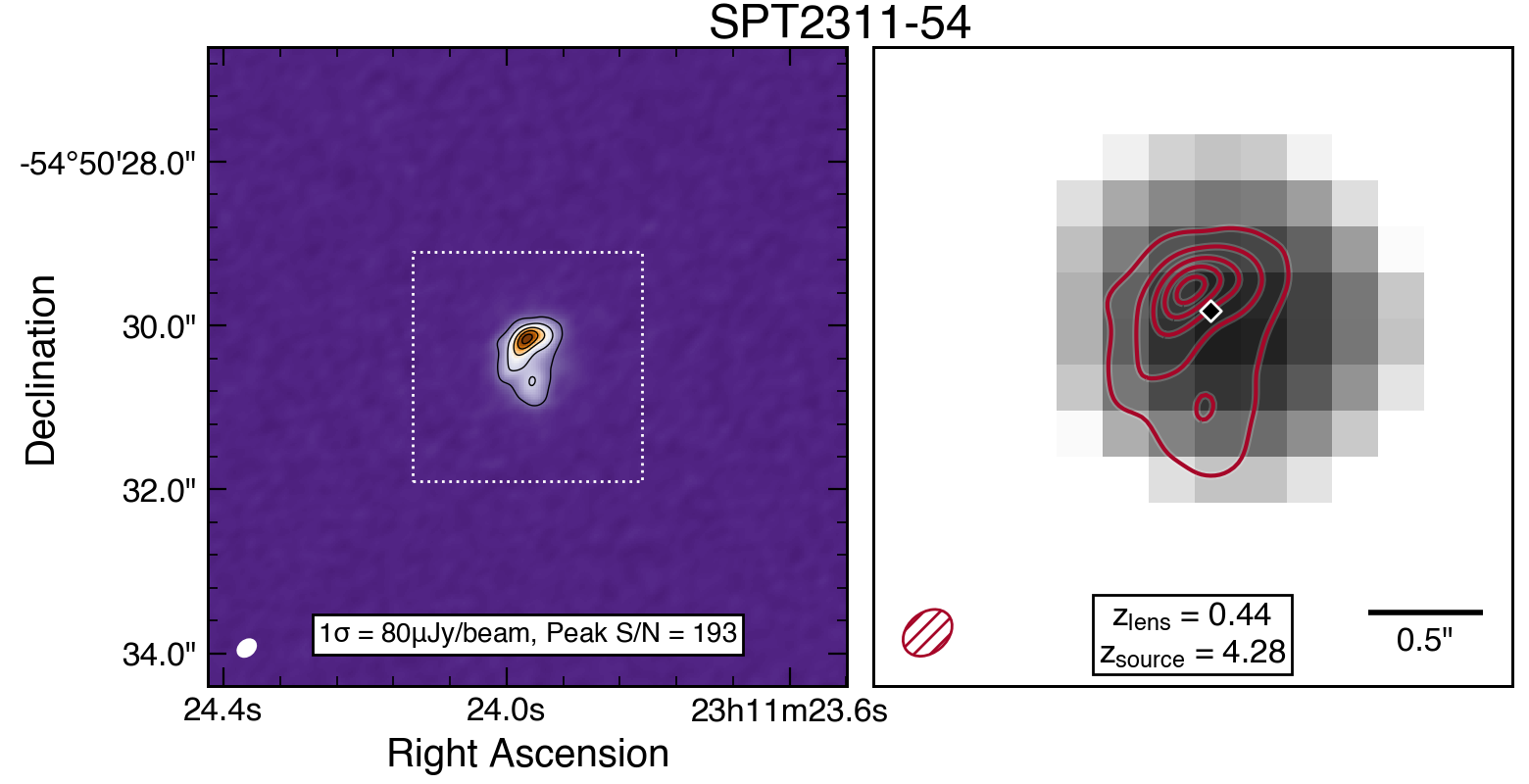}\\
\end{centering}
\caption{
\textit{Left:} Rest-frame 119\,\um ALMA continuum images of each sample object. Contours are drawn at 10, 30, 50, 70, and 90\% of the peak. Squares mark the zoomed regions in the right panels.
\textit{Right:} ALMA continuum contours overlaid on the best-available near-IR images from several different facilities \citep{spilker16}, which detect only the foreground lensing galaxies. Diamonds mark the best-fit positions of the lens(es); we do not use these images to constrain the lens positions because the astrometry is more uncertain than the uncertainties on the positions. Ellipses at lower left show the synthesized beam.
}\label{fig:contimages}
\end{figure*}

We generate images of each target using natural weighting of the visibilities, which maximizes sensitivity at the expense of spatial resolution. Natural weighting is also the closest approximation to the visibility weighting used in our subsequent visibility-based lens modeling procedure (Section~\ref{lensing}). We followed standard imaging procedures, manually applying a clean mask over regions with clear, high S/N emission, and stopped the image cleaning at 5 times the image noise level. Continuum images were created combining all available data, while image cubes of the OH lines were created with channel resolutions varying from 50--150\,\kms in order to maximize the S/N. To extract integrated OH spectra, we performed aperture photometry within the region where the continuum is detected at $>$3$\sigma$. We also performed a similar procedure on image cubes created by tapering the visibilities to resolutions $\sim$2--3 times lower than the full data and find no evidence that significant flux has been resolved out in our observations. Continuum images of each source are shown in Figure~\ref{fig:contimages} and integrated spectra in Figure~\ref{fig:OHspectra}.

\begin{deluxetable*}{lllhllcccc}
\tabletypesize{\small}
\tablecaption{Summary of ALMA Observations \label{tab:almaobs}}
\tablehead{
\colhead{Source} & 
\colhead{R.A.} & 
\colhead{Dec.} & 
\nocolhead{$z_{\mathrm{source}}$} & 
\colhead{$\nu_{\mathrm{obs}}$} &
\colhead{Program ID} & 
\colhead{t$_\mathrm{obs}$} & 
\colhead{Beam Size} & 
\colhead{$\sigma_{\mathrm{cont}}$} &
\colhead{$\sigma_{\mathrm{100\kms}}$} \\
\colhead{} & \colhead{} & \colhead{} & \nocolhead{} & \colhead{GHz} & \colhead{} & \colhead{min} & \colhead{arcsec} & \colhead{$\mu$Jy/beam} & \colhead{mJy/beam}
}
\startdata
SPT0202-61 & $02^\mathrm{h}02^\mathrm{m}58.86^\mathrm{s}$ & $-61^\circ21{}^\prime11.1\arc$ & 5.0180 & 417.8 & 2016.1.00089.S       & 43 & 0.37$\times$0.45 &  64 & 0.41 \\
SPT0418-47 & $04^\mathrm{h}18^\mathrm{m}39.67^\mathrm{s}$ & $-47^\circ51{}^\prime52.7\arc$ & 4.2248 & 481.2 & 2015.1.00942.S$^{a}$ & 12 & 0.11$\times$0.16 & 250 & 1.28 \\
           &                                              &                                &        &       & 2018.1.00191.S       & 87 & 0.32$\times$0.46 & 110 & 0.67 \\
SPT0441-46 & $04^\mathrm{h}41^\mathrm{m}44.08^\mathrm{s}$ & $-46^\circ05{}^\prime25.5\arc$ & 4.4770 & 459.1 & 2015.1.00942.S       & 11 & 0.14$\times$0.18 & 352 & 2.10 \\
           &                                              &                                &        &       & 2016.1.00089.S       & 17 & 0.25$\times$0.33 & 205 & 0.89 \\
           &                                              &                                &        &       & Combined             & 28 & 0.22$\times$0.30 & 178 & 0.84 \\
SPT0459-58 & $04^\mathrm{h}58^\mathrm{m}59.80^\mathrm{s}$ & $-58^\circ05{}^\prime14.3\arc$ & 4.8560 & 429.4 & 2019.1.00253.S       & 40 & 0.33$\times$0.40 &  82 & 0.69 \\
SPT0459-59 & $04^\mathrm{h}59^\mathrm{m}12.33^\mathrm{s}$ & $-59^\circ42{}^\prime20.6\arc$ & 4.7993 & 433.6 & 2019.1.00253.S       & 41 & 0.33$\times$0.41 & 120 & 0.56 \\
SPT0544-40 & $05^\mathrm{h}44^\mathrm{m}00.80^\mathrm{s}$ & $-40^\circ36{}^\prime31.1\arc$ & 4.2692 & 477.2 & 2019.1.00253.S       & 32 & 0.28$\times$0.31 & 103 & 1.00 \\
SPT2048-55 & $20^\mathrm{h}48^\mathrm{m}22.86^\mathrm{s}$ & $-55^\circ20{}^\prime21.3\arc$ & 4.0923 & 493.7 & 2018.1.00191.S       & 48 & 0.37$\times$0.44 & 118 & 0.72 \\
SPT2103-60 & $21^\mathrm{h}03^\mathrm{m}30.85^\mathrm{s}$ & $-60^\circ32{}^\prime40.5\arc$ & 4.4357 & 462.6 & 2016.1.00089.S       & 23 & 0.46$\times$0.49 & 238 & 1.10 \\
SPT2132-58 & $21^\mathrm{h}32^\mathrm{m}43.23^\mathrm{s}$ & $-58^\circ02{}^\prime46.2\arc$ & 4.7677 & 435.9 & 2015.1.00942.S       & 17 & 0.32$\times$0.50 & 345 & 1.64 \\
SPT2311-54 & $23^\mathrm{h}11^\mathrm{m}23.97^\mathrm{s}$ & $-54^\circ50{}^\prime30.2\arc$ & 4.2795 & 476.2 & 2015.1.00942.S       & 45 & 0.15$\times$0.20 & 157 & 1.27 \\
           &                                              &                                &        &       & 2018.1.00191.S       & 49 & 0.28$\times$0.37 &  91 & 0.92 \\
           &                                              &                                &        &       & Combined             & 94 & 0.23$\times$0.30 &  80 & 0.75 \\
\tableline
SPT2319-55$^{b}$ & $23^\mathrm{h}19^\mathrm{m}21.67^\mathrm{s}$ & $-55^\circ57{}^\prime57.8\arc$ & 5.2950 & 399.4 & 2016.1.00089.S & 30 & 0.27$\times$0.39 &  71 & 0.52 \\
\enddata
 \tablenotetext{a}{SPT0418-47 was observed in 2015.1.00942.S at much higher spatial resolution than requested. Given the large extent of the source and the short observing duration, these data are too low S/N to be usable, and are excluded from all figures and models. We include the observations in this table for completeness.}
 \tablenotetext{b}{Reproduced from \citet{spilker18a}.}
\tablecomments{All beam sizes and sensitivities are measured from naturally-weighted images. The spectral line sensitivity $\sigma_{\mathrm{100\kms}}$ is measured in a 100\,\kms channel near the upper OH rest frequency.}
\end{deluxetable*}

\subsection{Ancillary Data} \label{ancillary}

In addition to the ALMA OH observations that are our primary focus, we also use a variety of ancillary photometric and spectroscopic data to aid in the interpretation of the OH data.

The systemic redshift and line profile of gas within each galaxy is key to our interpretation of the OH spectra. Where available (5 sources), we use very high signal-to-noise ALMA \cii 158\,\um spectra, observed in program 2016.1.01499.S (see \citealt{litke19} for a representative object from this sample). While these data are fairly high spatial resolution, $\sim$0.3\arc, we use only the integrated \cii line profile extracted similarly to the OH spectra. For those sources without high-quality \cii data, we instead stack the spectra of all available transitions of CO for each source, weighted by the S/N of each line. These CO lines were observed with ALMA and the Australia Telescope Compact Array (ATCA), and were the primary features used to measure the redshift of each source. The CO lines include CO(2--1) and CO(5--4) for all sources and CO(4--3) for $z < 4.47$, each typically detected at S/N\,$\sim$\,5--10. For the sources with ALMA \cii spectra, we find no evidence for a difference in line width compared to the lower S/N CO lines.

We measure the IR (8--1000\,\um) and far-IR (40-120\,\um) luminosities by fitting to the available far-IR and submm photometry \citep{weiss13,strandet16,reuter20}, correcting for the lensing magnifications of each source as described further in Section~\ref{lensing}. For all sources, the available data include \textit{Herschel}/PACS and SPIRE data at 100, 160, 250, 350, and 500\,\um, APEX/LABOCA 870\,\um, the SPT 1.4, 2, and 3\,mm photometry, and ALMA 870\,\um and 3\,mm data. We find consistent results for the luminosities between simple modified blackbody fits and more complex modeling because the far-IR SED is very well sampled. We also make use of the PACS photometry at 100 and 160\,\um, which probes rest-frame mid-IR wavelengths $\sim15-30$\,\um, to constrain the contribution of hot dust heated by AGN activity.  We use the CIGALE SED fitting code \citep{burgarella05,boquien19} to place limits on the fractional AGN contribution to the total luminosity integrated over rest-frame 5--1000\,\um \fagn; no source shows strong evidence for AGN-related mid-IR emission. For the low-redshift comparison samples (Section~\ref{lowzcomp}) it is more common to measure \fagn using the rest-frame 30\um/15\um flux ratio assuming fixed mid-IR flux ratios for pure star formation and pure AGN emission. While we prefer the CIGALE fitting values for easier comparison with typical practice in the extragalactic literature, we have verified that we recover \fagn to within $\approx$0.2--0.3 using the mid-IR color definition. We note that it is possible that our target galaxies are optically thick at mid-IR wavelengths, which could hide very highly obscured AGN and result in less strict limits on \fagn than we adopt here \citep[e.g.][]{snyder13}.

Molecular gas masses were measured from observations of CO(2--1) using ATCA. Five of the 11 sources in our sample were published in \citet{aravena16}, while the remainder have been observed and analyzed using the same procedures and will be published elsewhere. We assume a line brightness temperature ratio of 0.9 between the CO(2--1) and (1--0) transitions and a CO--H$_2$ conversion factor $\alphaco = 0.8$\,\Msol (K\,\kms pc$^2$)$^{-1}$, both typical for highly star-forming DSFGs like our sample \citep{spilker14,spilker15,aravena16}.

\begin{deluxetable*}{lcccccccc}
\tabletypesize{\small}
\tablecaption{Summary of Sample Properties \label{tab:sample}}
\tablehead{
\colhead{Source} &
\colhead{$z_\mathrm{lens}$} & 
\colhead{$z_\mathrm{source}$} &
\colhead{$\mu$} &
\colhead{$\lir$ $(10^{12}\Lsol)$} & 
\colhead{$\lfir$ $(10^{12}\Lsol)$} &
\colhead{\fagn} &
\colhead{$\MHt$ $(10^{9}\Msol)$} & 
\colhead{$r_\mathrm{cont}$ (kpc)}
}
\startdata
SPT0202-61 &  ... & 5.0180 & 17.5 &  9.6$\pm$1.5 &  4.6$\pm$0.6 & $<$0.25 & 25.1$\pm$3.5 & 0.72 \\
SPT0418-47 & 0.26 & 4.2248 & 37.2 &  3.0$\pm$0.5 &  1.7$\pm$0.2 & $<$0.1  &  6.0$\pm$0.5 & 0.74 \\
SPT0441-46 & 0.88 & 4.4770 & 11.5 &  6.1$\pm$1.3 &  3.5$\pm$0.6 & $<$0.15 & 12.3$\pm$2.0 & 0.53 \\
SPT0459-58 &  ... & 4.8560 &  7.3 &  8.1$\pm$2.0 &  4.5$\pm$0.8 & $<$0.2  & 27.4$\pm$3.3 & 1.22 \\
SPT0459-59$^a$& 0.94& 4.7993& 3.1 & 18.1$\pm$5.7 &  9.9$\pm$2.0 & $<$0.3  & 79.9$\pm$7.0 & 3.99 \\
SPT0544-40 &  ... & 4.2692 & 10.5 &  7.3$\pm$1.1 &  4.3$\pm$0.6 & $<$0.2  & 46.6$\pm$4.3 & 0.69 \\
SPT2048-55 &  ... & 4.0923 & 10.8 &  4.5$\pm$1.2 &  2.6$\pm$0.5 & $<$0.2  & 16.0$\pm$2.6 & 0.67 \\
SPT2103-60 & 0.76 & 4.4357 & 20.9 &  2.9$\pm$0.5 &  1.7$\pm$0.3 & $<$0.15 &  9.8$\pm$1.7 & 1.02 \\
SPT2132-58 &  ... & 4.7677 &  5.7 & 11.3$\pm$4.3 &  6.2$\pm$1.4 & $<$0.25 & 27.6$\pm$2.6 & 0.78 \\
SPT2311-54 & 0.44 & 4.2795 &  2.5 & 29.8$\pm$7.9 & 16.2$\pm$2.9 & $<$0.45 & 63.5$\pm$4.9 & 1.08 \\
\tableline
SPT2319-55$^b$& 0.91& 5.2943& 5.8 &  7.9$\pm$3.0 &  4.3$\pm$0.8 & $<$0.3  & 11.8$\pm$2.1 & 0.92 \\
\enddata
 \tablenotetext{a}{Excludes the faint source southwest of the lensed source (see Fig.~\ref{fig:contimages}).}
 \tablenotetext{b}{Reproduced from \citet{spilker18a}.}
 \tablecomments{\lir and \lfir are integrated over rest-frame 8-1000 and 40-120\,\um, respectively. All values have been corrected for the lensing magnification $\mu$; we estimate uncertainties of $\sim$15\% on the magnification. \fagn is the fractional contribution of AGN-heated dust to the rest-frame 5--1000\,\um luminosity; upper limits are 1$\sigma$. \MHt from \citet{aravena16} and Aravena \etal in prep. using updated magnifications from this work. Intrinsic dust continuum sizes $r_\mathrm{cont}$ are circularized radii of the regions where the continuum is detected at S/N$>$5 in the lensing reconstructions. This table is available in machine-readable format at \url{https://github.com/spt-smg/publicdata}.}
\end{deluxetable*}

\subsection{Literature Comparison Samples} \label{lowzcomp}

Throughout this work we compare to a number of studies of OH absorption in low-redshift galaxies performed by the \textit{Herschel}/PACS instrument. While detailed sensitivity metrics (e.g., the typical fractional contrast compared to the continuum reached per spectral element) are generally not given, the published spectra of these sources appear to be of broadly similar quality to our own. A brief description of these samples follows, and a few relevant quantities are summarized in Figure~\ref{fig:sample}.

\citet{veilleux13} present OH spectra of 43 nearby galaxy mergers, mainly consisting of ultra-luminous infrared galaxies (ULIRGs) and IR-luminous QSOs. The sample supplements 23 galaxies observed as part of the SHINING key program \citep{sturm11} with 15 additional sources selected to have higher values of \fagn and a further 5 chosen to be less IR-luminous than the full sample. \citet{spoon13} present an analogous sample of 24 ULIRGs from the HERUS program \citep{farrah13} that largely overlap in source properties. \citet{calderon16} and \citet{herreracamus20} further expand on the aforementioned samples, presenting a combined total of 9 ULIRGs at slightly higher redshifts ($z\sim0.25$) and correspondingly higher typical luminosity ($\log \lir/\Lsol \sim 12.5-13.5$) to ensure sufficiently bright continuum fluxes. We collectively refer to these samples as `ULIRGs and QSOs' throughout this work.

\citet{stone16} present OH observations of a sample of 52 nearby hard X-ray selected AGN. These sources are the least similar to the other samples or to the $z>4$ SPT DSFGs, consisting of objects more than an order of magnitude less luminous ($\lir \lesssim 10^{11}$\,\Lsol) than those in the other samples but whose bolometric luminosities are dominated by the AGN power (typically $\fagn > 0.8$). Over half of this sample shows OH only in emission, a common feature among AGN-dominated systems, and only 17 show OH in absorption, which Stone \etal use to define their outflow detection rate.

For these low-redshift samples, we adopt the sample properties as published by the original authors. Although for maximal consistency we would re-measure, for example, \lir and \fagn using the same methods as our own sample, we prefer the literature values because they are publicly and centrally available. We spot-checked a few sources from the various samples using far-IR photometry from the literature and found results consistent with the published values.

A handful of additional high-redshift objects besides our sample have published OH 119\,\um spectra. \citet{george14} present OH data for SMM~J2135-0102 (the `Eyelash' DSFG) at $z=2.3$. Those authors argue that the OH absorption is associated with one of several spatio-kinematic `clumps' seen in the dust continuum emission in early interferometric data, and is outflowing with respect to that clump. Those clumps have recently been shown to be false \citep{ivison20}, so it is unclear how to interpret the OH spectrum in light of the new understanding of the source structure. By our adopted definition (Section~\ref{whatoutflow}) this source would not be classified as an outflow because the absorption is fully contained within the cores of the bright CO and \cii emission lines. Additionally, \citet{herreracamus20} present ALMA spectroscopy of a $z=6.1$ quasar with a tentative $3\sigma$ detection of blueshifted OH absorption. Given the tentative nature of this detection we also consider this case to be inconclusive. Finally, \citet{zhang18a} detect OH 119\,\um in a stack of 45 lensed DSFGs at $1<z<3.6$ and $\sim$3$\sigma$ detections in two individual objects, but the spectral resolution was too low to measure any velocity shifts or search for blueshifted line wings.

\section{Analysis} \label{analysis}

\subsection{Spectral Analysis} \label{specfitting}

\begin{figure*}[t]
\begin{centering}
\includegraphics[width=0.245\textwidth]{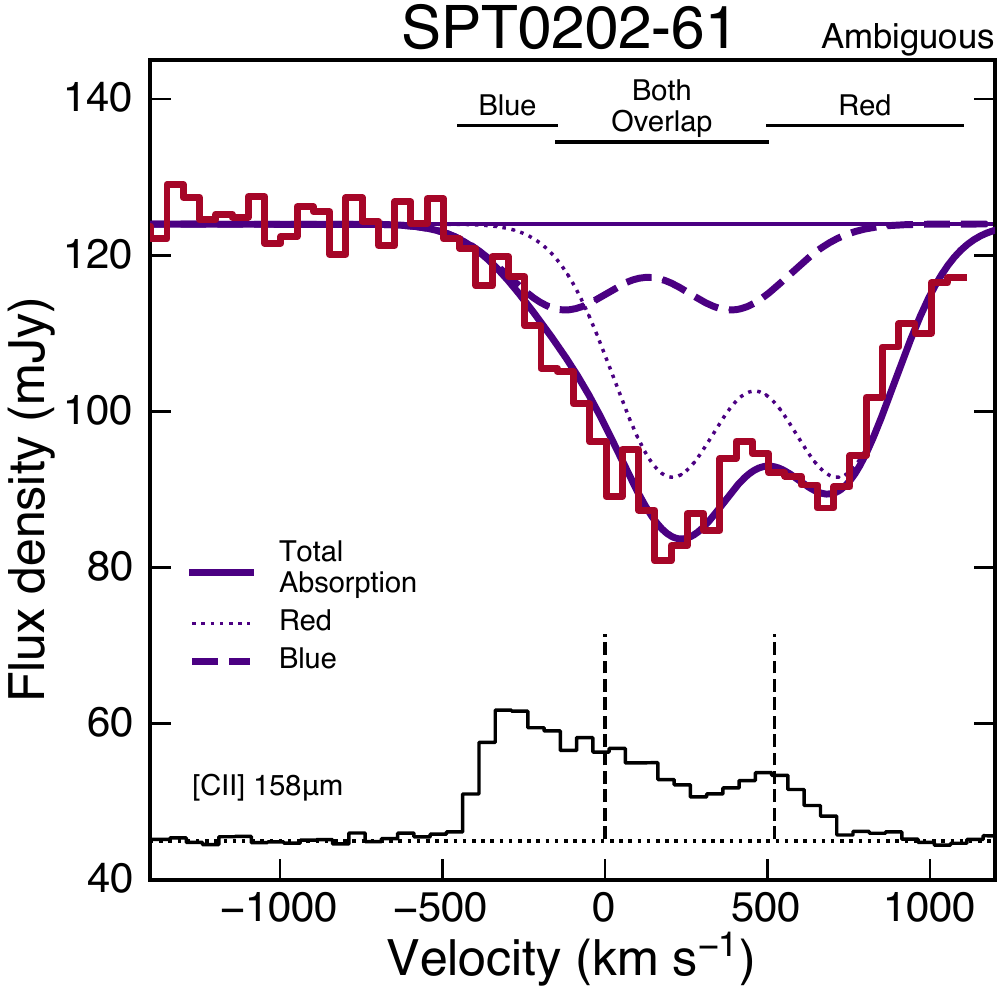}
\includegraphics[width=0.245\textwidth]{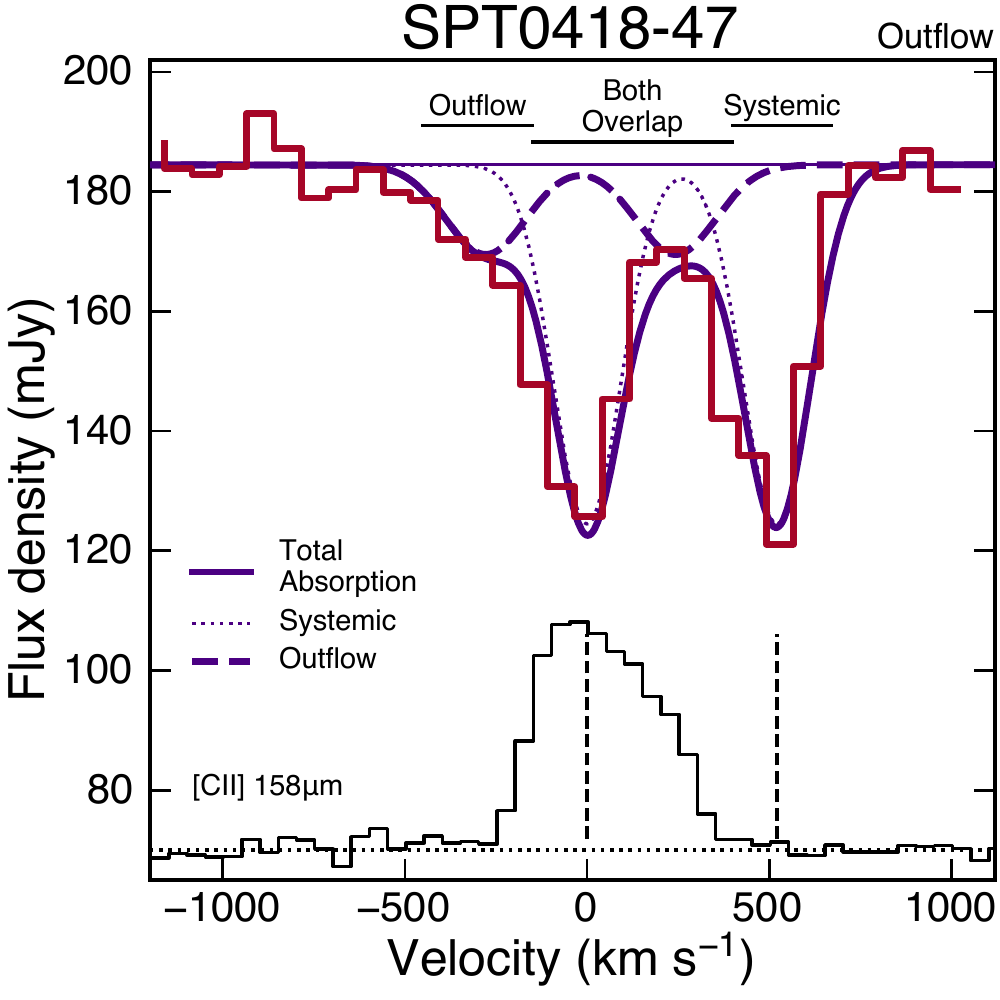}
\includegraphics[width=0.245\textwidth]{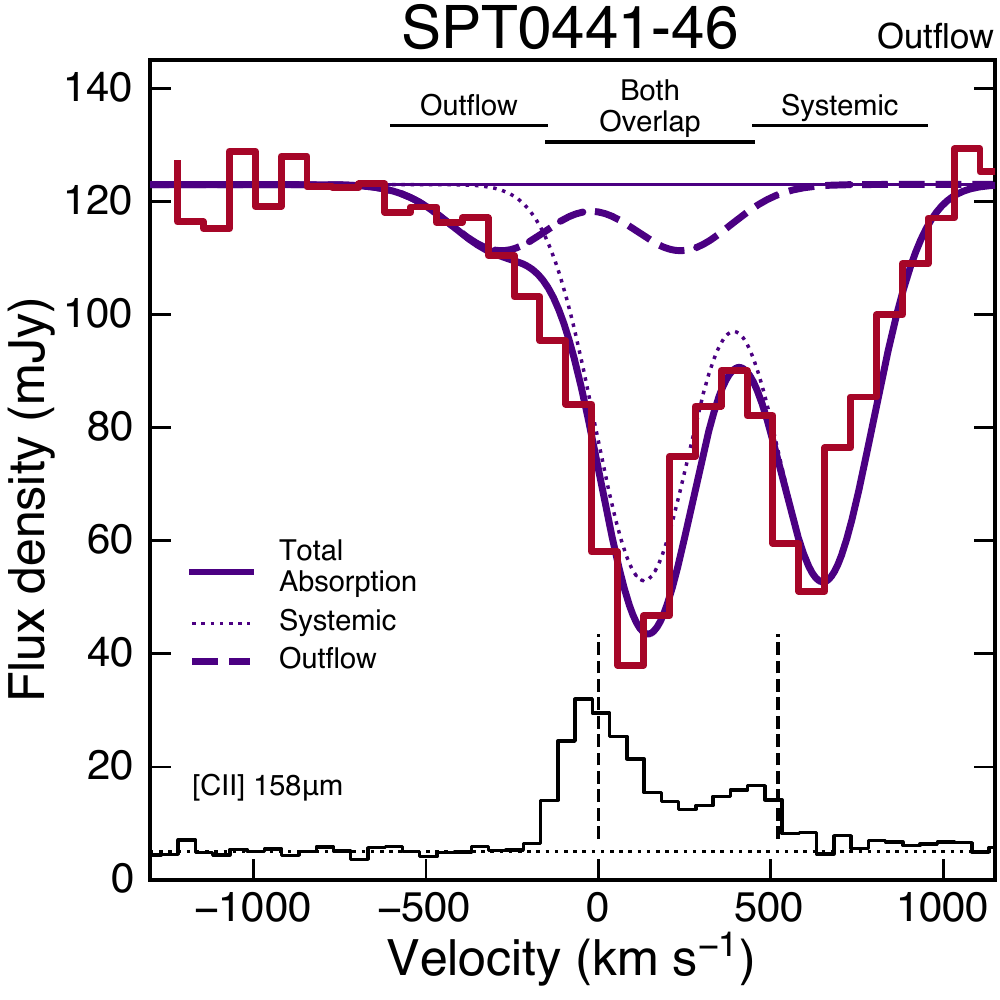}
\includegraphics[width=0.245\textwidth]{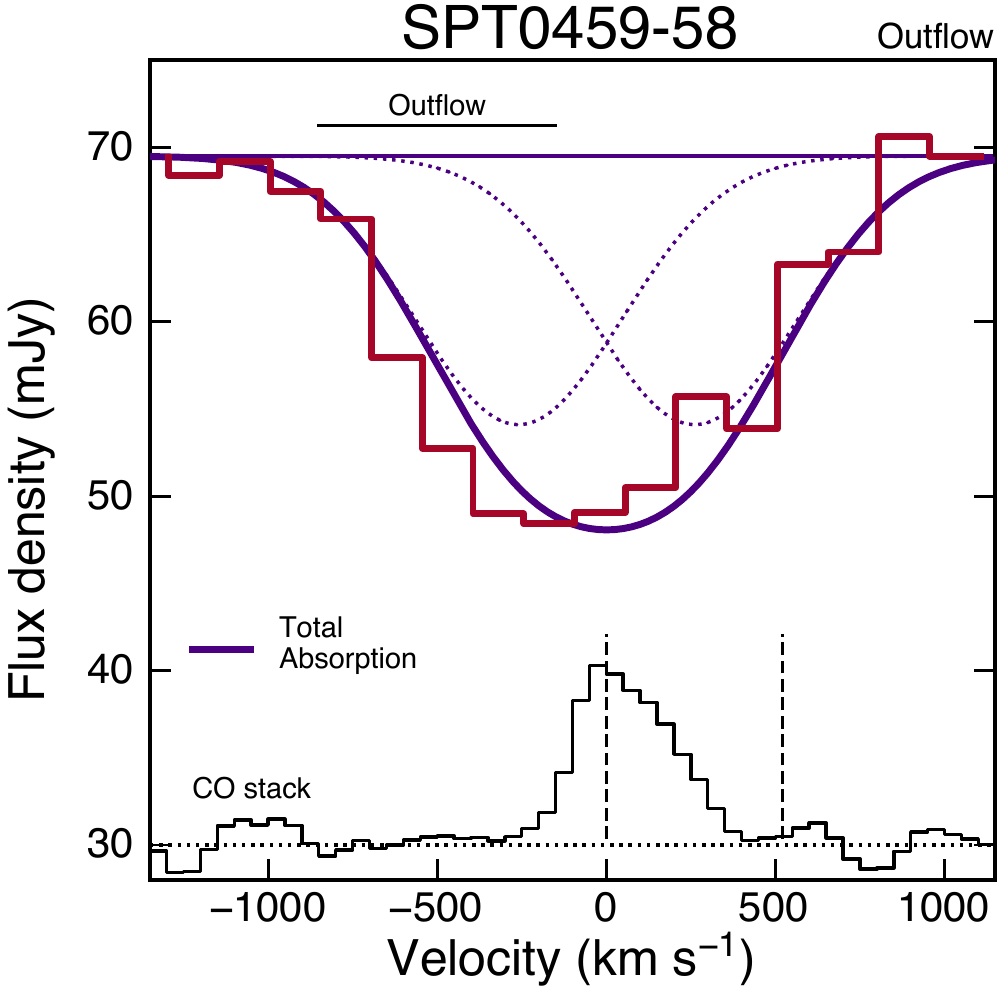}
\includegraphics[width=0.245\textwidth]{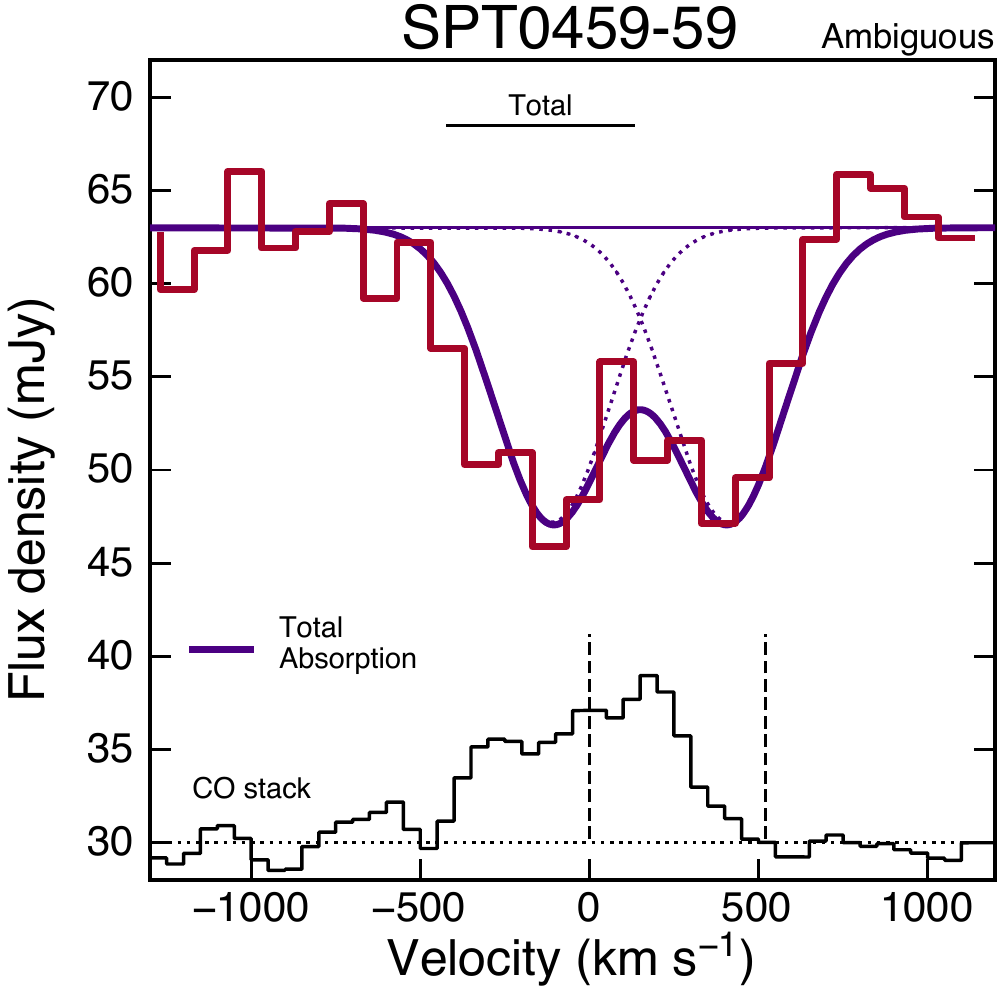}
\includegraphics[width=0.245\textwidth]{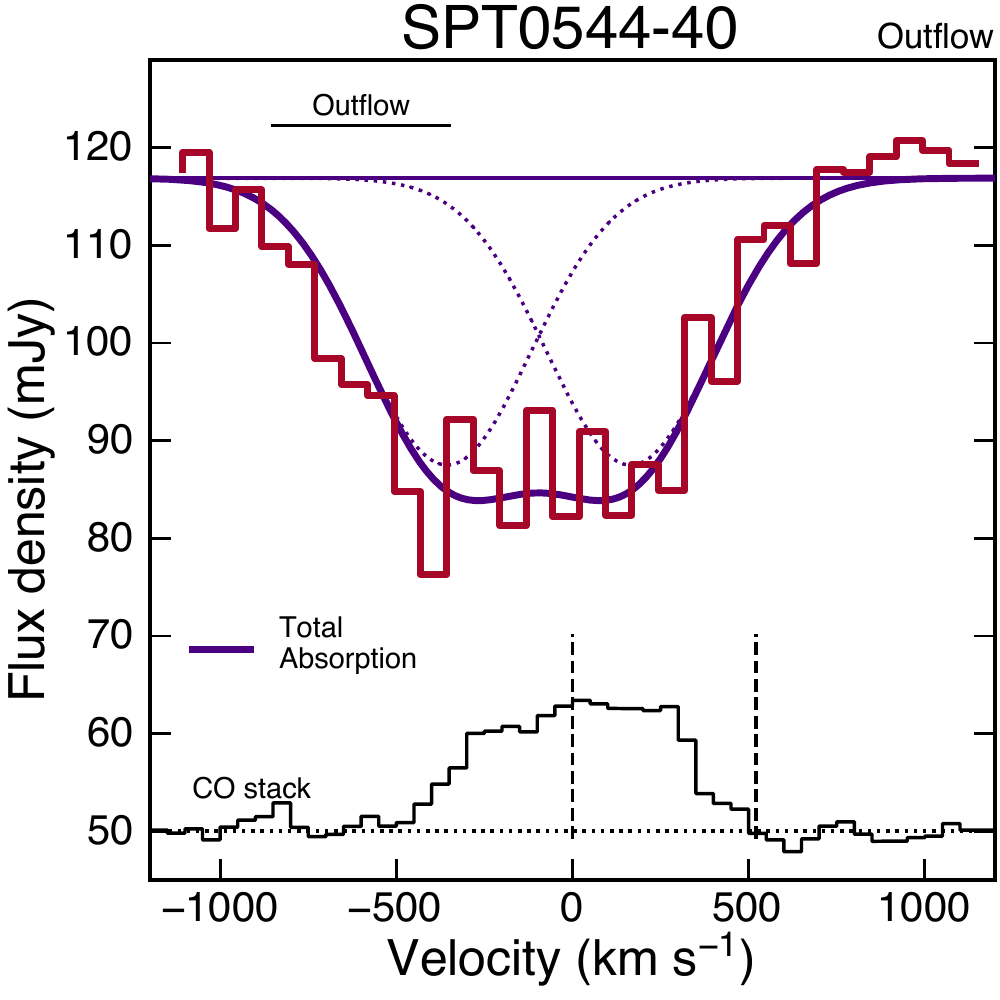}
\includegraphics[width=0.245\textwidth]{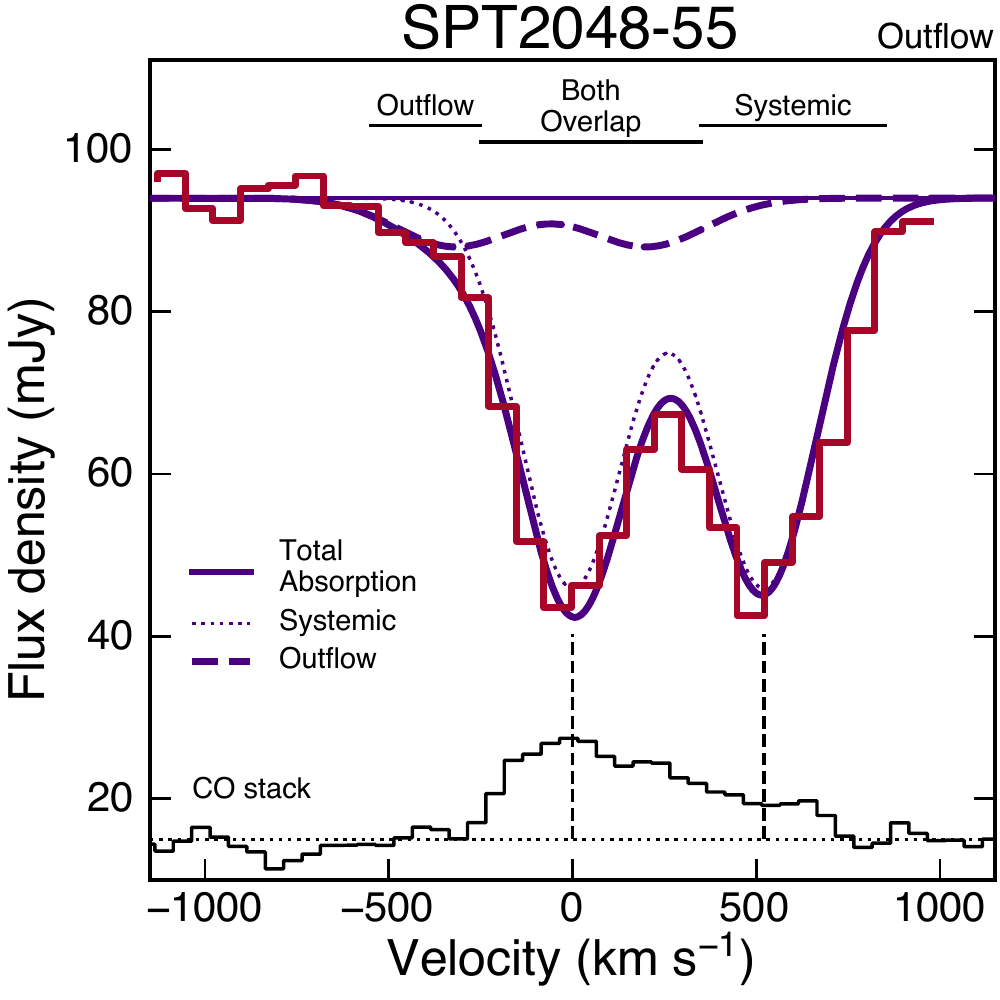}
\includegraphics[width=0.245\textwidth]{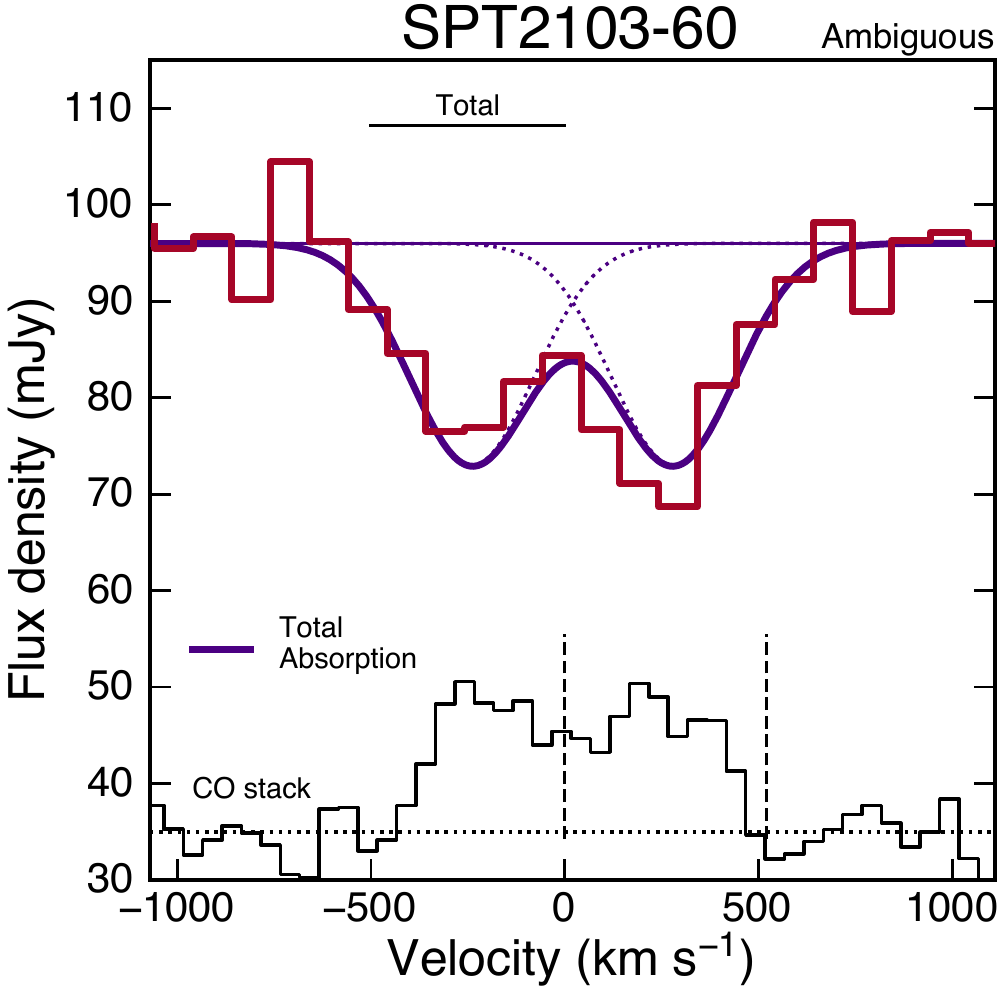}
\includegraphics[width=0.245\textwidth]{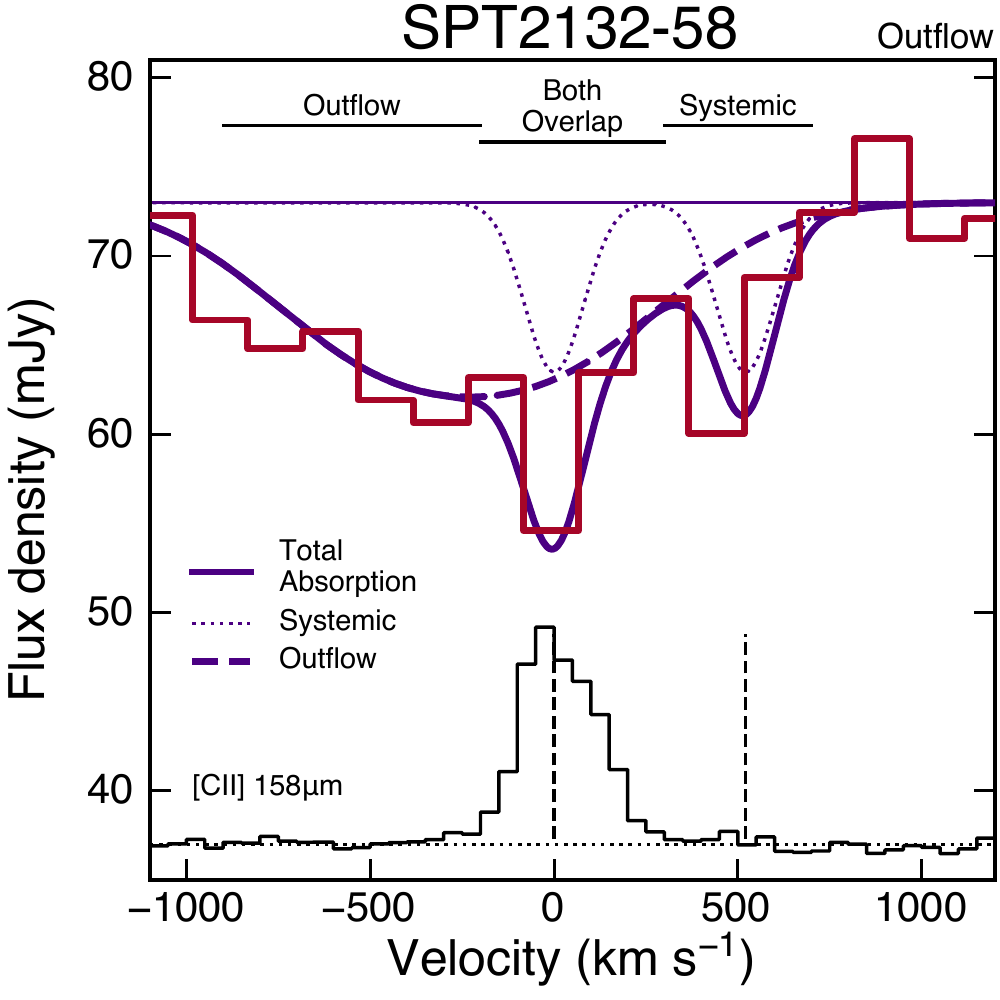}
\includegraphics[width=0.245\textwidth]{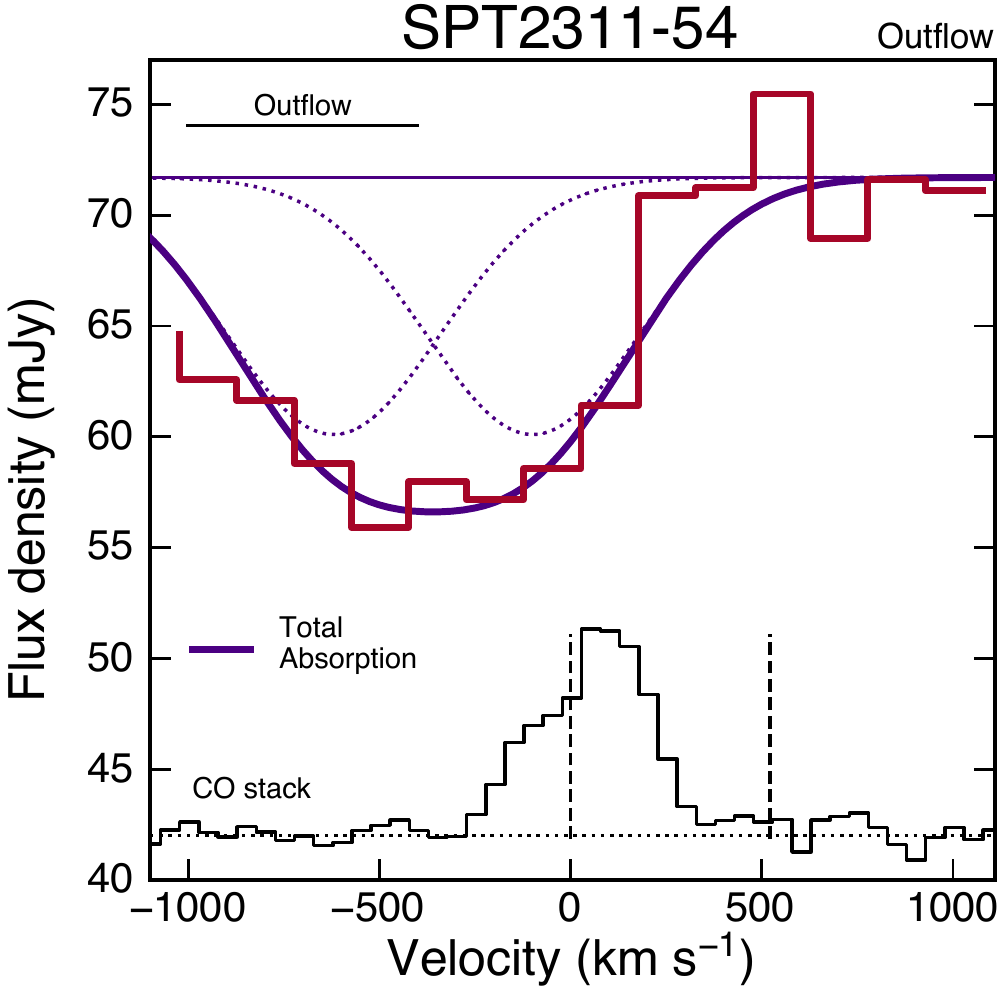}\\
\end{centering}
\caption{
OH 119\,\um spectra of each sample target (red), not corrected for lensing magnification. Also overplotted are fits to the spectra using either one or two pairs of Gaussians (navy lines), as detailed in Section~\ref{specfitting}. Vertical dashed lines show the rest velocities of the two OH doublet components, where we assign the higher-frequency transition to zero relative velocity. To help interpret the complex OH doublet spectra, we also show a `reference' line profile of \cii or CO expected to be dominated by gas internal to the galaxies. Horizontal bars at top label the velocity ranges we selected for lens modeling, chosen to be dominated by single velocity components (we also label the velocity ranges where multiple velocity components overlap, where applicable). There is clearly a large diversity of line profile shapes, but at least 7/10 of these sources host molecular outflows as defined by absorbing components more blueshifted than the reference line emission. The remaining 3 sources typically show broad reference line profiles with multiple peaks themselves, making interpretation of the OH line profiles difficult.
}\label{fig:OHspectra}
\end{figure*}

While OH is clearly detected in absorption toward each source, the integrated spectra in Figure~\ref{fig:OHspectra} show a diverse range of absorption depths and line profiles. In about half the sample the OH doublet lines are sufficiently narrow to be individually resolved, while in the rest they are wide enough to be blended, creating a single wide absorption profile. Some sources show obvious signs of multiple velocity components contributing to the overall line profile (e.g. SPT0418-47), while others are adequately fit by a single velocity component (e.g. SPT0459-58).

To help interpret these line profiles, we fit the integrated spectra with one or two pairs of Gaussians depending on the complexity of the line profile for each source. These fits are by no means unique, but do capture the information available in the spectra. As a $\Lambda$ doublet, the two 119\,\um OH lines are expected to have equal amplitude, and have a fixed separation in velocity of $\approx$520\,\kms. The free parameters are the continuum flux density and either one or two of each of the absorption depth, the velocity offset relative to the systemic, and Gaussian line width. 

We assume a constant continuum flux density across the relatively narrow bandwidth of these observations.  In a handful of sources there are few absorption-free channels to constrain the continuum level (e.g. SPT0544-40), or there is blueshifted absorption that extends out to and beyond the edge of the band (e.g. SPT2311-54). For these sources we also make use of the continuum data in the alternate sideband of the ALMA data and/or a global fit to the long-wavelength SED. Rest-frame 119\,\um is near the peak of the dust SED where the slope is nearly flat, and we estimate that taking the continuum level from the alternate sideband introduces at most a few percent uncertainty in the OH sideband continuum level due to the sideband wavelength separation. We tested this procedure using the sources with sufficient line-free bandwidth in both sidebands and find no appreciable differences in the OH fit parameters.

We derive several other parameters from the best-fit Gaussian profiles. The OH equivalent width is straightforward to derive from the fitting results. We also report several velocity-related quantities to facilitate comparison with literature samples, including \vfifty and \vef, the velocities above which 50\% or 84\% of the absorption takes place, and \vmax, the estimated terminal outflow velocity. As in \citetalias{spilker18a}, here we take \vmax to be the velocity above which 98\% of the absorption occurs. Various definitions of \vmax have been used in the literature and its value depends on the signal-to-noise of the data (and, for SPT2311-54, the assumption that the absorption continues to follow a Gaussian profile beyond the edge of the ALMA bandwidth). We note that these fit parameters are largely immune to gravitational lensing, since both the continuum and absorption must be magnified by nearly the same factor. The line profile best-fit and derived parameters are given in Tables~\ref{tab:absfits} and \ref{tab:absderived}. 

Finally, we use the fits to the OH line profiles and the ancillary spectral data to define velocity ranges dominated by either outflowing or systemic absorption that we later use in our lensing reconstructions to measure the structure of the absorbing components. The wide variety of OH line profiles makes it difficult to define these ranges with unambiguous criteria, and we have no way to separate low-velocity outflowing material from high-velocity systemic material. For sources with multiple velocity components, we take the systemic and outflowing components to be those velocity ranges where each component dominates the total absorption profile. For sources with single broad absorption profiles, we use the fits to try to avoid double-counting gas at the same source-frame velocity due to the $\Lambda$ doublet splitting. In both cases, we prefer to define the outflows as beginning at the velocity where we no longer detect CO or \cii emission, but this is not always possible given the relatively weak absorption seen in some sources. In total, we are confident that the velocity ranges we select are at least dominated by outflowing or systemic gas, although in some cases not exclusively so.

\begin{deluxetable*}{lclcccccc}
\tabletypesize{\small}
\tablecaption{Spectral fitting results and lens model velocity ranges\label{tab:absfits}}
\tablehead{
\colhead{Source} &
\colhead{$S_{119\um}$} & 
\colhead{Component} &
\colhead{$v_{\mathrm{cen}}$} &
\colhead{$S_{\mathrm{abs}}$} &
\colhead{FWHM} &
\colhead{Eq. Width} &
\colhead{Model $v_{\mathrm{cen}}$} &
\colhead{Model $\Delta$$v$} \\
\colhead{} & \colhead{mJy} & \colhead{} & \colhead{\kms} & \colhead{mJy} & \colhead{\kms} & \colhead{\kms} & \colhead{\kms} & \colhead{\kms}
}
\startdata
SPT0202-61     & $124.4 \pm 0.6$ & Red      & $+200 \pm 30$ & $-31.7 \pm 2.9$ & $410 \pm  50$ & $120 \pm  9$ & $+$800 & 600 \\
               &                 & Blue     & $-130 \pm 80$ & $-10.9 \pm 2.8$ & $400 \pm 120$ &  $34 \pm  9$ & $-$300 & 300 \\
SPT0418-47     & $184.4 \pm 1.2$ & Systemic & 0             & $-58.8 \pm 2.7$ & $220 \pm  20$ &  $83 \pm  8$ & $+$535 & 270 \\
               &                 & Outflow  & $-280 \pm 30$ & $-14.8 \pm 2.9$ & $260 \pm  90$ &  $18 \pm  5$ & $-$300 & 300 \\
SPT0441-46     & $123.0 \pm 1.5$ & Systemic & $+130 \pm 10$ & $-68.0 \pm 2.8$ & $330 \pm  30$ & $197 \pm 20$ & $+$700 & 500 \\
               &                 & Outflow  & $-280 \pm 80$ & $-11.7 \pm 2.6$ & $350 \pm 160$ &  $38 \pm 10$ & $-$375 & 450 \\
SPT0459-58     & $69.6  \pm 1.1$ & Outflow  & $-260 \pm 20$ & $-15.4 \pm 1.3$ & $730 \pm 120$ & $176 \pm 22$ & $-$500 & 700 \\
SPT0459-59     & $63.0  \pm 0.6$ & Total    & $-110 \pm 20$ & $-15.8 \pm 1.3$ & $400 \pm  30$ & $107 \pm  8$ & $-$145 & 550 \\
SPT0544-40     & $116.9 \pm 1.5$ & Outflow  & $-360 \pm 20$ & $-29.4 \pm 2.5$ & $560 \pm  60$ & $165 \pm 27$ & $-$600 & 500 \\
SPT2048-55     & $94.0  \pm 1.3$ & Systemic & 0             & $-48.3 \pm 2.2$ & $340 \pm  20$ & $213 \pm  8$ & $+$600 & 500 \\
               &                 & Outflow  & $-320 \pm 50$ & $-6.1  \pm 3.2$ & $380 \pm 160$ &  $26 \pm  5$ & $-$400 & 300 \\
SPT2103-60     & $96.2  \pm 1.2$ & Total    & $-240 \pm 20$ & $-23.6 \pm 2.2$ & $380 \pm  30$ &  $97 \pm  7$ & $-$250 & 500 \\
SPT2132-58     & $73.1  \pm 1.9$ & Systemic & 0             & $-9.5  \pm 3.0$ & $190 \pm  40$ &  $19 \pm  7$ & $+$500 & 400 \\
               &                 & Outflow  & $-490 \pm 70$ & $-7.7  \pm 1.6$ & $750 \pm 320$ &  $99 \pm 23$ & $-$550 & 700 \\
SPT2311-54     & $71.7  \pm 1.2$ & Outflow  & $-620 \pm 40$ & $-11.6 \pm 1.9$ & $660 \pm 160$ & $118 \pm 12$ & $-$700 & 600 \\
\tableline
SPT2319-55$^a$ & $52.1  \pm 0.5$ & Systemic & 0             & $-7.8  \pm 1.2$ & $330 \pm  80$ &  $52 \pm 15$ & $+$500 & 400 \\
               &                 & Outflow  & $-440 \pm 50$ & $-7.0  \pm 1.2$ & $450 \pm  60$ &  $64 \pm 14$ & $-$450 & 500 \\
\enddata
 \tablenotetext{a}{Reproduced from \citet{spilker18a}.}
 \tablecomments{OH spectral component fits are labeled as in Fig.~\ref{fig:OHspectra}. Velocities are relative to the higher-frequency OH doublet transition. Systemic profiles centered on 0\,\kms were fixed to the systemic redshift of those sources. Equivalent widths are given for only one of the OH doublet transitions (i.e. should be multiplied by 2 for the total equivalent width). The final two columns give the center velocity and width that we use for lens modeling, selected to be dominated by each absorbing component (see Figure~\ref{fig:OHspectra}).}
\end{deluxetable*}

\begin{deluxetable}{lcCCC}
\tabletypesize{\scriptsize}
\tablecaption{OH absorption profile characteristics \label{tab:absderived}}
\tablehead{
\colhead{Source} &
\colhead{Outflow?} &
\colhead{\vfifty} &
\colhead{\vef} &
\colhead{\vmax} \\
\colhead{} & \colhead{} & \colhead{\kms} & \colhead{\kms} & \colhead{\kms}
}
\startdata
SPT0202-61 & --- & +135\pm10 & -115\pm25  & -370\pm55 \\
SPT0418-47 &  Y  &  -35\pm10 & -230\pm50  & -430\pm80 \\
SPT0441-46 &  Y  & +100\pm10 & -120\pm45  & -440\pm75 \\
SPT0459-58 &  Y  & -260\pm20 & -560\pm60  & -890\pm100 \\
SPT0459-59 & --- & -110\pm20 & -280\pm25  & -460\pm40 \\
SPT0544-40 &  Y  & -360\pm20 & -590\pm90  & -830\pm170 \\
SPT2048-55 &  Y  &  -25\pm10 & -210\pm30  & -480\pm85 \\
SPT2103-60 & --- & -240\pm20 & -400\pm20  & -570\pm30 \\
SPT2132-58 &  Y  & -360\pm60 & -750\pm120 & -1110\pm220 \\
SPT2311-54 &  Y  & -620\pm35 & -900\pm65  & -1200\pm105^a \\
SPT2319-55 &  Y  & -215\pm40 & -525\pm40  & -760\pm60  \\
\enddata
 \tablenotetext{a}{For SPT2311-54 the given \vmax is an extrapolation of our fit to the spectrum because the absorption profile continues beyond the end of the ALMA bandwidth.}
 \tablecomments{
See Section~\ref{whatoutflow} for our metrics for whether or not a given source shows a definite molecular outflow. 
The quantities \vfifty, \vef, and \vmax refer to the velocities above which 50, 84, and 98\% of the total absorption take place, see Section~\ref{specfitting}.}
\end{deluxetable}

\subsection{Molecular Outflow Classification} \label{whatoutflow}

The wide diversity in OH absorption profiles among our sample raises the obvious question of how to determine whether or not a particular source has a molecular outflow. Various definitions to answer this question have been used in the literature. Perhaps the most common method is to classify any source with $\vfifty < -50$\,\kms as an outflow, following \citet{rupke05} and subsequently adopted by several studies of OH in low-redshift galaxies with \textit{Herschel} \citep[e.g.][]{veilleux13,herreracamus20}. For a variety of reasons we find this definition unsatisfying for our sample. First, it is clear that many of our sources have very broad and/or double-peaked \cii or CO lines that are hundreds of \kms wide, which makes the redshift used to define systemic (zero) velocity somewhat arbitrary. In other words, the relevant metric is not whether the absorption appears to be blueshifted based on the assigned systemic velocity, but whether the absorption is blueshifted relative to the emission line profiles of the gas within the galaxies. Second, this definition ignores the fact that the total absorption profiles are often a superposition of a component at systemic velocities and a second blueshifted component with a typically much weaker absorption depth. There is no reason to expect that the presence or strength of absorption at systemic velocities has any bearing on whether or not an outflowing component is also present. Objects with extremely deep systemic absorption (e.g. SPT0441-46) have \vfifty biased by this very strong systemic absorption since \vfifty is measured from the total absorption profile. 

Instead, we define a source as containing an outflow if it shows OH absorption more blueshifted than the detected \cii or CO emission, which we expect to be a conservative definition. The benefit of this definition is that outflows defined this way are unambiguous -- no fitting technique or analysis method can change the fact that a source shows absorption blueshifted more than any gas in the host galaxy. This definition has the drawback of being signal-to-noise dependent, which may exclude weak outflows (although this is also true of outflow classification based on \vfifty). It may also exclude sources in which the outflow shows strong emission in \cii or CO. However, studies of local objects in multiple outflow tracers typically find that the high-velocity emission outside the line cores is indeed very weak \citep{lutz20} and we see no evidence for high-velocity wings of emission in any of our sources. In such cases additional information (such as high spatial resolution kinematics) would be needed to determine whether the bright emission is associated with gas in the host galaxy, a merging partner, or a bona fide outflow. With this classification, we consider 8/11 sources in our sample to show unambiguous signs of molecular outflow (detailed further in Section~\ref{detection}).

Compared to some literature classifications, our outflow detection rate could be considered a conservative limit. In particular at high redshift we would call the outflows in the $z=6.1$ quasar ULAS~J1319+0950 \citep{herreracamus20}, the $z=2.3$ DSFG SMM~J2135-0102 \citep{george14}, and the $z=5.7$ DSFG SPT0346-52 \citep{jones19} ambiguous cases instead of confirmed outflows. In the former case the OH spectrum is too low signal-to-noise to be confident in its classification (and was also noted as tentative by those authors), while in the latter two cases the absorption lines used to claim outflow are fully contained within the bright \cii line profile cores. In the case of SPT0346-52, \citet{litke19} suggest that the galaxy is actually a major merger based on modeling of the \cii data. The absorption is well aligned with one of the \cii line peaks and is instead most likely simply systemic absorption within one of the merging pair. This particular object highlights that in such cases high spatial resolution kinematics can clarify whether or not a given absorption profile is indicative of an outflow.

We note that using an alternative outflow definition, $\vfifty < -50$\,\kms, would still result in 7/11 sources being classified as showing outflows, with 5 sources exhibiting outflows both by this metric and our preferred definition. The three sources that we identify as showing outflows but with $\vfifty \ge -50$\,\kms, as expected, all have very strong absorption at systemic velocities (or even slightly redshifted from systemic), which results in a biased value of \vfifty. On the other hand, the two sources that would be classified as outflows based on the \vfifty criterion that we label as ambiguous are sources with broad CO emission, where the OH absorption profile is still fully contained within the bright emission from gas inside the galaxies. The OH could simply be absorption internal to the galaxies, and we are not confident enough to label them outflows despite their \vfifty values. This comparison highlights the value of having high-quality reference emission line spectra and the peril of accepting the results of OH spectral fitting in the absence of additional information.

\subsection{Lens Modeling Methods and Tests} \label{lensing}

We create gravitational lens models to reconstruct the intrinsic structure of each source utilizing the pixellated modeling code described in detail in \citet{hezaveh16}, as in \citetalias{spilker18a}. For each source we fit to the available continuum data, consisting of the data in the line-free sideband of the ALMA data. For SPT2132-58, which has no useful alternate sideband data due to atmospheric opacity, we instead simply use the full OH-containing sideband.  Once the best-fit parameters of the lensing potential have been determined following the above procedure, we then use these parameters to reconstruct the OH absorption components using the velocity ranges shown in Figure~\ref{fig:OHspectra} and listed in Table~\ref{tab:absfits}. In principle a joint fit to the continuum and absorption components would provide the optimal constraints on the lens model parameters, but this becomes computationally expensive due to the large number of visibilities.

Briefly, the code fits directly to the interferometric visibilities, which we average temporally unless doing so would cause a binned visibility to span more than 10m in the $uv$ plane. While the code also has the ability to marginalize over residual time-variable antenna-based phase calibration errors, we neglect this capability for computational efficiency. In any case, the phase self-calibration performed as part of the data reduction largely supplants the need for further control over the antenna phases. We fit for the lensing potential using a Markov Chain Monte Carlo (MCMC) sampling algorithm. In practice, we first use a code that represents the source plane with one or more simple parametric light profiles \citep{spilker16} to get a reliable estimate of the lens parameters before re-fitting with the pixellated code in order to minimize the number of MCMC iterations required for the chains to converge.

The lensing potential is described by one or more singular isothermal ellipsoid (SIE) mass profiles \citep[e.g.][]{kormann94}. Each SIE potential is described by two positional coordinates, a strength related to the lensing mass, and two orthogonal components of the lens ellipticity. Where necessary, we allow for additional angular structure in the lensing potential with external shear and low-order multipoles in the main lens (up to $m=4$), as parameterized in \citet{hezaveh16}. The best-fit lens model parameters are given in Appendix~\ref{applensing}.

The source plane is represented by a grid of pixels that is regularized by a linear gradient prior on the source, which minimizes pixel-to-pixel variations in the source plane in order to avoid over- or under-fitting the data \citep{warren03,suyu06,hezaveh16}. The best-fit strength of the regularization is determined by maximizing the Bayesian evidence given a fixed set of parameters for the lensing potential. Because the regularization strength is only fit for a fixed set of lens parameters, we perform an iterative process of fitting for the regularization strength, MCMC fitting for the lens parameters, and re-fitting for the regularization strength until all parameters have converged.

\subsubsection{Model Selection and Tests} \label{lenstests}

We generally begin each modeling process assuming a simple lens potential parameterization, adding complexity where necessary. Our models begin under the assumption that the lensing potential can be described adequately by a single SIE profile and an external shear component. If this model does not satisfactorily reproduce the data, we introduce additional complexity as needed. For SPT0459-58 and SPT2103-60, the morphology of the lensed images is clearly inconsistent with a simple SIE mass profile, in agreement with the near-IR imaging that shows multiple plausible lensing galaxies near the main lens. For these sources, we fit models using two (SPT0459-58) or three (SPT2103-60) lensing galaxies. SPT0441-46 also shows a second object $\sim$1\arc west of the main lens that may influence the lensing potential, but our models do not \textit{require} the inclusion of a second lens to reproduce the data.

We use several metrics to determine whether these simple models are sufficient to capture the information in the data or whether further complexity is warranted. We first compare the deviance information criterion of different models \citep{spiegelhalter02}, preferring the models with greater likelihood if the additional free parameters from additional lens potential complexity legitimately provide a better fit to the data. Second, in reality we know that the dust continuum source reconstructions should be positive, but nothing in our methodology forces positivity. If a given model yields a source-plane reconstruction with large negative `bowls', we take this as an indication that the parameterization of the overall lensing potential probably requires additional complexity to exclude an unphysical source reconstruction. In practice we flagged models where the peak negative pixels in the source reconstruction had an absolute value $\gtrsim$10\% of the peak positive pixels for additional scrutiny.

We perform extensive tests of the effective sensitivity and resolution of the source reconstructions. It is not straightforward to infer an effective source-plane resolution or sensitivity from the observed (image plane) data. For example, the effective resolution and sensitivity vary with location in the source plane based on the local lensing magnification, and the source regularization strength depends on both the resolution and sensitivity of the original data. This becomes even more complicated when considering absorption components because the detection of absorption requires the presence of continuum emission, but the continuum brightness, absorption depth, and effective resolution and sensitivity all vary across the source plane. Following \citetalias{spilker18a}, we perform a series of reconstructions of mock data to test the resolution and sensitivity of the source reconstructions. 

Briefly, we create many mock observations of pointlike background sources tiled across the source plane, analyzing this fake data identically to the real data. The intrinsic flux density of the artificial sources is set such that the total apparent (magnified) flux density matches that of the real sources (since these sources were selected in part based on their apparent brightness). We then fit the source reconstructions of each set of fake data with a two-dimensional Gaussian, taking the FWHM as an empirical estimate of the resolution of the reconstruction. We also measure the differences between the input and best-fit positions of the artificial sources, which are small in all cases except when the input source lies very near the lensing caustics with magnifications $\gtrsim$50. We associate these failed solutions with too-poor resolution in the source- and image-plane pixel grids, as the pixel sizes of these grids were not optimized for such extreme cases of high magnification and compact source sizes. Finally, we repeat this entire procedure but change the input source flux density to be a factor of 2--3 weaker than the faintest component in the real absorption reconstructions in order to test the ability to recover input fluxes fainter than those actually observed. This final test shows that even these weak signals are recoverable to $\approx$25\% accuracy.

In summary, these tests lead us to conclude that the structures seen in the source reconstructions are real and that absorption signals a few times weaker than those actually observed can be successfully recovered. We find no evidence that the lensing reconstruction procedure introduces artificial clumpy structure. The sources in our sample are resolved over $\approx$5--20 independent resolution elements, in agreement with the image-plane data (Figure~\ref{fig:contimages}). 

For the purposes of Figure~\ref{fig:recon} we illustrate the effective source-plane resolution with an ellipse based on the fits to the artificial data at the position corresponding to the peak of the actual reconstructed source continuum emission. Full maps of the effective resolution are provided in Appendix~\ref{applensing}.

\subsection{Notes on Individual Sources} \label{snowflakes}

Our sample shows very diverse characteristics both in terms of lensing geometry and OH 119\,\um line profiles, in some cases requiring special treatment. Here we give a brief summary of these particularities and comment on the conclusions we draw from the OH spectra.

\subsubsection{SPT0202-61}
The OH spectrum of this source clearly requires two velocity components to reproduce the data. The \cii profile of this source is also very broad and shows two peaks, likely indicating a major merger. The deepest OH absorption is redshifted compared to one of the peaks and blueshifted relative to the other. It is thus possible that we are seeing a molecular outflow (launched from the fainter \cii component) or a molecular \textit{inflow} (falling towards the brighter \cii peak), or simply strong systemic absorption from the interaction/overlap region. The \cii data will be analyzed in future work and a detailed comparison between the extended \cii and the OH absorption (confined to the continuum emitting region by definition) is complicated, but initial modeling does not conclusively point to an outflow. We thus consider this source ambiguous and do not claim a molecular outflow.

Additionally SPT0202-61, nearly uniquely among the SPT DSFG sample, shows submillimeter emission at the center of the Einstein ring of background source emission. This is most clearly visible in Figure~\ref{fig:contimages}. The available data make clear that the central emission is not a (demagnified) lensed image of the background source. There is an additional unlensed continuum source located $\approx$6.5'' southwest of the lensed source, also noted by \citet{spilker16}. These sources will be explored in more detail in future work. The pixellated reconstruction tool we employ does not have the ability to simultaneously model lensed and unlensed emission, so we model and subtract the lens and secondary source before lens modeling. The lens emission does not cleanly separate from the lensed background emission, so the subtraction is imperfect. To mitigate this, we carefully define the image-plane and source-plane pixel grids in the lensing code such that no source-plane pixels map to the center of the Einstein ring. We have verified that this choice has no impact on our reconstructions, apart from leaving residuals in the model-subtracted data from the imperfectly-subtracted lens emission.

\subsubsection{SPT0418-47}
This source shows a clear but not especially deep outflow component extending well beyond the high signal-to-noise \cii emission. We note that this source was observed in ALMA projects 2015.1.00942.S and 2018.1.00191.S. The 2015.1.00942.S data were taken with the array in a much more extended configuration than we originally requested, yielding a synthesized beam $\sim$0.15''. These data proved too shallow given their high spatial resolution, and we excluded them from all of our analysis.

\subsubsection{SPT0441-46}
The OH absorption at systemic velocities is deeper in this source than any other in our sample, but the spectrum is better fit with an additional blueshifted velocity component. While the \cii spectrum of this source is also broad and double-peaked like SPT0202-61, unlike that source the OH absorption continues beyond the bluest \cii emission. We can thus unambiguously confirm that this source hosts a molecular outflow. The deepest OH absorption is slightly redshifted compared to the \cii peak, possibly indicating a molecular inflow towards the \cii peak.

The lens model of this source required low-order multipoles to adequately reproduce the data. The best near-infrared image of the lens galaxy, from \hst/WFC3, shows a second source $\sim$1'' west of the main lens galaxy. We do not know if this second source is associated with the main lens galaxy, but it may be the cause of the additional lens model complexity required to fit the data.

\subsubsection{SPT0459-58}
The OH spectrum of this source shows very deep blueshifted absorption, and does not require multiple velocity components to fit the spectrum (mostly because the absorption is so broad that any additional velocity components are indistinct).  

The continuum image of this source shows a morphology that is clearly inconsistent with a single simple lensing potential, which went unrecognized in earlier analysis due to the much lower sensitivity and resolution of the earlier data \citep{spilker16}. The northernmost lensed image in particular requires that a second lens potential be placed in its vicinity in order to reproduce the image splitting in that location. Unfortunately the best-available near-IR image of this source, from VLT/ISAAC, is too poor quality to confirm or refute optical counterparts to the best-fit lens potentials we find.

\subsubsection{SPT0459-59}
The OH spectrum of this source shows no obvious evidence of additional blueshifted absorption beyond the broad CO emission, and we do not classify it as an outflow source.  In addition to the lensed galaxy, this source shows at least two other weakly-lensed sources, one just south of the lensed emission and the other $\sim$1.5'' southwest of the lensed emission. We do not detect OH absorption from either of these sources, but they are faint in the continuum so it is unclear whether they are physically associated with the lensed source. High spatial resolution observations of an emission line could clarify the structure of the galaxy and allow for a better characterization of the OH absorption. In our modeling of this source, we subtracted the southwestern source prior to lens modeling in order to allow for more computationally feasible image- and source-plane pixel grids. This also enables better source-plane regularization, since the fitting for the regularization strength need not be influenced by both the strongly-lensed main source and the very weakly-lensed southwestern source.

\subsubsection{SPT0544-40}
The OH spectrum of this source shows a straightforward, albeit broad, blueshifted absorption profile. While this galaxy appears to be a standard quadruply-imaged background source at first glance, both the flux ratios and the spatial extent of the images make clear that it must instead contain two continuum components. It is possible that these two continuum components both contribute to cause the overall broad absorption line profile.

\subsubsection{SPT2048-55}
Much like SPT0441-46, this source shows very strong systemic absorption and a weak blueshifted absorption component, and the stacked CO lines used as a reference are not particularly high S/N. The outflow in this source is the weakest of those we consider unambiguous.

\subsubsection{SPT2103-60}
Similarly to SPT0459-59, the OH absorption in this source does not show an unambiguous outflowing component. While the absorption troughs are blueshifted compared to the flux-weighted mean redshift of the CO emission, it is still possible that the absorption is simply at systemic velocities with respect to the bluest part of the CO emission.

This source is also known to be lensed by a small group of galaxies, three of which are required in order to reproduce the data \citep{spilker16}. We note that we do not require the positions of these lens galaxies to align with galaxies detected in the near-IR in either absolute or relative astrometry, since the baryonic and dark matter masses can become spatially decoupled in overdense environments. In particular the best-fit mass and position of the southwestern lens are degenerate.

\subsubsection{SPT2132-58}
This source shows very broad blueshifted absorption as well as narrow systemic absorption, only barely reaching back to the continuum level at the blue end of the ALMA bandwidth. This source was also studied in detail by \citet{bethermin16}, who found a high excitation in the CO(12--11) transition that could be due to the presence of an AGN.

\subsubsection{SPT2311-54}
This source shows very broad and extremely blueshifted absorption. This galaxy hosts the fastest outflow of our sample, although the maximum outflow velocity is obviously uncertain since the limited ALMA bandwidth does not extend far enough to capture the full absorption profile.

\section{Results} \label{results}

\subsection{Molecular Outflow Detection Rate} \label{detection}

We detect 119\,\um OH absorption in 100\% of our sample, and in 8/11 cases associate this absorption with an unambiguous outflow (including SPT2319-55 published by \citetalias{spilker18a}). The remaining 3 sources all have broad and/or double-peaked CO or \cii line profiles that make it difficult to interpret the OH absorption. Assuming binomial statistics, the outflow detection rate is 73$\pm$13\%. No source shows evidence of OH in emission at systemic velocities; our ability to detect redshifted emission as in the classic P Cygni profile is limited due to the small ALMA bandwidth.

The overall high outflow detection rate demonstrates that molecular winds in these highly star-forming objects are very common. Additionally, because OH as an outflow tracer manifests in absorption and because the absorption is highly optically thick \citep[e.g.][]{gonzalezalfonso17}, our high detection rate also implies that the opening angle of outflowing material must also be high (otherwise the outflows would not be detectable in absorption as most lines of sight would not intersect outflowing gas). Our detection rate is therefore a lower limit on the true occurrence rate of molecular winds in $z>4$ DSFGs because there are presumably some sources driving outflows that do not intersect the line of sight towards the galaxy continuum.  

Even without performing any lensing reconstructions or other spatial analysis, then, we can infer some details about the spatial structure of these outflows. In one possible scenario, large galaxy-scale outflows are being driven with a high opening angle so that most sources have a wind detectable in absorption due to the high covering fraction of outflowing material.  Alternatively, it may be that the outflows are launched such that they are viewable along most lines of sight even with a small covering fraction. For example, if spherical outflows are preferentially launched from the nuclear regions of the host galaxies, we would nearly always detect an outflow even if the molecules are halted or destroyed before reaching kiloparsec-scale distances from the nucleus, resulting in low covering fractions but a high detection rate. These two scenarios are distinguishable using our lensing reconstructions of the wind absorbing material.

\begin{figure}
\includegraphics[width=\columnwidth]{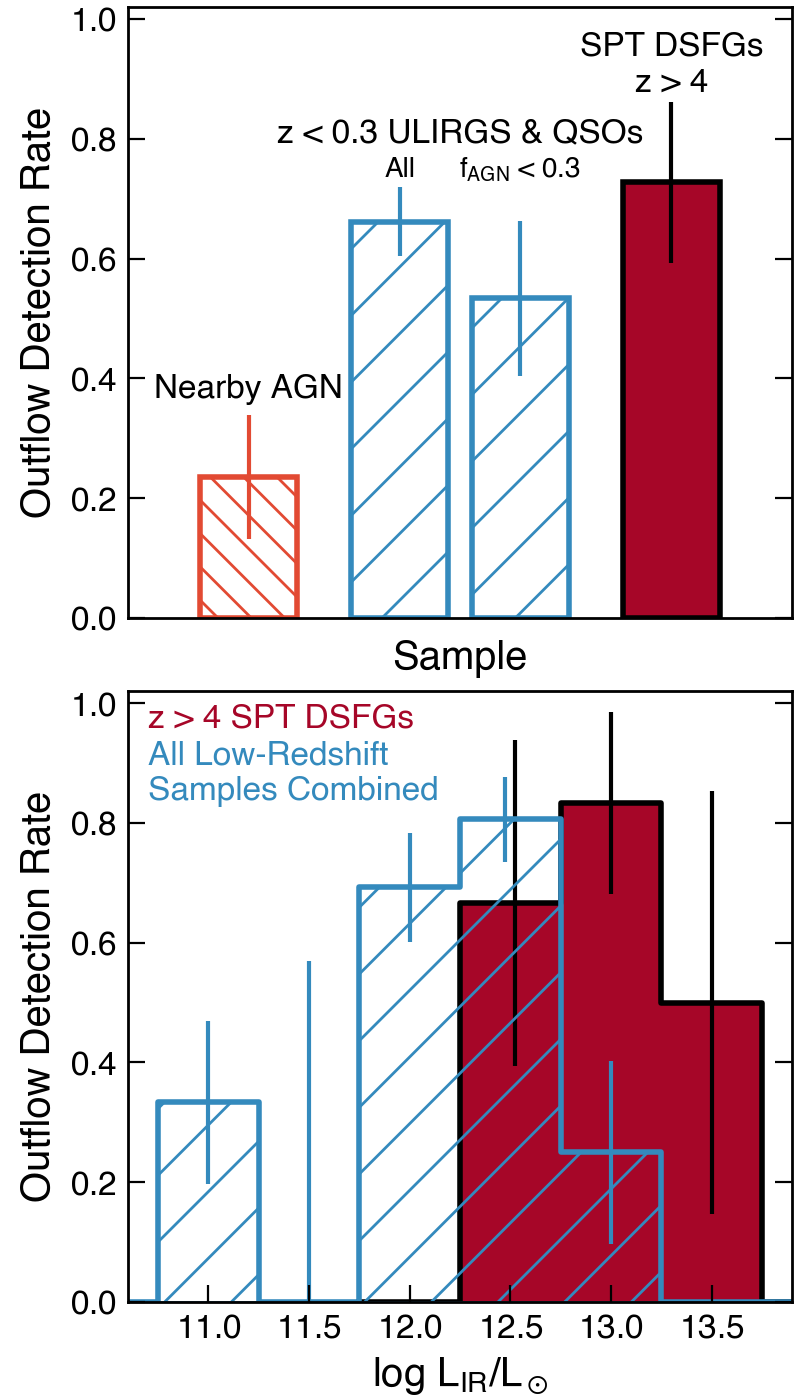}
\caption{
\textit{Top:} Comparison of outflow detection rates via blueshifted OH absorption between our high-redshift DSFGs and a compilation of low-redshift samples (Section~\ref{lowzcomp}). The high detection fraction we find is very similar to that in low-redshift ULIRGs. Uncertainties assume binomial statistics.
\textit{Bottom:} Outflow detection rates as a function of \lir. Although the number of sources is small, we do not find a decrease in the detection rate in the most luminous galaxies.
}\label{fig:detrates}
\end{figure}

Figure~\ref{fig:detrates} compares the outflow detection rate of our $z>4$ DSFG sample with the low-redshift comparison samples (Section~\ref{lowzcomp}). Our detection rate is very similar to that of $z<0.3$ IR-luminous galaxies, but much higher than that of lower-luminosity AGN. We find a slightly higher detection rate when considering only the low-redshift ULIRGs with the lowest AGN fraction, $\fagn < 0.3$, although this difference is not statistically significant. Figure~\ref{fig:detrates} also shows the outflow detection rate as a function of \lir, combining all available low-redshift OH samples. We do not find a decrease in the detection rate among the most luminous galaxies, a marginally significant difference compared to the low-redshift samples. Although the high outflow detection rate in our very luminous high-redshift sample was not unexpected, our observations place the first statistical constraints on the outflow occurrence rate in the early universe. Following \citet{veilleux13} in assuming all galaxies in our sample have a biconically-expanding outflow, our detection rate would correspond to an opening angle of $\sim$150$^\circ$, again very similar to the 145$^\circ$ inferred by \citet{veilleux13} or the 125$^\circ$ we infer for the low-\fagn subsample of all local objects.

Finally, we note that no source shows unambiguous evidence for molecular \textit{inflows}, which would manifest as redshifted absorption profiles. Two sources show $\vfifty \ge +50$\,\kms, sometimes used to classify sources as showing evidence for inflows; our criticisms of this metric when applied to outflows also hold here (Section~\ref{whatoutflow}). In our sample, both sources have clearly double-peaked \cii profiles, often a sign of mergers. It is unclear if the slightly redshifted absorption we see is due to inflow toward one of the velocity components, outflow from the other, or simply systemic absorption from the putative interaction region between them. The inflow detection rate in low-redshift ULIRGs is $\sim$10\% \citep{veilleux13,herreracamus20} using the \vfifty metric, far lower than the outflow detection rate. Given the ambiguity in our data and small sample size, we cannot draw strong conclusions on this point, but there is no immediately obvious difference between the inflow detection rates at $z\sim0$ and $z>4$.

\subsection{Basic OH Absorption Properties} \label{basics}

We detect OH in absorption in all 10 target DSFGs (11 including the source in \citetalias{spilker18a}); no source shows OH in emission or P Cygni profiles. While observational restrictions preclude us from detecting P Cygni profiles, the lack of OH in emission in any source is interesting because OH 119\,\um appears in emission primarily in AGN-dominated galaxies ($\fagn \gtrsim 0.8$; e.g. \citealt{veilleux13,stone16,runco20}). For example, in the \citet{stone16} sample of nearby AGN-dominated galaxies, $>$70\% of the detected objects showed either pure emission or emission/absorption composite spectra, with 60\% purely in emission. A similar conclusion applies to the more IR-luminous ULIRGs and QSOs: objects with $\fagn > 0.8$ typically show OH in emission \citep{veilleux13}. This is presumably because the dense nuclear regions are able to excite the 119\,\um energy levels in spite of the very high gas densities required for collisional excitation, $\nht \sim 10^8$\,\percc \citep{spinoglio05,runco20}. This may also be an evolutionary effect, where wide-angle outflows have cleared the sightline to the dense nuclear region and already subsided \citep[e.g.][]{veilleux13,stone16,falstad19}. The fact that we do not see OH in emission in any of our sources is a secondary empirical indication that AGN are probably not dominant in these galaxies, unless the column densities are so high as to be optically thick in the mid-IR in the direction of the emitting regions.

\begin{figure*}
\centering
\includegraphics[width=\textwidth]{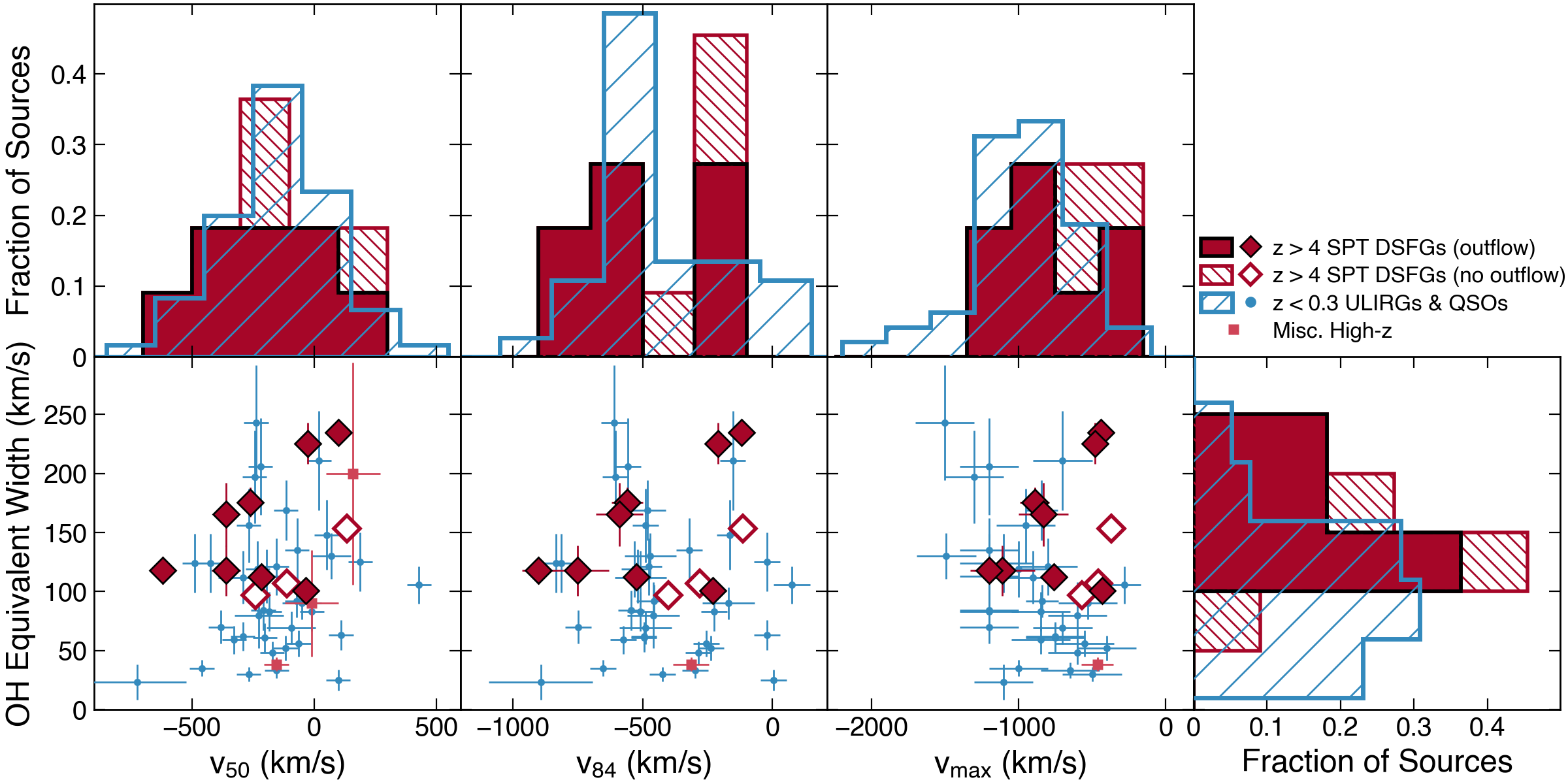}
\caption{
Distributions of the OH absorption median, 84th percentile, and maximum velocities and the equivalent widths of the OH doublet lines for our sample and literature sources.  Histogram bins have been slightly offset between the samples for clarity. The limited ALMA bandwidth would prevent us from detecting the fastest outflows seen in the low-redshift samples. Not all literature studies report all quantities in this figure, so there are some discrepancies between the histograms and the scatter plots. In particular \citet{spoon13} (a subset of the `$z<0.3$ ULIRGs and QSOs') do not provide equivalent widths; this sample is noticeably absent in the plot of equivalent width vs. \vmax compared to the \vmax histogram.
}\label{fig:veleqw}
\end{figure*}

Figure~\ref{fig:veleqw} shows histograms of three different metrics of the OH absorption velocity, \vfifty, \vef, and \vmax, for our sample and literature sources. We exclude those sources for which OH was detected only in emission. We find very similar distributions of these quantities between the low- and high-redshift samples. A two-sided KS test confirms that we cannot reject the hypothesis that the two samples are drawn from the same underlying distribution in any of the three velocity metrics. While it appears that the low-redshift sample has a tail to extremely fast maximum outflow velocities $\vmax \lesssim -1400$\,\kms that is not present in our data, the limited ALMA bandwidth means we could not probe such high velocities if they were present in our sample. Indeed, two of our sources show absorption that continues essentially all the way to the edge of the observed bandwidth and could plausibly reach faster outflow velocities than the extrapolations from our fits to the spectra suggest, depending on the true line profiles.

Figure~\ref{fig:veleqw} also shows the equivalent width distributions of these samples and the relationships between the equivalent widths and the velocity metrics. The equivalent widths are the total values (systemic plus any red- or blueshifted absorption) for one of the two doublet lines (i.e. should be multiplied by 2 for the total equivalent widths). The equivalent width distribution does show a significant difference between the low- and high-redshift samples; a two-sided KS test rejects the hypothesis that the two samples are drawn from the same underlying distribution ($p=0.004$). This appears to be because the SPT DSFGs show stronger OH absorption than the low-redshift samples, being under-represented at low equivalent widths and over-represented at high equivalent widths. 

This difference cannot be explained by simple observational selection effects. While it is plausible that weak absorption would be more easily detected in the bright nearby samples than our very distant targets, OH was strongly detected in every SPT source. We could have detected equivalent widths $\sim$5$\times$ lower than even the weakest absorption actually seen; the dearth of weak OH absorption in the high-redshift sample is genuine. The differences in typical \lir and \fagn between the two samples also do not explain the different distributions. The same two-sided KS test returns $p=0.02$ when considering only the low-redshift sources with higher \lir than our least-luminous source, $\log \lir/\Lsol \gtrsim 12.5$, and $p=0.04$ for only the low-redshift sources with $\fagn < 0.4$ (the number of low-redshift objects that are \textit{both} very luminous and have low \fagn is too small for a meaningful comparison). We conclude that OH absorption at high redshift does appear to be genuinely stronger than in similar low-redshift galaxies, but a more complete sampling of objects at lower \lir and/or higher \fagn at high redshift is required to make a more thorough comparison.

\subsection{Continuum and Absorption Reconstructions} \label{structure}

Figure~\ref{fig:recon} shows the lensing reconstructions of the continuum and the OH absorption component(s) for each source, presenting the absorption maps in terms of both integrated absorbed flux and equivalent width. While absorption is a multiplicative and not an additive process, our ability to detect absorption is not. Consider for example a region with continuum S/N$\sim$3: we would likely not confidently detect absorption even in the case of 100\% absorption, while far weaker absorption is detectable where the continuum is brightest. We mask the absorbed flux maps in regions where the continuum is detected at S/N$<$5. Even above this threshold in most sources we find some pixels in the absorption maps with positive flux (or negative equivalent width), which is simply a reflection of differencing two moderately uncertain measurements. These positive regions are mostly eliminated if we change the threshold to continuum S/N$>$10, which we consider to be an empirical indication of the level where the absorption reconstructions are most reliable.

\begin{figure*}
\begin{centering}
\includegraphics[width=0.656\textwidth]{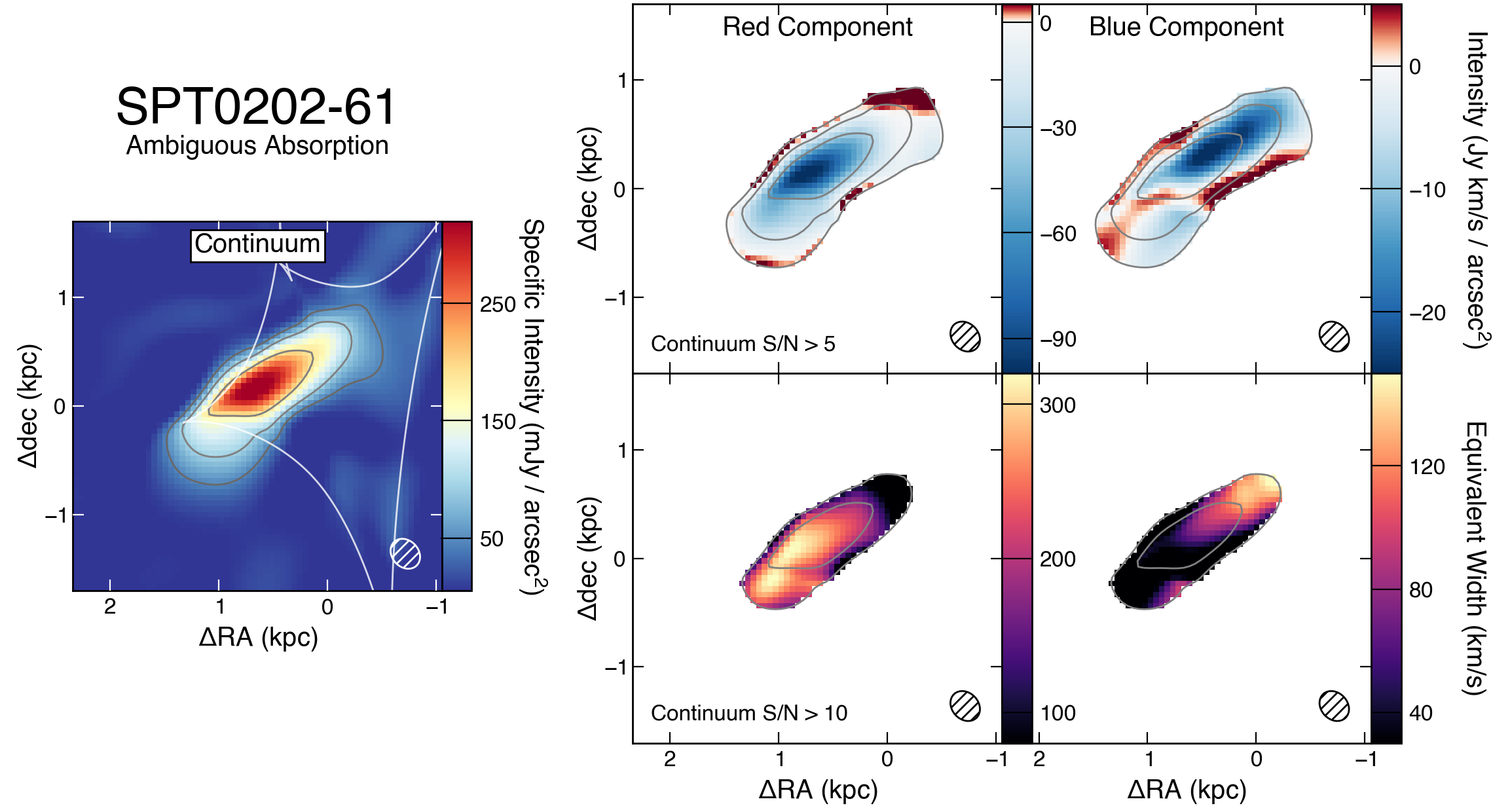}
\includegraphics[width=0.656\textwidth]{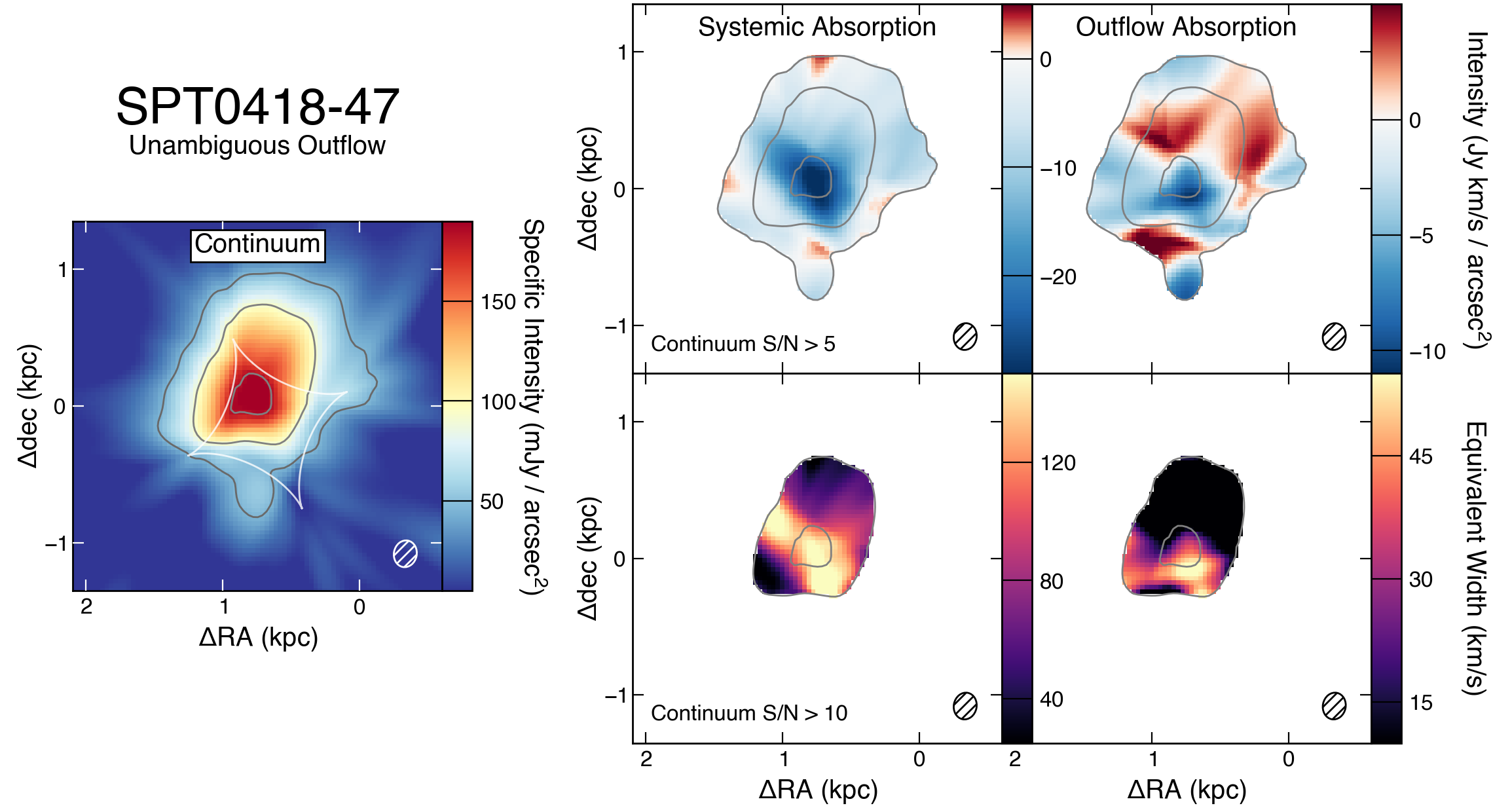}\\
\end{centering}
\caption{
Lensing reconstructions of the OH data for each target DSFG. Axes are relative to the ALMA phase center. Left shows the dust continuum emission with grey contours overlaid at signal-to-noise of 5, 10, and 20; these contours are repeated in the other panels. The lensing caustics are shown in white. Right panels show the reconstructions of the OH absorption component(s) as indicated in Figure~\ref{fig:OHspectra} and listed in Table~\ref{tab:absfits}. Upper rows show maps of the integrated absorbed flux density while lower rows show maps of the equivalent width. We mask the upper rows for continuum S/N $<$ 5 and the lower rows for continuum S/N $<$ 10. At lower right in each panel we show an ellipse representing the effective spatial resolution of the reconstructions at the peak of the reconstructed continuum emission. The spatial resolution varies across the source plane; see Section~\ref{lenstests} and Appendix~\ref{applensing}.
}\label{fig:recon}
\end{figure*}

\begin{figure*}
\figurenum{6}
\begin{centering}
\includegraphics[width=0.656\textwidth]{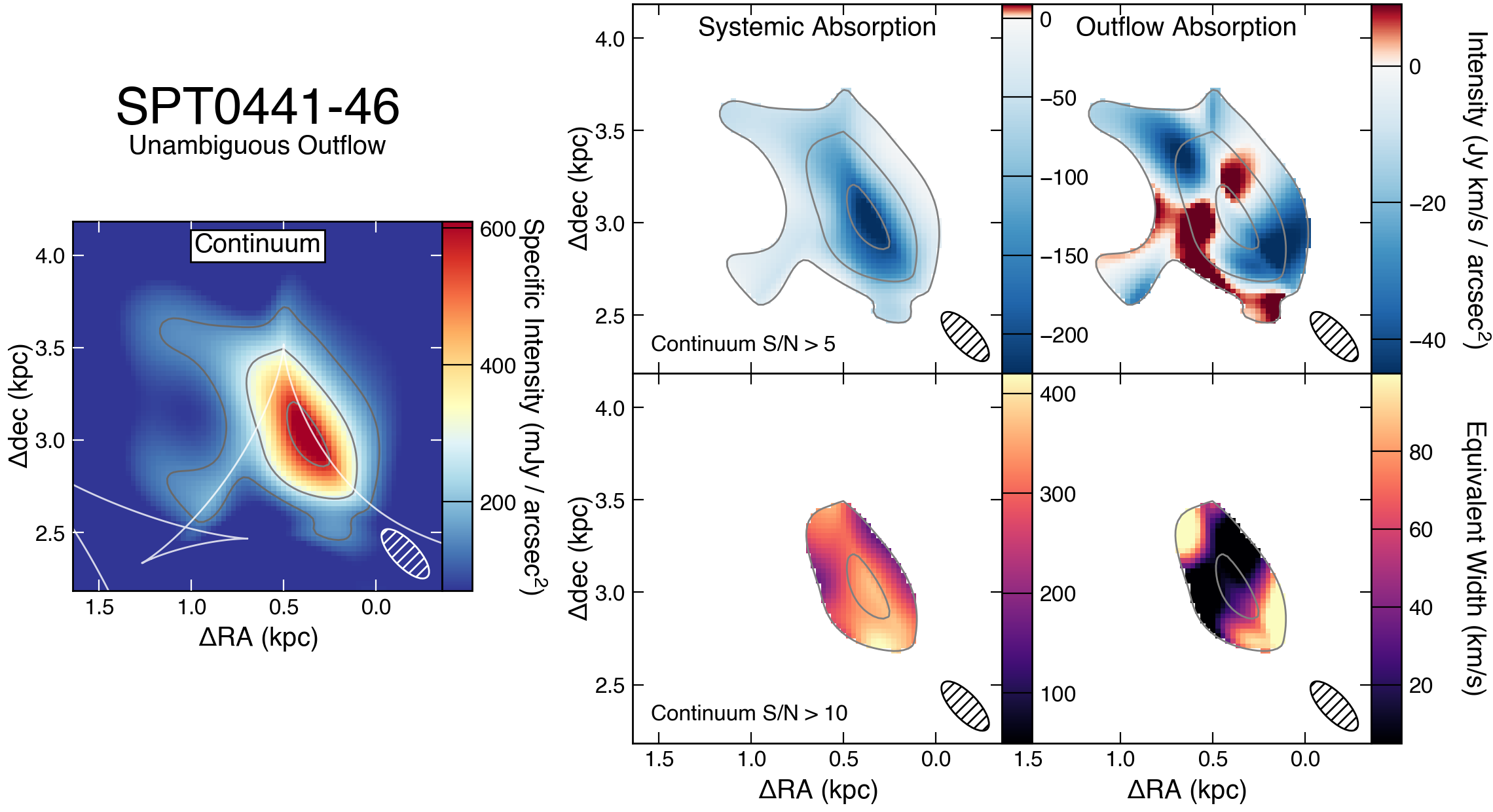}\\
\includegraphics[width=0.495\textwidth,trim=0in 0in 2.8in 0in,clip]{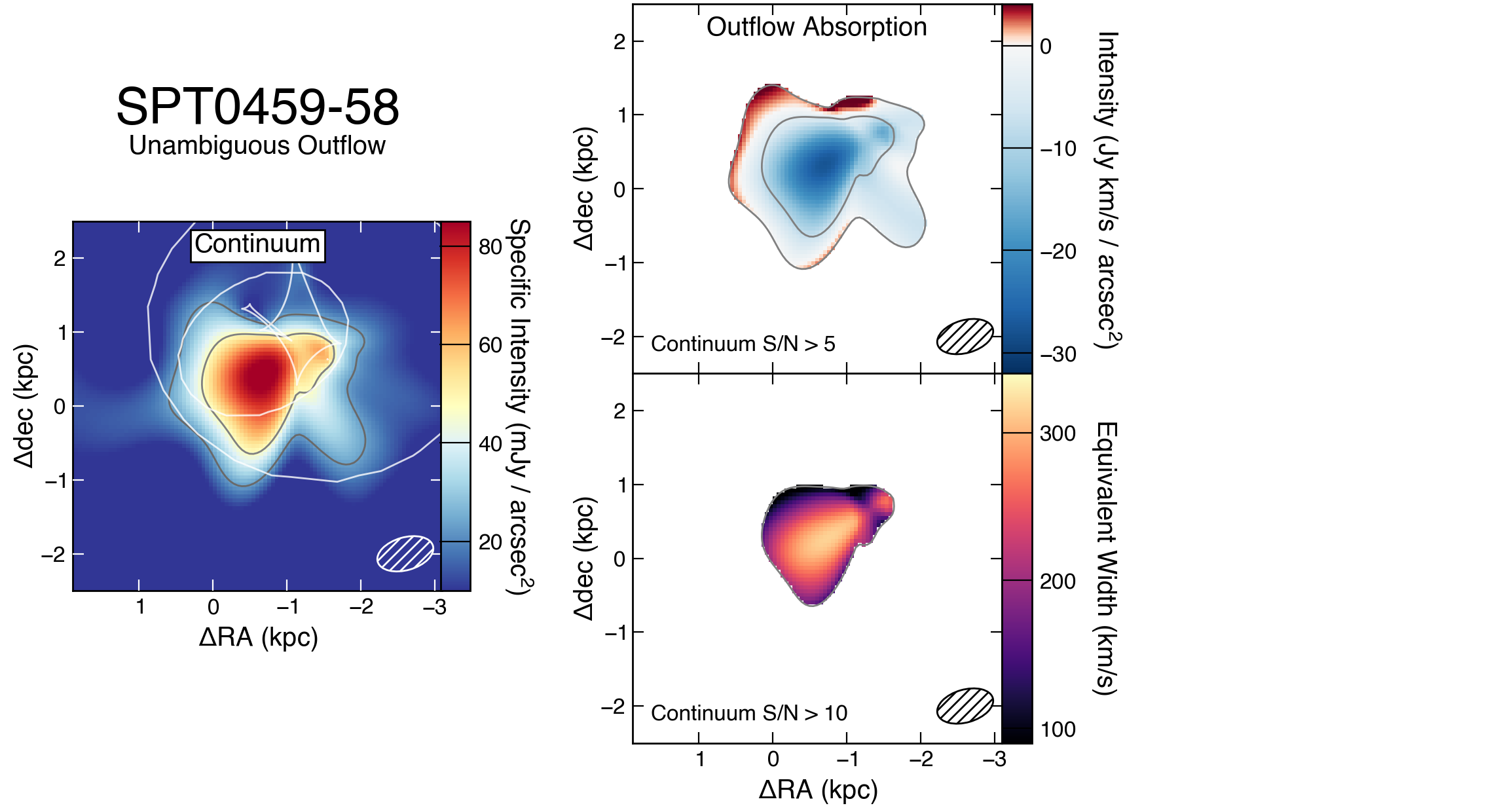}
\includegraphics[width=0.495\textwidth,trim=0in 0in 2.8in 0in,clip]{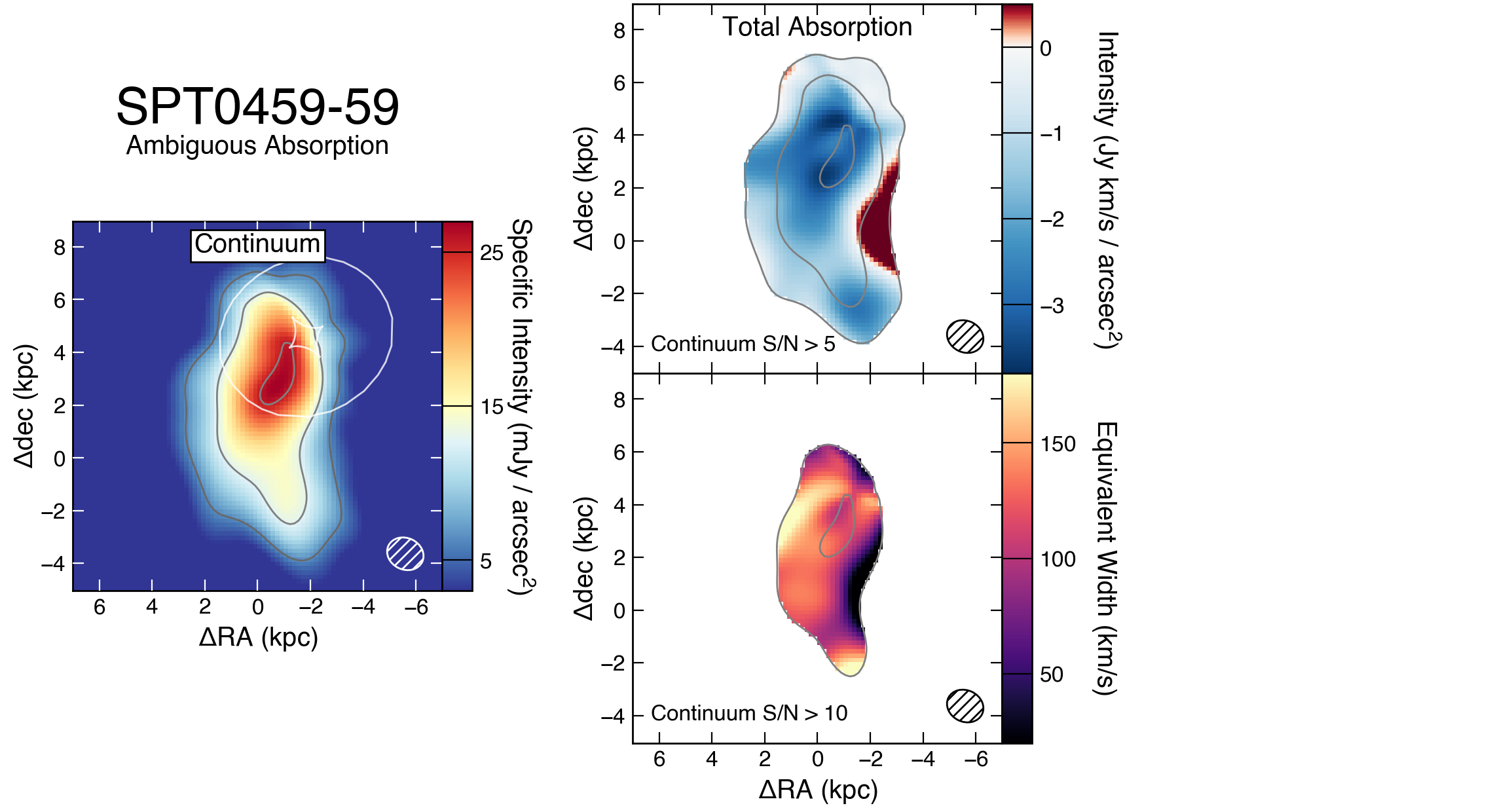}
\includegraphics[width=0.495\textwidth,trim=0in 0in 2.8in 0in,clip]{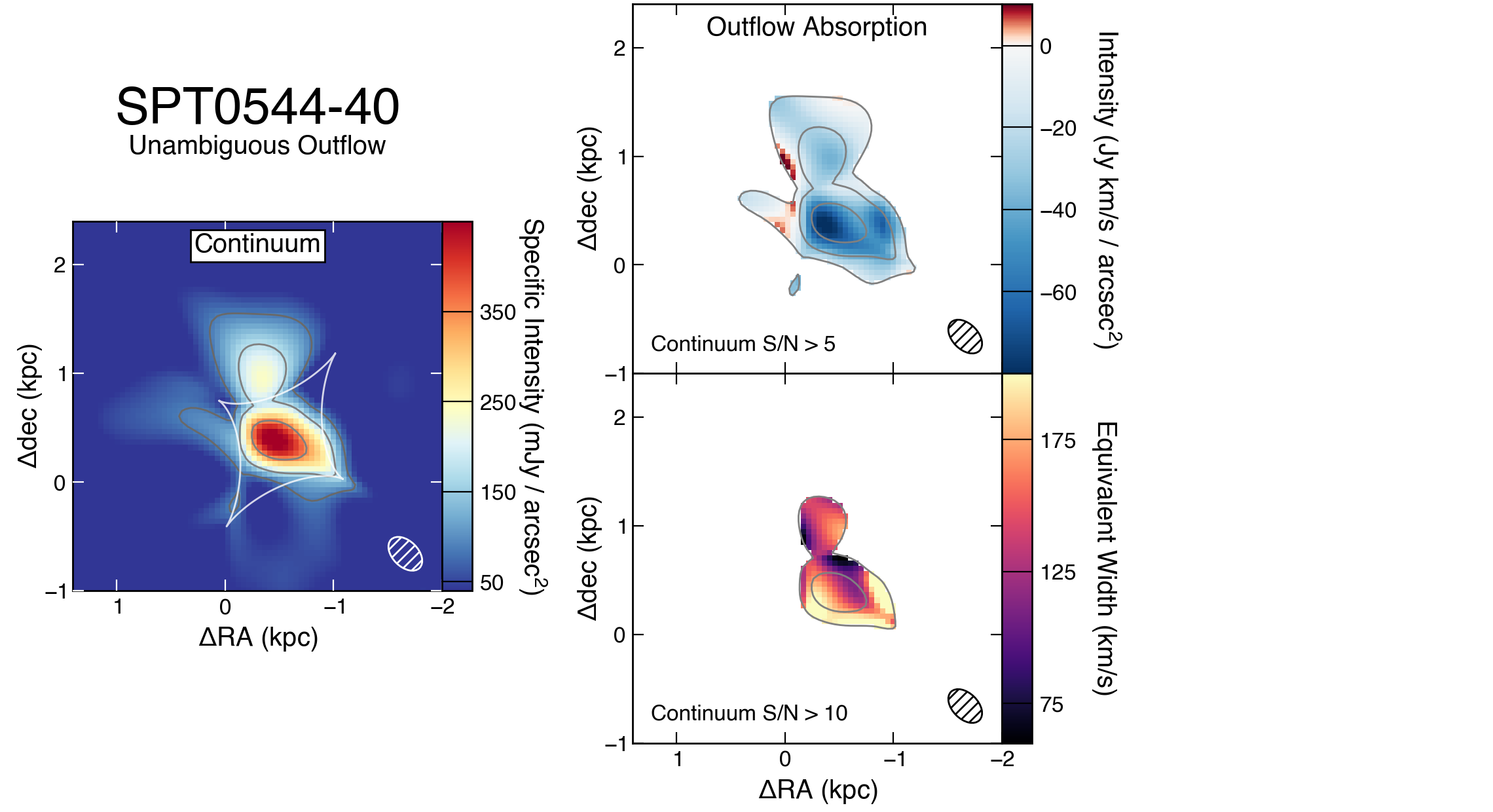}
\includegraphics[width=0.495\textwidth,trim=0in 0in 2.8in 0in,clip]{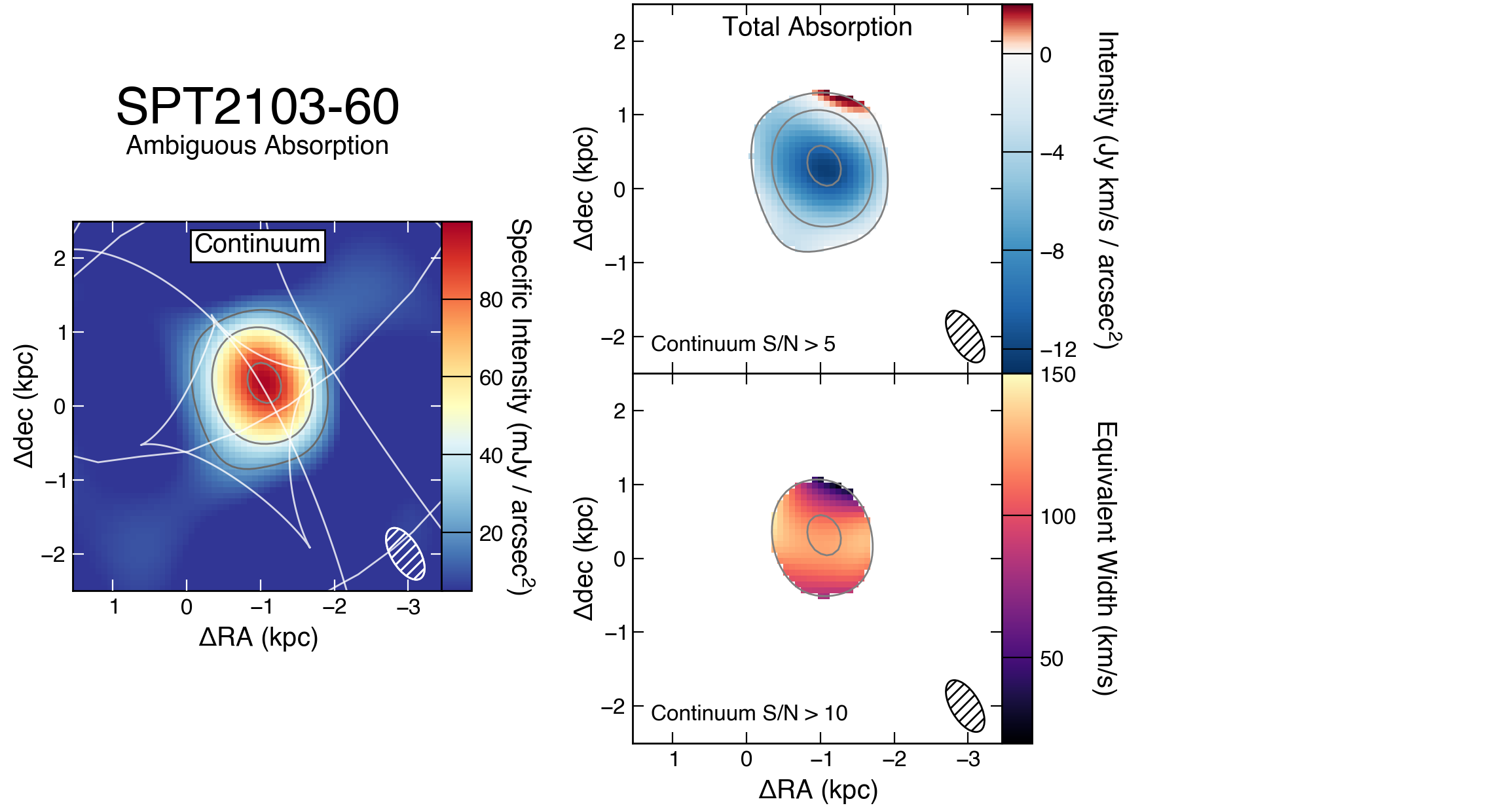}
\end{centering}
\caption{Continued.
}\label{fig:recon2}
\end{figure*}

\begin{figure*}
\figurenum{6}
\begin{centering}
\includegraphics[width=0.656\textwidth]{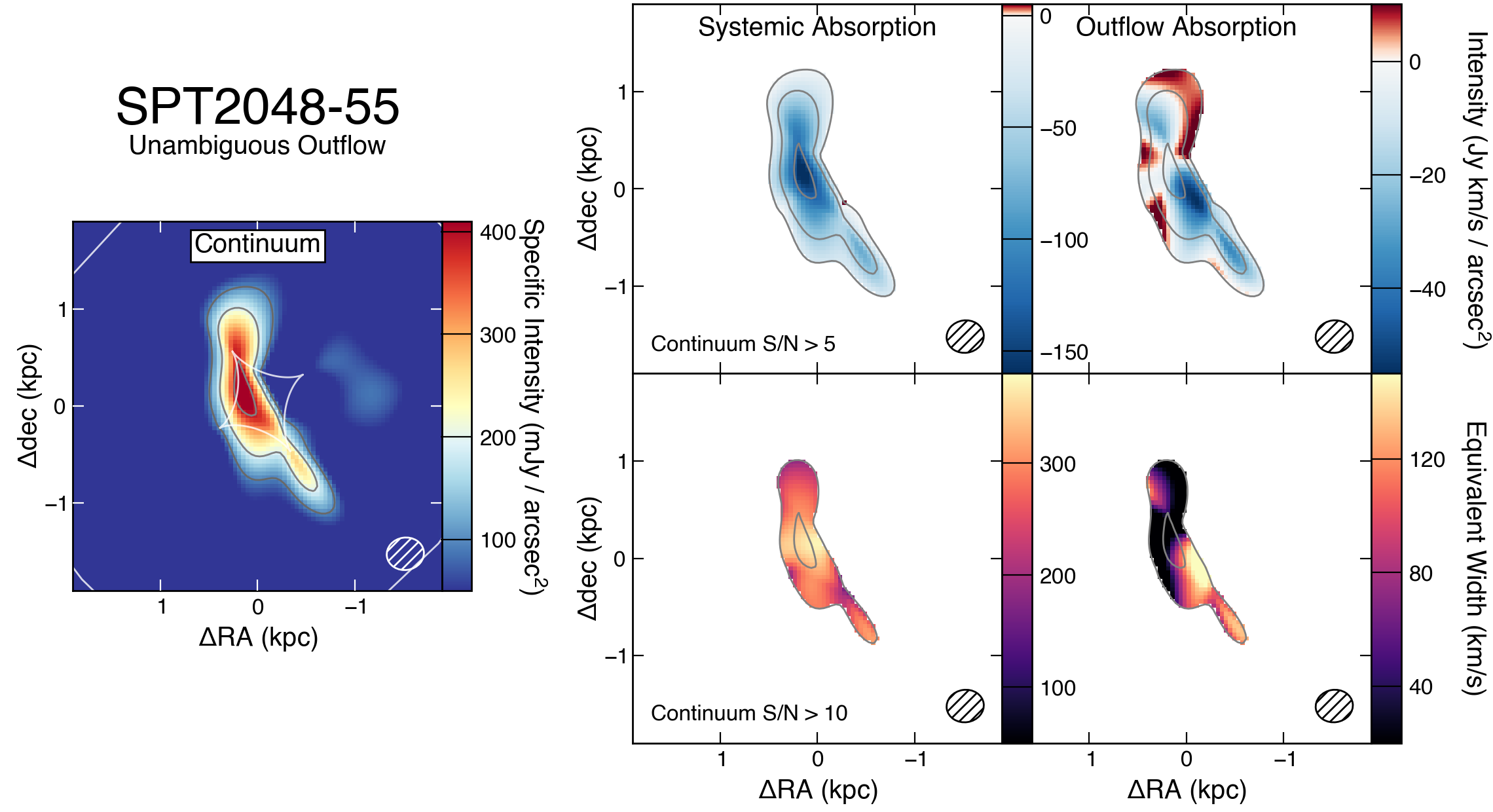}\\
\includegraphics[width=0.656\textwidth]{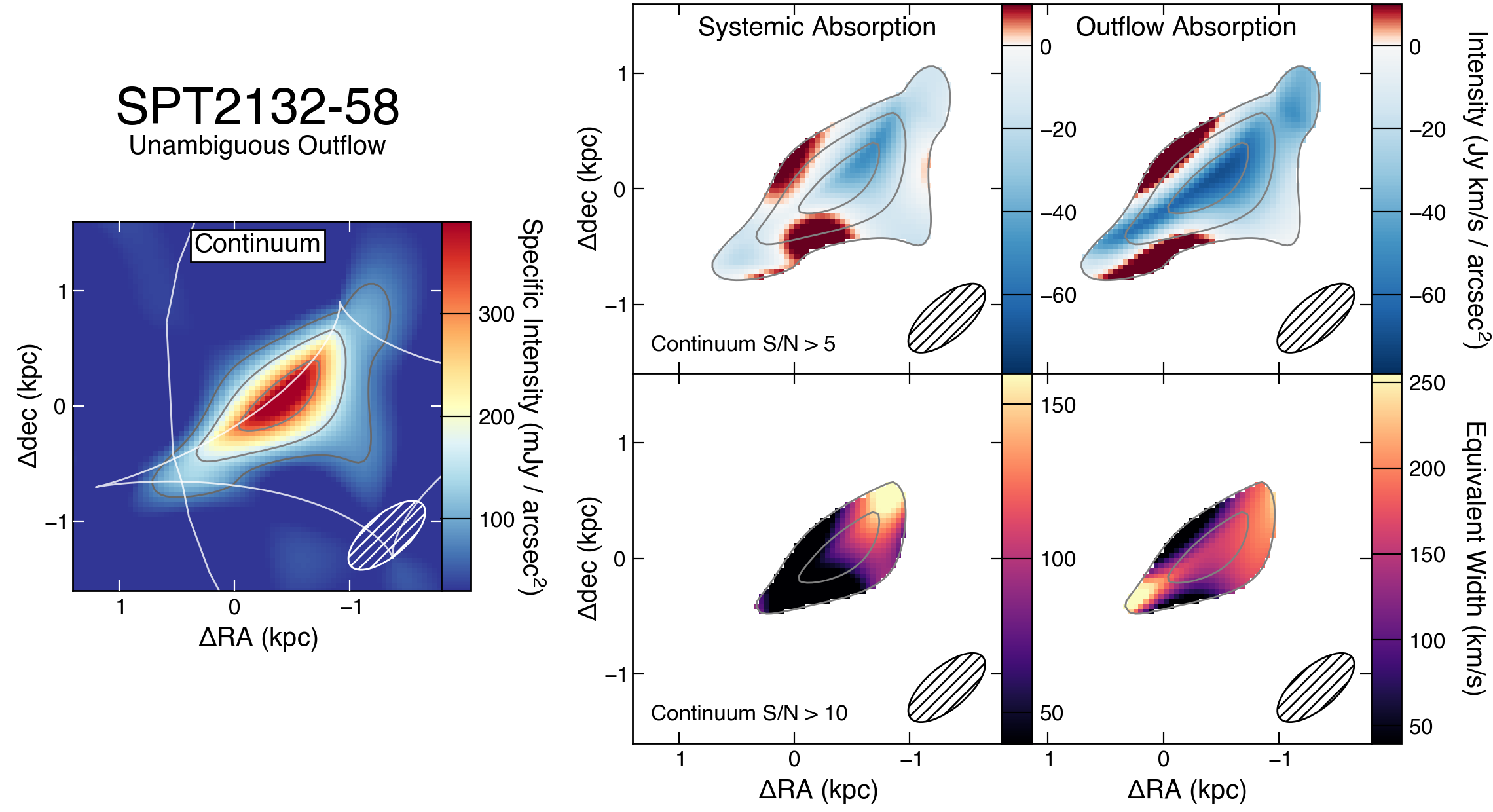}\\
\includegraphics[width=0.495\textwidth,trim=0in 0in 2.8in 0in,clip]{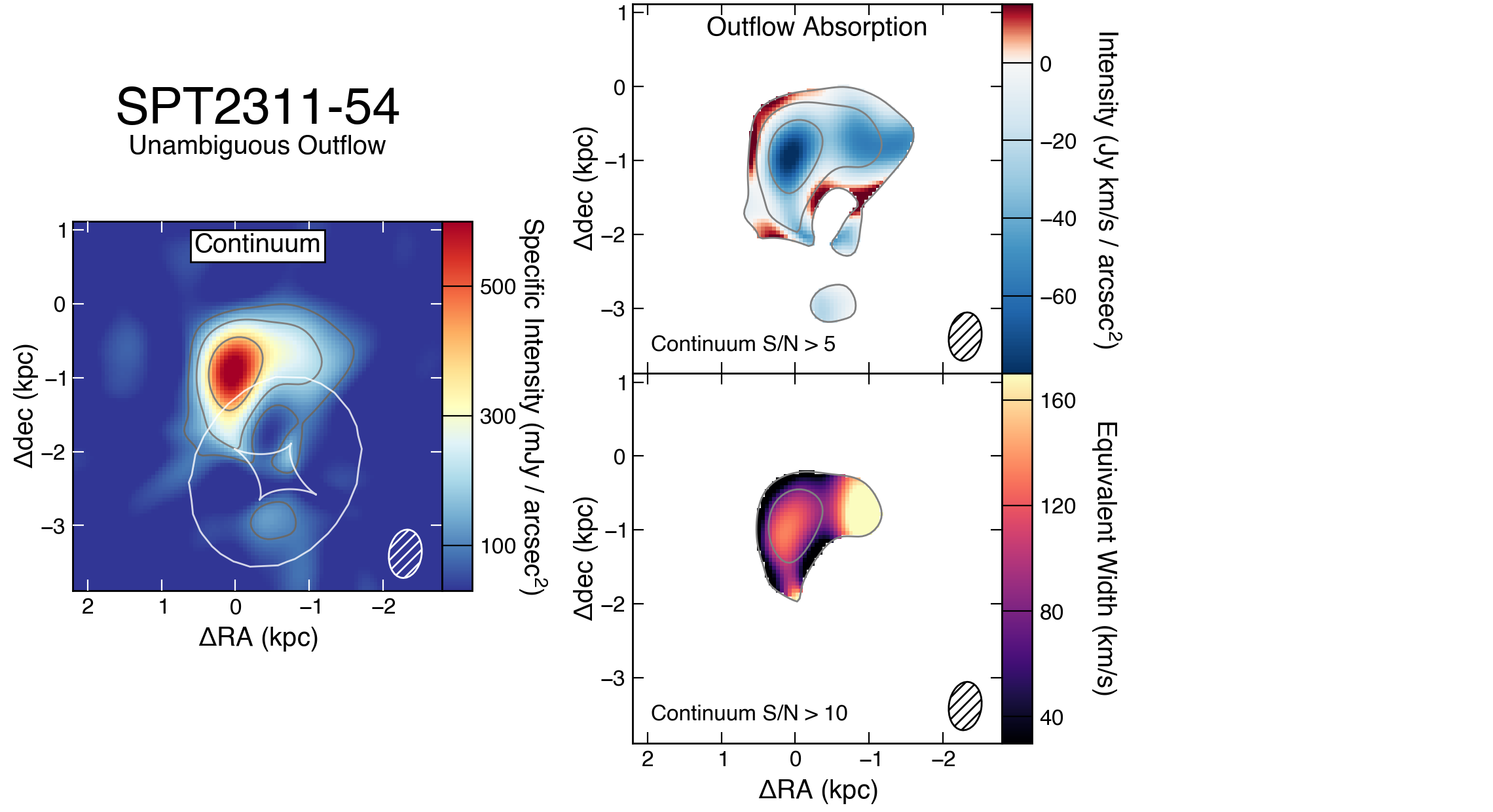}\\
\end{centering}
\caption{Continued.
}\label{fig:recon3}
\end{figure*}

The continuum emission in our sources is mostly smooth at the resolution of these observations, with effective circularized radii $r_\mathrm{cont} \approx 0.5-4$\,kpc (the largest source, SPT0459-59, clearly consists of multiple sources poorly resolved by the present data). These sizes are similar to or somewhat smaller than those found by \citet{hodge16} in a sample of unlensed DSFGs at lower redshift but typical for the lensed SPT DSFGs \citep{spilker16}, although the size estimate methodology differs between these works. A few sources show multiple peaks in the continuum reminiscent of mergers, but we see little evidence for distinct sub-kpc substructure. Recent ALMA observations of lensed and unlensed galaxies have demonstrated that identifying clumps in dust emission can be precarious even in data with S/N$\gtrsim$30 because the relevant metric is the contrast between the clumps and the underlying smooth emission \citep[e.g.][]{hodge16,hodge19,rujopakarn19,ivison20}. Our observations typically reach peak S/N values far higher than these studies, but the substantial freedom afforded by the pixellated lensing reconstructions reduces the effective peak S/N in the reconstructions to $\sim$25--40 for all sources. That is, there is not a direct correspondence between image- and source-plane S/N because our lensing reconstructions fully account for the uncertainties in the data rather than directly map from image to source. We rule out significant 500pc-scale clumpiness in our sample at $\gtrsim$10\% of the total flux density as seen in some ALMA studies \citep{hodge19} but cannot rule out weaker and/or smaller clumps.

In sources with systemic absorption, the absorption generally peaks near the same location as the continuum, with an equivalent width distribution that can be either centrally-concentrated or more widespread and uniform. The systemic absorption tends not to show signs of distinct clumps, like the continuum emission. This is altogether unsurprising, given that OH is sensitive to even fairly small column densities of molecular gas, satisfied over large regions of these gas-rich galaxies but subject to line-of-sight geometric effects \citep[e.g.][]{gonzalezalfonso17}.

The outflow maps show more interesting structure when visualized either in terms of absorbed flux or equivalent width. In contrast to the systemic absorption, the blueshifted outflowing gas rarely peaks in equivalent width at the locations of the continuum peaks, instead often showing the highest equivalent widths significantly offset from the brightest continuum emission. Also in contrast to the continuum and systemic absorption, distinct clumps or small-scale substructures are common in the molecular outflows. Combined, these two facts may indicate that the high-velocity gas has already moved away from the nuclear regions and now appear offset (if indeed they were launched from the nuclear regions), similar to the offsets commonly seen between the core and high-velocity CO emission in low-redshift outflows \citep[e.g.][]{lutz20}. Alternatively, it could be that high-velocity outflows are more easily launched in regions with lower column densities than the very dense nuclear regions \citep[e.g.][]{thompson16a,hayward17}.

\begin{figure*}
\begin{centering}
\includegraphics[width=\textwidth]{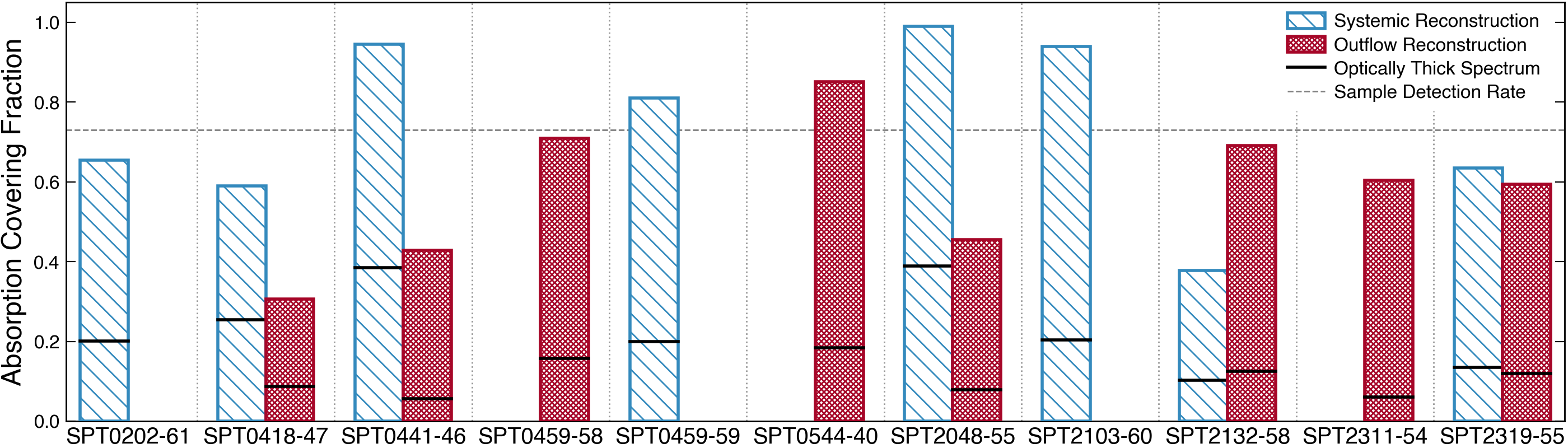}
\end{centering}
\caption{
Covering fractions of OH absorption from the reconstructions of each source, defined as the fraction of pixels with detectable absorption where the continuum S/N$>$10. Horizontal black lines show the expected covering fraction derived from the absorption depth over the velocity range of each modeled component, assuming optically thick absorption; these represent hard lower bounds on the covering fractions. The dashed grey line shows the sample-averaged detection rate of 73\% (which is the same for the outflows and systemic absorption). The differences between these values may imply the presence of substructure in the absorption on scales several times smaller than the resolution our data provides; see Section~\ref{smallclumps}.
}\label{fig:covfracs}
\end{figure*}

Figure~\ref{fig:covfracs} shows the absorption covering fraction for the reconstructions in each source, defined as the fraction of pixels in the equivalent width reconstructions with detectable absorption where the continuum S/N$>$10. We find generally high covering fractions ranging from 30--85\% for the outflows or 40--95\% for the systemic absorption. These values are typically somewhat lower than the overall sample detection rate of 73\% for either component, but significantly higher than the fractional absorption depths in the apparent OH spectra. This may be somewhat surprising given that OH 119\,\um absorption is expected to be highly optically thick. We return to this comparison in Section~\ref{smallclumps}.

\begin{figure*}
\begin{centering}
\includegraphics[width=0.495\textwidth]{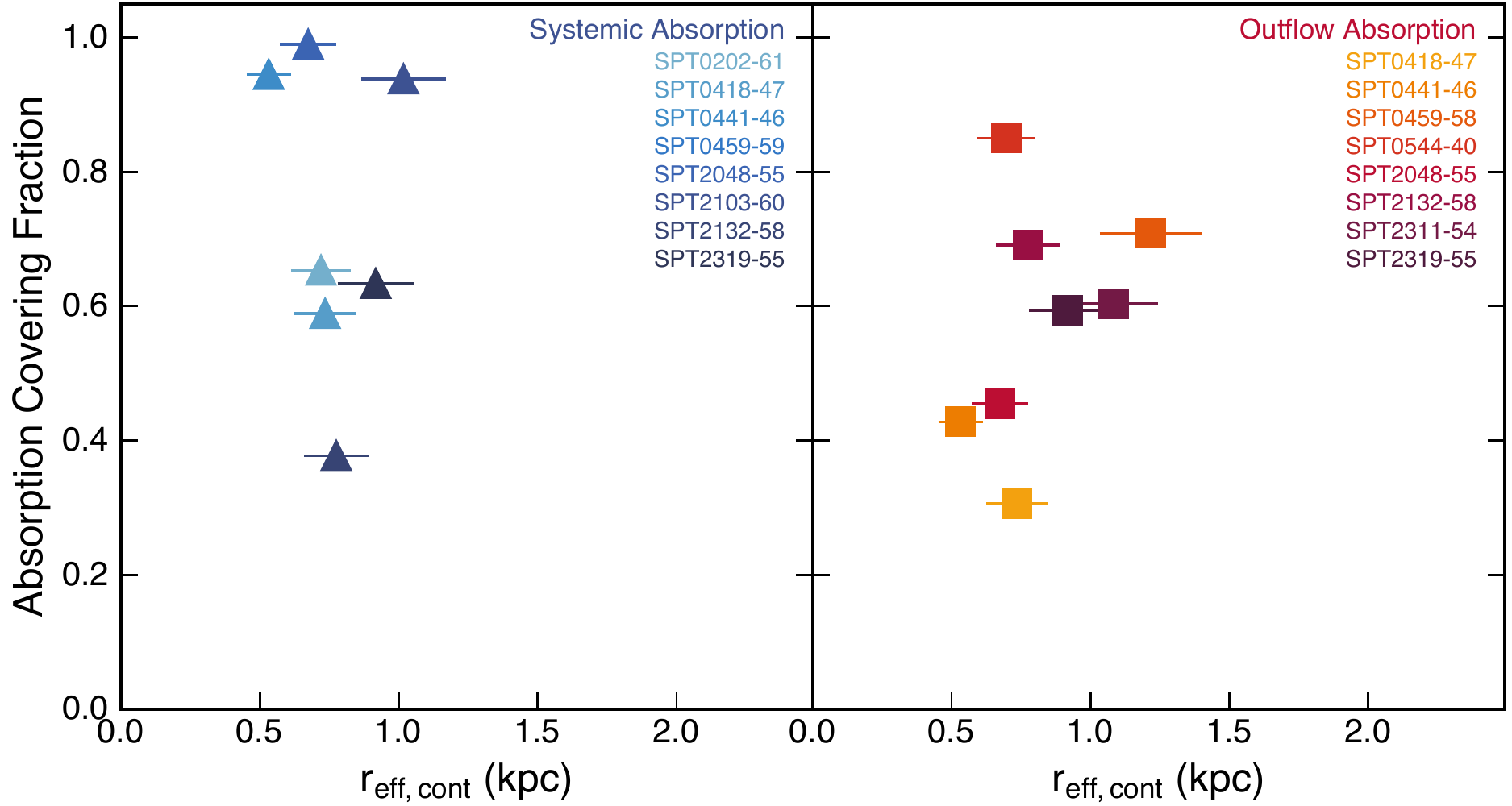}
\includegraphics[width=0.495\textwidth]{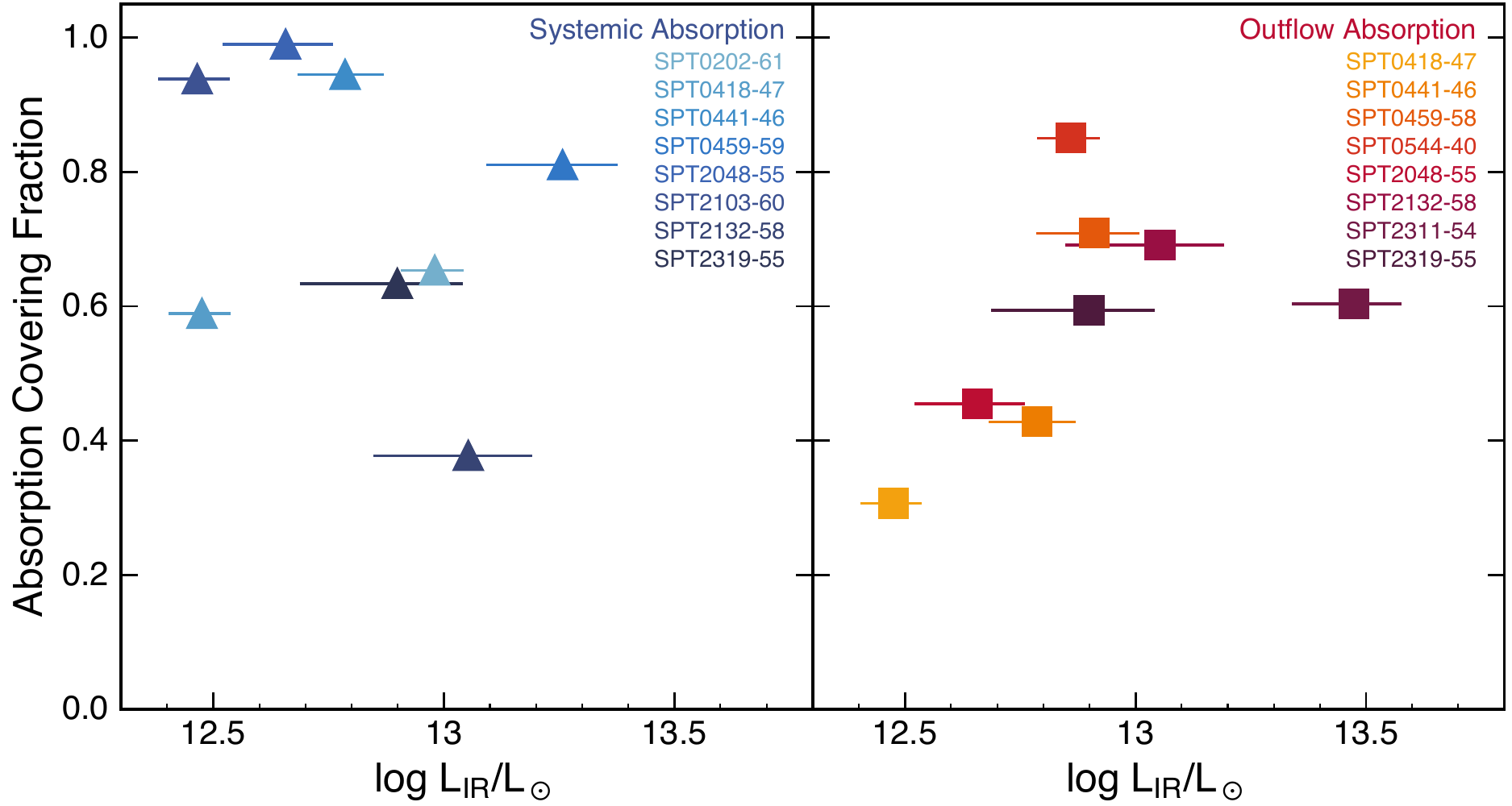}
\end{centering}
\caption{
The most IR-luminous sources in our sample have higher outflow covering fractions, implying that these galaxies are able to launch more widespread outflows. The figure shows the OH absorption covering fractions from the lensing reconstructions against the effective dust continuum emitting sizes (left) and \lir (right). The lack of an inverse correlation with the continuum size rules out that the absorbing structures have a typical constant physical size, but there is an intriguing correlation between covering fraction and \lir.
}\label{fig:covfracs_trends}
\end{figure*}

Figure~\ref{fig:covfracs_trends} searches for trends of absorption covering fraction with select other quantities we have measured. First, we find no correlation between the covering fraction of either systemic or outflow components with the actual physical size of the dust continuum emitting region (Fig.~\ref{fig:covfracs_trends} left). This lack of correlation rules out that the absorbing material has a typical constant physical size. If this were so, we would expect the covering fraction and continuum emitting size to be inversely correlated, with the typically-sized absorbing regions covering a larger fraction of intrinsically smaller galaxies.

Second, we do find intriguing evidence for a correlation between the outflow covering fraction and \lir, with the most luminous sources showing higher covering fractions (Fig.~\ref{fig:covfracs_trends} right).  Taken at face value, this result implies that more luminous galaxies are able to launch more widespread outflows, either with larger opening angle or more and/or larger clumps. At some level this result is also consistent with the trend of rising outflow detection rate with \lir, which extends over a larger dynamic range in \lir than probed by our sample alone (Fig.~\ref{fig:detrates}). In order to see such a trend, one can invoke either a rising  outflow occurrence rate or an increasing outflow covering fraction with \lir, or both. The same trend does not exist for the systemic absorption components, implying (unsurprisingly) that the covering fraction of gas at rest is dependent on the line-of-sight geometry but not necessarily any intrinsic physical properties of the galaxies.

\begin{figure}
\begin{centering}
\includegraphics[width=\columnwidth]{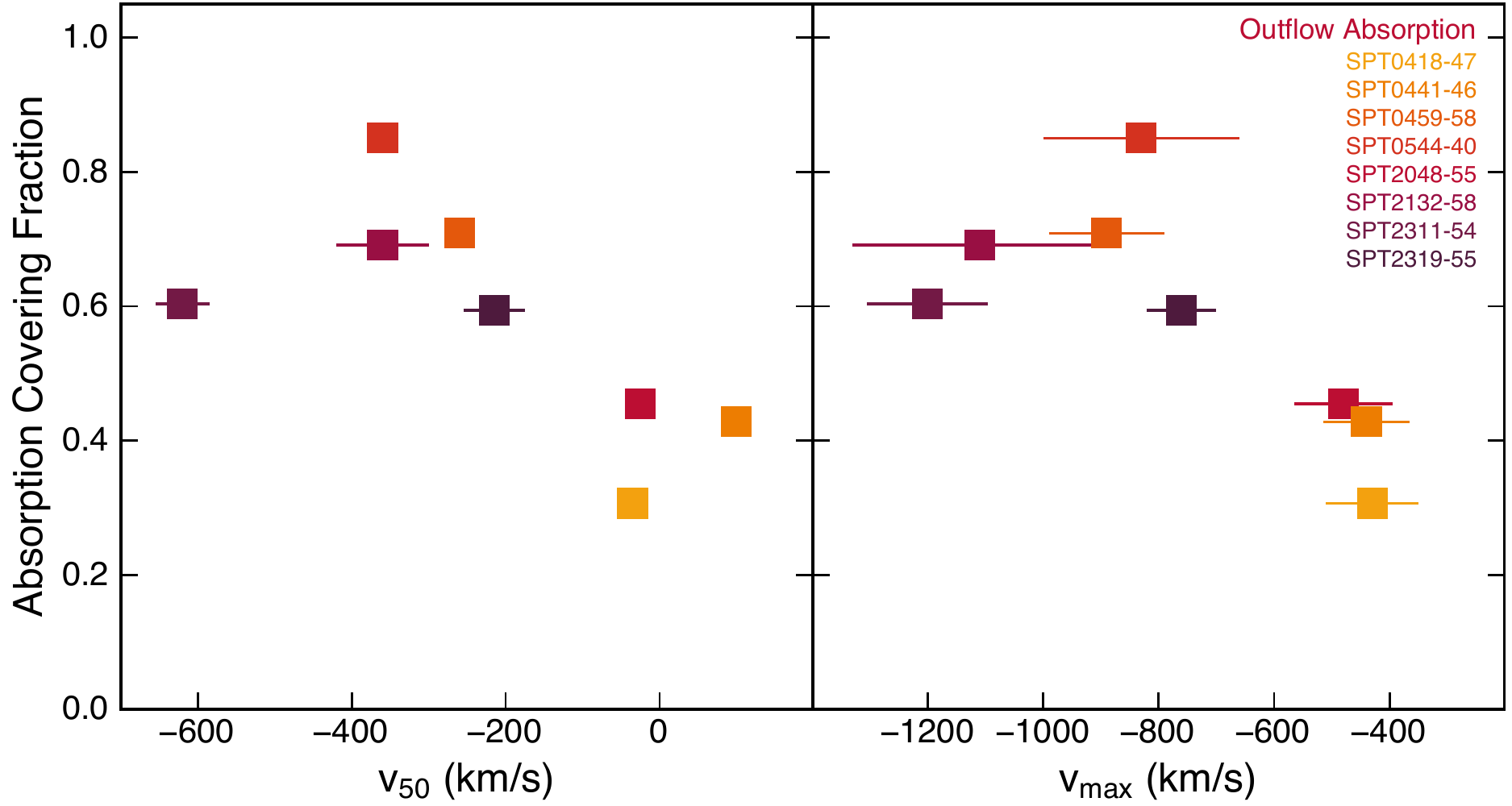}
\end{centering}
\caption{
The galaxies with the fastest outflows also show the highest covering fractions or most widespread outflows. We show the lens model covering fractions against \vfifty and \vmax, plotting only those sources with unambiguous outflows.
}\label{fig:covfracs_vel}
\end{figure}

For the sources that show unambiguous outflows only, in Figure~\ref{fig:covfracs_vel} we additionally find correlations between the outflow covering fraction and the outflow velocities \vfifty and \vmax. In addition to being driven by the most luminous sources, the galaxies with high covering fractions also show the fastest outflows. Some of this is no doubt driven by underlying trends we find between \lir and \vfifty or \vmax for the sources in our sample with outflows, as we explore in more detail in \citetalias{spilker20a}. It is intriguing nonetheless that the outflow covering fraction appears to be correlated with either/both quantities. Whether these trends hold with increased sample sizes that span a wider dynamic range in host properties is an interesting motivation for future investigation.

\subsection{Ubiquitous Clumpy Molecular Outflows} \label{clumpiness}

In an effort to further quantify the degree of small-scale `clumpiness' in the reconstructions, we borrow several metrics from the extensive literature on morphological analyses of clumps and mergers in high-redshift galaxies \citep[e.g.][]{lotz04,law07,forsterschreiber11,wuyts12,guo18}. Specifically, for the continuum and absorption component(s) of each source, we calculate: (1) the Gini coefficient $G$ \citep{abraham03}, which quantifies how uniformly the pixel values in an image are distributed, ranging from 0 if all pixels are equal to 1 if a single pixel contains all the flux; (2) the second-order moment of the 20\% brightest pixels $M_{20}$ \citep{lotz04}, which measures how centrally-peaked the pixels in a source are, with high values indicating the presence of off-nuclear clumps; and (3) the `multiplicity' $\Psi$ \citep{law07}, sensitive to the presence of multiple clumps of flux with higher dynamic range than $M_{20}$, where higher values indicate that an image contains more and/or brighter clumps. As a reference for the dynamic range of these quantities, $G$ is by definition confined to the range 0 to 1, $M_{20}$ typically ranges from -3 to -1 in optical SDSS images of nearby galaxies, and $\Psi$ ranges from 0 to $\approx$30 in rest-UV images of z$\sim$2 galaxies \citep{lotz04,law07}. 

We prefer these metrics as opposed to other methods for several reasons. First, they do not rely on subjective visual identification of clumps, and are deterministic and reproducible. Second, they are non-parametric and do not rely on further modeling of the reconstructions (which are themselves already models of the original data). Third, it is straightforward to compare these quantities between the absorption reconstructions and the continuum emission.

We calculate each of these quantities for the absorption and continuum reconstructions, measured from the equivalent width maps and for pixels where the continuum S/N$>$10 (our conclusions are unchanged if we use the absorbed flux maps instead). We calculate uncertainties for these measurements through a Monte Carlo procedure, adding normally-distributed noise to the reconstructions and re-measuring $G$, $M_{20}$, and $\Psi$, and taking the 16--84th percentile range of the distributions as the uncertainty on the measured values. We measure the difference between the absorption and the continuum for each of these quantities to mitigate effects from the varying spatial resolution across the source plane, because the same pixels are considered for both absorption and continuum.

\begin{figure*}
\begin{centering}
\includegraphics[width=\textwidth]{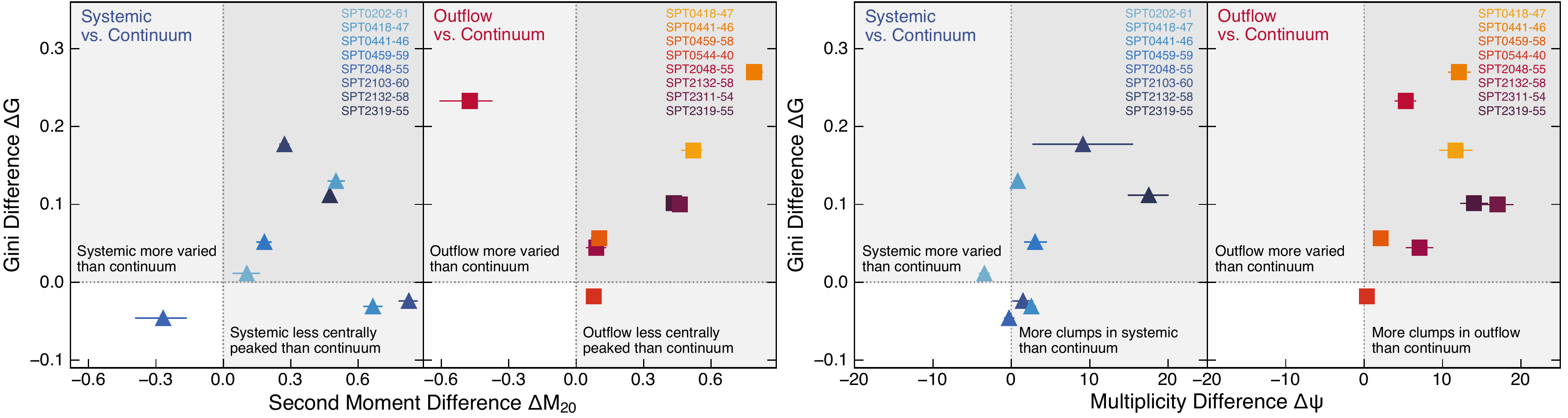}
\end{centering}
\caption{
The molecular outflows in our sample are clumpier than the continuum emission, and generally clumpier than the systemic absorption as well. We show the difference (absorption minus continuum) of the Gini coefficient $G$, second-moment parameter M$_{20}$, and multiplicity $\Psi$ for all sources. Qualitative descriptions of these differences are stated on the plots; sources in the upper-right quadrant can be considered to have clumpier structure in the absorption than the continuum by multiple metrics. The molecular outflows are significantly clumpier in general than the continuum, with 6/8 sources considered so by all three metrics. The systemic absorption, on the other hand, is generally less clumpy, with only 3/8 sources showing significant differences in all three metrics.
}\label{fig:clumpiness}
\end{figure*}

Figure~\ref{fig:clumpiness} shows the results of this procedure for the systemic and outflow absorption components of each source. This figure confirms what was apparent by eye from the reconstructions: the molecular outflows are significantly clumpier than the continuum emission. All outflows would be considered as clumpier by at least two metrics, and 6 of 8 outflows by all three metrics. The systemic absorption components, on the other hand, are less clearly clumpier than the continuum: only 3/8 sources show significant differences in all three metrics. The multiplicity $\Psi$ shows the largest differences between outflow and systemic absorption, with 6/8 outflows showing large differences $\Delta \Psi > 5$ between the outflow and continuum, compared to only 2/8 of the systemic reconstructions.

From this analysis, we conclude that the molecular outflows in $z>4$ DSFGs are much more irregular on 500pc scales than the continuum emission. These metrics and the reconstructions themselves make clear that the outflows are not uniform as expected in a simple spherically-expanding geometry. The equivalent width maps show that the absorption is rarely strongest where the continuum is brightest, even though absorption is easiest to detect in those regions. Instead the equivalent width often reaches its maximum values in small clumps near the outskirts of the continuum emission, as might be expected in the case of either a clumpy expanding shell or a scenario in which the lowest column densities in the ISM are more easily removed \citep[e.g.][]{thompson16a,hayward17}.

\section{Discussion} \label{discussion}

\subsection{Expectations and Prospects for Detecting Small-Scale Clumpiness} \label{smallclumps}

We have presented the first sample of spatially-resolved molecular outflows in the early universe, reaching typical spatial resolutions $\approx$200-800\,pc. The physics of OH absorption allow some predictions to be made for the smaller-scale structure of the outflows, beyond our current resolution limits.

OH 119\,\um absorption is highly optically thick in local galaxies, $\tau_\mathrm{OH119} \gtrsim 10$ even in the line wings \citep[e.g.][]{fischer10}; in effect, wherever OH is present along the line of sight $\sim$100\% of the 119\,\um continuum light is absorbed. Alternatively, if we observe that, say, 10\% of the continuum light has been absorbed from the OH spectra, we can infer that $\approx$10\% of the continuum source is covered by absorbing OH molecules.  For our sample, the outflow absorption components show depths of 5--20\%, which implies that only 5--20\% of the continuum emitting regions are covered by outflowing gas. These values are much smaller than either the overall detection rate (73$\pm$13\%) or the covering fractions from the lensing reconstructions.\footnote{There is some wiggle room at the factor-of-2 level allowed in this comparison if one allows significant turbulent velocity dispersion in the absorbing material at a given covering fraction, which effectively redistributes some amount of absorption to adjacent velocities; see \citet{gonzalezalfonso17}, their Figure~3.} Figure~\ref{fig:covfracs} compares these three quantities. How can these values be reconciled?

The difference with the outflow detection rate is easiest to consider: if molecular outflows typically have some fortuitous geometry, then the detection rate and the covering fractions need not be similar. Consider a simple scenario in which small spherical outflows are launched from the nuclear regions of a population of galaxies. In this case an outflow would be detected in virtually every source and from every viewing angle even if the molecular outflows cover only a small fraction of the galaxies before being decelerated or destroyed. While our reconstructions do not necessarily support this geometry, the argument is the same. There need not be a correspondence between the typical outflow covering fraction and a sample-averaged detection rate.

The comparison between the covering fractions inferred from the OH spectra and what we find from the lensing reconstructions has more interesting potential implications because now two different covering fraction estimates are being compared for individual galaxies. The differences between the two estimates in our sample span factors of 2--10, with the lensing reconstructions always showing larger covering fractions. The reconstructions also show that the outflowing gas tends to be distributed in several distinct clumps at the spatial resolutions we achieve, but the individual clumps themselves are generally not spatially resolved. While we cannot conclusively reconcile these two covering fraction methods with the data at hand, we briefly consider a few possibilities.

First, it is important to remember that the lensing reconstructions use data that span a wide range of velocities, using 300--700\,\kms of bandwidth from the ALMA spectra.  The reconstructions could thus represent the superposition of absorbing clouds at many different velocities that cover only small fractions of the sources at any individual velocity but a much larger fraction of the source when considered as an ensemble. This scenario would be eminently testable with more sensitive data at the same spatial resolution as we currently have, allowing more detailed lensing reconstructions that subdivide the blueshifted line wings into narrower velocity bins.

Second, the continuum could be `filled in' by dust emission from within the outflows, particularly relevant if a large fraction of the total dust mass is contained in the outflows and/or if the dust in the outflows is substantially warmer than in the galaxies. Cold dust has been detected in a handful of local outflows \citep[e.g.][]{leroy15,melendez15,barcosmunoz18}, comprising $\lesssim$10\% of the total dust mass in the galaxies when it has been possible to separate the contributions. This is insufficient to explain the gap between the outflow covering fractions from our OH spectra and reconstructions unless the dust in the outflows is much warmer than in the galaxies. \citet{melendez15} find that either warmer dust temperatures or steeper dust emissivity index could explain the differences they observe between the dust emission in the disk compared to the outflow in the nearby galaxy NGC4631. However, even under the assumption that the emissivity plays no role, the dust in the outflow is only $\sim$10\% warmer than the dust in the disk, well short of the $\sim$50--100\% warmer temperatures that would be required to reconcile the covering fraction estimates in our sample. Thus, while we expect that some portion of the absorption profiles has been filled in by emission from dust in the outflows, it is not likely to be enough to reconcile the difference in covering fractions we find.

Finally, it is very likely that the outflows we have observed contain substructure on spatial scales smaller than our current resolution limits. Most of the clumpy structures in the lensing reconstructions are not individually spatially resolved; we only identify distinct clumps because they are separated from each other by more than our effective resolution. Thus, we would expect that if these outflows were observed at higher spatial resolution that the covering fractions in the lensing reconstructions would decrease, most likely in conjunction with any decrease resulting from modeling narrower velocity bins in the outflows as well. Note that this implicitly assumes that the spectra-based covering fractions are the `true' values.

\begin{figure}
\begin{centering}
\includegraphics[width=\columnwidth]{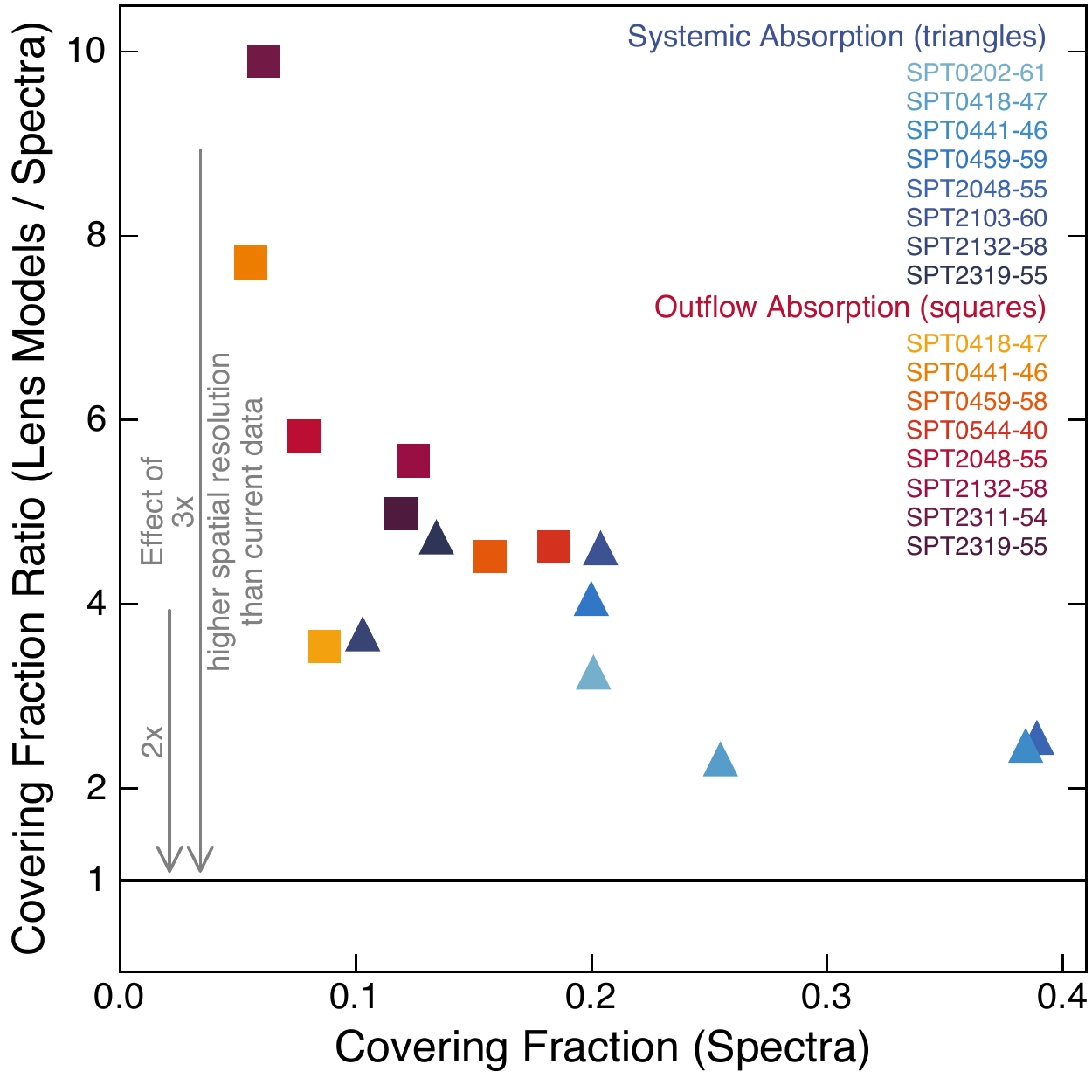}
\end{centering}
\caption{
We estimate that the substructures in the molecular outflows would be directly resolved at $\gtrsim$3$\times$ better resolution than the current data provide, corresponding to 50--200pc scales. The ratio of our measured covering fractions from the lensing reconstructions to those from the OH spectra provides an estimate of the spatial resolution that would be required to directly resolve the absorbing substructures because OH is expected to be highly optically thick. We implicitly assume that the spectra-based covering fractions are the `true' values and neglect any emission from dust in the outflows, which would move the points downward on this plot.
}\label{fig:covfracs_hires}
\end{figure}

Ignoring any re-emission of continuum from dust in the outflows, the ratio between the two covering fraction estimates corresponds to the expected size scale required to resolve the substructures in the outflows. If we take the spectrum-based covering fractions as the `true' values, then the differences with our lens modeling results must be due to insufficient spatial resolution; higher-resolution observations would then yield lower covering fractions until in agreement with the spectrum-based values. This is shown more clearly in Figure~\ref{fig:covfracs_hires}. This Figure shows the ratio of the covering fractions derived from the lens models to those from the spectra (or the ratio of the bars to black lines in Figure~\ref{fig:covfracs}). Assuming the OH absorption is optically thick, this ratio is a proxy for the spatial resolution that would be required to directly resolve the absorbing structures. We estimate that linear spatial resolutions 2--3 times better than the current data would be capable of resolving the absorption.\footnote{That is, doubling or tripling the spatial resolution resulting in a factor of 4--9 decrease in the beam area.} Given the resolution and covering fractions measured in the current data, we would predict that the true size scale of clumps in the molecular outflows we have observed is of the order 50--200\,pc. We note that a similar estimate can be obtained simply by combining the covering fractions from the OH spectra with the observed continuum sizes. Even such small spatial scales are accessible with ALMA in lensed objects given sufficient observational investment.

\subsection{\cii is Not a Reliable Molecular Outflow Tracer} \label{ciicomparison}

We have detected molecular outflows in a large fraction of our sample, but OH is an inconvenient outflow tracer. Because the outflows are detected in absorption, galaxies bright in the FIR continuum are required, limiting us to dusty objects. The rest frequency of the 119\,\um doublet feature is also rather inconvenient because the atmospheric transmission limits observations to particular redshift ranges for $z<6$ galaxies and is inaccessible from the ground for $z\lesssim2$.

One potential alternative outflow tracer is \cii 158\,\um, where the outflow manifests as excess emission in broad line wings at high relative velocities, or as emission that does not follow the overall galactic rotation curve if high-resolution data is available. \cii is excited in a wide variety of gas conditions and arises from warm ionized, neutral atomic and molecular gas \citep[e.g.][]{langer14}. In the nearby universe \citet{janssen16} find tentative correlations between the OH outflow velocity and the broad \cii line width and between the mass outflow rates derived from each tracer in a sample of ULIRGs and QSOs, suggesting that \cii could prove a useful tracer of outflows at high redshift as well.

In the distant universe, a \cii outflow has been reported in an individual $z=6.4$ quasar \citep{maiolino12,cicone15}, although such bright \cii line wings as in those works have not been replicated in any other individual high-redshift source or confirmed through deeper observations of the original target. Now that ALMA observations of \cii at $z>4$ are fairly routine, several groups have stacked the \cii spectra of various galaxy samples in an effort to detect broad line wings indicative of outflows, with mixed success. \citet{gallerani18,bischetti19,ginolfi20} each report \cii outflows in stacked spectra of, respectively, 9 $z\sim5.5$ UV-selected star-forming galaxies from \citet{capak15}, 48 $4.5<z<7.1$ quasars assembled from the ALMA archive, and 50 $z\sim5$ star-forming galaxies from the ALPINE survey \citep{lefevre19,bethermin20,faisst20}. Other quasar studies, however, have found no or only marginal evidence for \cii outflows despite many overlapping sources with the Bischetti \etal sample \citep[e.g.][]{decarli18,stanley19}, suggesting that perhaps residual systematic uncertainties in the stacking make the detection and interpretation of broad \cii wings less straightforward. \cii wings are also not seen towards an individual $z\sim6$ quasar with a tentative OH outflow, although the \cii sensitivity is too low to be conclusive \citep{shao17,herreracamus20}.

\begin{figure*}[t]
\begin{centering}
\includegraphics[width=\textwidth]{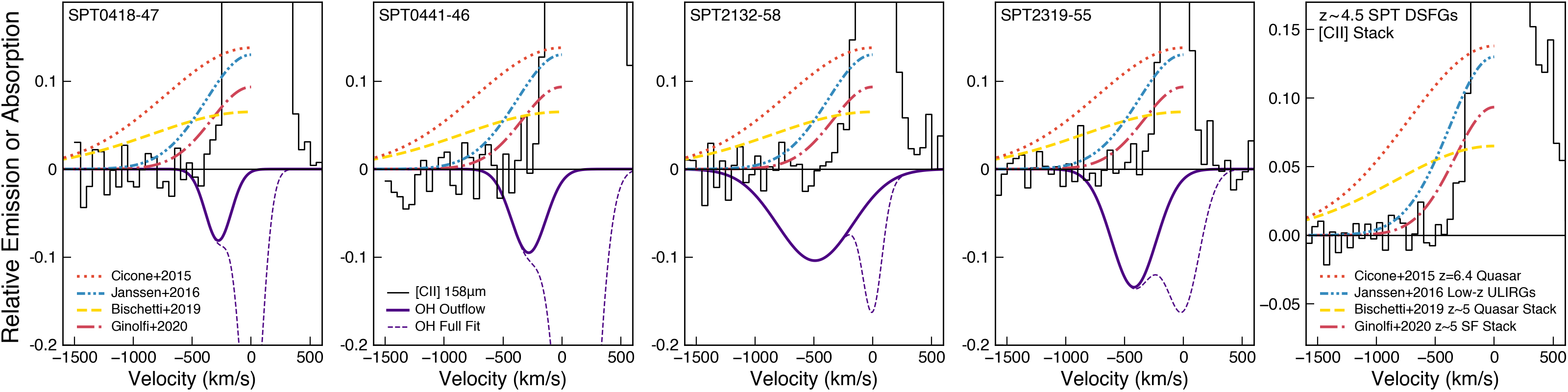}
\end{centering}
\caption{
\cii is at best an unreliable indicator of whether molecular outflows are present. We compare \cii and OH for the four sources in our sample with unambiguous OH outflows and high-quality ALMA \cii data (left four panels), and the stacked \cii spectra of these four sources (right-most panel). The \cii spectra have been normalized to peak at 1. The OH best-fit absorption profiles have been normalized by the 119\,\um continuum level and show only one of the doublet components for clarity. 
Colored lines show several claimed outflows seen as broad \cii line wings in a $z=6.4$ quasar \citep{cicone15}, the median result for a sample of 22 nearby ULIRGs \citep{janssen16}, a stack of 22 $z\sim5.5$ quasars \citep{bischetti19}, and a stack of 25 $z\sim5$ star-forming galaxies \citep{ginolfi20}. We find no evidence for excess \cii emission at the velocities with clear blueshifted OH absorption. The signal-to-noise in our data rules out \cii wings similar to these studies in individual sources and strongly excludes them in the stacked spectrum.  }\label{fig:ciioh}
\end{figure*}

Four galaxies in our sample with unambiguous OH outflows also have high-quality ALMA observations of \cii (i.e. signal-to-noise that rivals or exceeds that of the stacking results above). These data will be analyzed in detail in future work. Figure~\ref{fig:ciioh} compares the OH absorption and \cii emission profiles of these four objects. No individual source shows evidence of excess \cii emission at the velocities where we detect OH outflows. Moreover, \cii wings similar to those reported in the literature would have been easily detected in all individual targets,\footnote{To make the most fair comparison in each case, we compare to the median properties of the \citet{janssen16} sources, the 2'' aperture fit of \citet{cicone15}, an average of the FWHM$>$400\,\kms subsets of \citet{bischetti19}, and the SFR$_\mathrm{median}=50$\,\Msol/yr result of \citet{ginolfi20}.}
with the possible exception of the \citet{ginolfi20} stacking result; the median SFR of those objects is more than an order of magnitude lower than our sample. 

A still more stringent constraint comes from stacking the \cii spectra of our four targets, which again shows no evidence for high-velocity \cii emission, strongly excluding line wings similar to those reported in the literature (Figure~\ref{fig:ciioh}, right). We construct this stack by subtracting a linear continuum  from the integrated spectrum of each source excluding velocities $|v| < 500$\,\kms, regridding each spectrum to a common velocity frame, and normalizing each spectrum by the peak of the \cii emission before averaging. We tried a variety of other stacking procedures and continuum subtraction methods and found that our result is not sensitive to these details because the individual input spectra are all very high S/N.

At the very least, then, we must conclude that even if \cii can be used as a molecular outflow tracer, it does not do so reliably given our non-detections in sources with clear OH outflows. We note that any vagaries of gravitational lensing or differential magnification cannot explain the lack of high-velocity \cii emission: each source shows high-velocity OH absorption which must necessarily be down-the-barrel and share essentially the same magnification as the dust continuum. Even if the \cii traces a more extended or diffuse outflow component, the portion along our line of sight would still have been detectable just as the OH was. We also note that the high-velocity \cii found by \citet{ginolfi20} in stacked images is compact and centered on the galaxies, while the core systemic emission is more extended, as expected for a shell-like geometry. We consider a few other possible explanations for our lack of evidence for \cii outflows. 

First, none of the literature stacking results targeted nor necessarily include DSFGs such as our objects; no clean comparison result for DSFGs is currently available. The \citet{ginolfi20} stacking result is the most difficult to rule out with our sample due to the intrinsically broader line profiles of our sources. The galaxies in that sample are both lower-mass and lower-SFR than our targets, and it seems unlikely such objects would drive higher outflow rate or brighter outflows than our more extreme targets. Meanwhile, the quasars of \citet{bischetti19} are probably fairly similar to our target galaxies in mass, but we have no evidence for AGN in any of our targets. Perhaps the additional energy and momentum from an accreting black hole are required to boost the \cii line wings to observable fluxes.

Second, most recent \cii outflow detections have relied on stacking low S/N spectra of many objects, which introduces additional complexity and uncertainty in the results. Given the faint line wings to be searched for and heterogeneity in the input objects, great care must be taken in the details of the stacking, including continuum subtraction, relative scaling, non-Gaussian line shapes in the input spectra, and the treatment of differing line widths among input sources. These details may also explain the differences in the quasar sample stacking results in the literature  even though these studies have many individual sources in common \citep{decarli18,bischetti19,stanley19}. It could also be that \cii is rarely apparent in outflows but when it is present it is unusually bright, such that the stacked signal seen in literature studies is dominated by a few atypical objects.

Third, OH is in principle a far more sensitive tracer of weak outflows than \cii, mostly due to the fact that it is an absorption tracer sensitive to column (rather than volume) density and the very high Einstein $A$ coefficients of the doublet lines. While we consider it highly unlikely given the wealth of evidence for high line opacities at low redshift, if the OH absorption were optically thin, the OH data are sensitive to outflow masses as low as $\sim2-6\times10^7$\,\Msol \citepalias{spilker20a}. On the other hand, following the calculations of \citet{haileydunsheath10}, our \cii data are sensitive to outflow masses $\sim5-15\times10^7$\,\Msol, or $\lesssim$5\% of the total molecular gas masses of these galaxies. These values are highly uncertain due to the unknown abundances of OH and C and the ionization fraction of C$^+$ but would reconcile the OH and \cii observations. More importantly, it is almost certainly not true that OH is optically thin (even the \cii may have non-negligible line opacity, though still far lower than expected for OH; e.g. \citealt{gullberg15}).

Finally, \cii is often stated to be a tracer of neutral atomic outflowing gas, and not necessarily of molecular material (although \cii is known to be emitted in both these phases as well as ionized gas). Perhaps, then, the molecular gas dominates the mass budget of these outflows, with atomic gas a smaller contribution to the total outflow mass. In \citetalias{spilker20a}, we find best estimates for the molecular outflow masses for these four sources $\sim40-80\times10^7$\,\Msol. If we assume the outflow mass sensitivity from \cii in the previous paragraph is solely for the atomic phase, this would require molecular-to-atomic outflow mass ratios of $\sim5-10$ to reconcile the \cii and OH observations. While the distribution of mass across outflow phases is completely unknown for high-redshift DSFGs from observations, this range is not infeasible based on estimates from chemical modeling in idealized hydrodynamic simulations \citep{richings18} or the few lower-redshift observational estimates \citep{feruglio10,rupke13,herreracamus19}.

Regardless of the reason, our data make clear that \cii should not be considered a reliable molecular outflow tracer. We find no evidence for excess \cii emission at the velocities where we see blueshifted OH absorption or excess emission at the level seen in literature stacking experiments. Further work is needed to determine exactly why \cii does not always appear in outflows.

\section{Conclusions} \label{conclusions}

In this work we have presented the first statistical sample of molecular outflows observed in $4 < z < 5.5$ DSFGs using ALMA observations of the ground-state OH 119\,\um doublet lines towards 11 DSFGs selected from the SPT DSFG sample. Our target galaxies, all of which are gravitationally lensed, are IR-luminous, $\log \lir/\Lsol > 12$, but do not show obvious signs of AGN activity in rest-frame mid-IR data. We detect OH in absorption in all objects. The observations also spatially resolve the targets, and we create source reconstructions of the rest-frame 120\,\um dust continuum emission and the OH line absorption. These galaxies represent the largest sample at $z>4$ with constraints on the molecular outflow properties, as well as the largest (and so far only) sample to spatially resolve the structure of said outflows in the early universe. Our main conclusions can be summarized as follows:

\begin{itemize}

\item We find unambiguous evidence for molecular outflows in 8 of the 11 sample targets; the remaining sources have broad CO or \cii line profiles that make interpretation of the OH absorption complicated. No source shows unambiguous evidence for molecular inflows. The outflow detection rate, 73$\pm$13\%, is similar to that found from OH observations of nearby ULIRGs and QSOs and significantly higher than for lower-luminosity AGN-dominated galaxies at low redshift. Because outflows with OH are only detectable in absorption, this detection rate is a lower limit on the true occurrence rate of molecular outflows in $z>4$ DSFGs; molecular outflows must be ubiquitous and widespread in such objects (Figure~\ref{fig:detrates} and Section~\ref{detection}).

\item The outflows reach maximum velocities 430--1200\kms and possibly faster. The distribution of outflow velocities in the $z>4$ sample is indistinguishable from that of low-redshift IR-luminous galaxies (Figure~\ref{fig:veleqw} and Section~\ref{basics}).

\item Using our lensing reconstructions of the sources, we measure the structure of the outflows on $\sim$500\,pc spatial scales. The covering fraction of the outflowing molecular gas is correlated with both \lir and outflow velocity, such that more luminous sources and those with the fastest outflows also host the outflows with the highest covering fractions (Figures~\ref{fig:covfracs_trends} and \ref{fig:covfracs_vel}).

\item The molecular outflows show significantly more clumpy structure than the (generally smooth) dust continuum emission on scales of $\sim$500\,pc, which we quantify with metrics borrowed from the literature on high-redshift star formation (Figure~\ref{fig:clumpiness} and Section~\ref{clumpiness}). While the clumps are not directly resolved, from optical depth arguments we expect that higher-resolution observations of the outflows would reveal substantially more clumpy structure on 50--200\,pc spatial scales (Figure~\ref{fig:covfracs_hires} and Section~\ref{smallclumps}).

\item We find no evidence of high-velocity wings in the \cii line profiles of the four sources with high-quality \cii spectra and obvious OH outflows; \cii is at best an unreliable indicator of molecular outflows. We strongly rule out \cii line wings at the level reported in several literature stacking results (though none explicitly targeted high-redshift DSFGs similar to our sample). This may be due to lingering systematics in the stacking results or genuine differences between the \cii outflow properties of DSFGs like our sample and the variety of objects for which \cii outflows have been claimed (Figure~\ref{fig:ciioh} and Section~\ref{ciicomparison}).

\end{itemize}

This work has largely focused on the structural properties of the OH outflows we have detected. In the second paper in this series we explore the physical properties of these outflows in much greater detail, including outflow rates and masses and implications for the outflow driving mechanisms. Moreover, the outflow occurrence rate, velocity distributions, and detailed structural properties we find place novel observational constraints on simulations of galactic feedback and winds in the early universe. While clearly a larger high-redshift sample that spans a wider range in galaxy properties will be required to fully understand galactic winds in the distant universe, this sample represents a first step towards understanding outflow properties among the general high-redshift galaxy population and the utility of OH absorption for characterizing these properties in statistical samples.

\acknowledgements{
We thank the referee for a thorough and constructive report that improved the quality of this paper.
JSS thanks the McDonald Observatory at the University of Texas at Austin for support through a Harlan J. Smith Fellowship and the Texas Advanced Computing Center (TACC) for providing high-performance computing resources. JSS is supported by NASA Hubble Fellowship grant \#HF2-51446  awarded  by  the  Space  Telescope  Science  Institute,  which  is  operated  by  the  Association  of  Universities  for  Research  in  Astronomy,  Inc.,  for  NASA,  under  contract  NAS5-26555. 
K.C.L., S.J., D.P.M., K.P., and J.D.V. acknowledge support from the US NSF under grants AST-1715213 and AST-1716127.
This work was performed in part at the Aspen Center for Physics, which is supported by National Science Foundation grant PHY-1607611.

This paper makes use of the following ALMA data: ADS/JAO.ALMA\#2015.1.00942.S, ADS/JAO.ALMA\#2016.1.00089.S, ADS/JAO.ALMA\#2016.1.01499.S, ADS/JAO.ALMA\#2018.1.00191.S, and ADS/JAO.ALMA\#2019.1.00253.S. ALMA is a partnership of ESO (representing its member states), NSF (USA) and NINS (Japan), together with NRC (Canada), MOST and ASIAA (Taiwan), and KASI (Republic of Korea), in cooperation with the Republic of Chile. The Joint ALMA Observatory is operated by ESO, AUI/NRAO and NAOJ. The National Radio Astronomy Observatory is a facility of the National Science Foundation operated under cooperative agreement by Associated Universities, Inc.

This research has made use of NASA's Astrophysics Data System.
}

\facility{ALMA}

\software{
CASA \citep{mcmullin07},
\texttt{visilens} \citep{spilker16},
\texttt{ripples} \citep{hezaveh16},
\texttt{astropy} \citep{astropy18},
\texttt{matplotlib} \citep{hunter07}}

\clearpage
\bibliographystyle{aasjournal}

\restartappendixnumbering
\appendix

\section{Supplementary Lens Model Results} \label{applensing}

This appendix provides additional results and diagnostic plots from the lens modeling procedure. 

Table~\ref{tab:lensparams} lists the best-fit lens model parameters. As described in Section~\ref{lensing}, the lensing potential in each source is parameterized as one or more SIE mass profiles with optional external tidal shear and low-order angular multipoles (see \citealt{hezaveh16} for a more thorough description of these parameters).

\begin{figure*}
\begin{centering}
\includegraphics[width=0.495\textwidth]{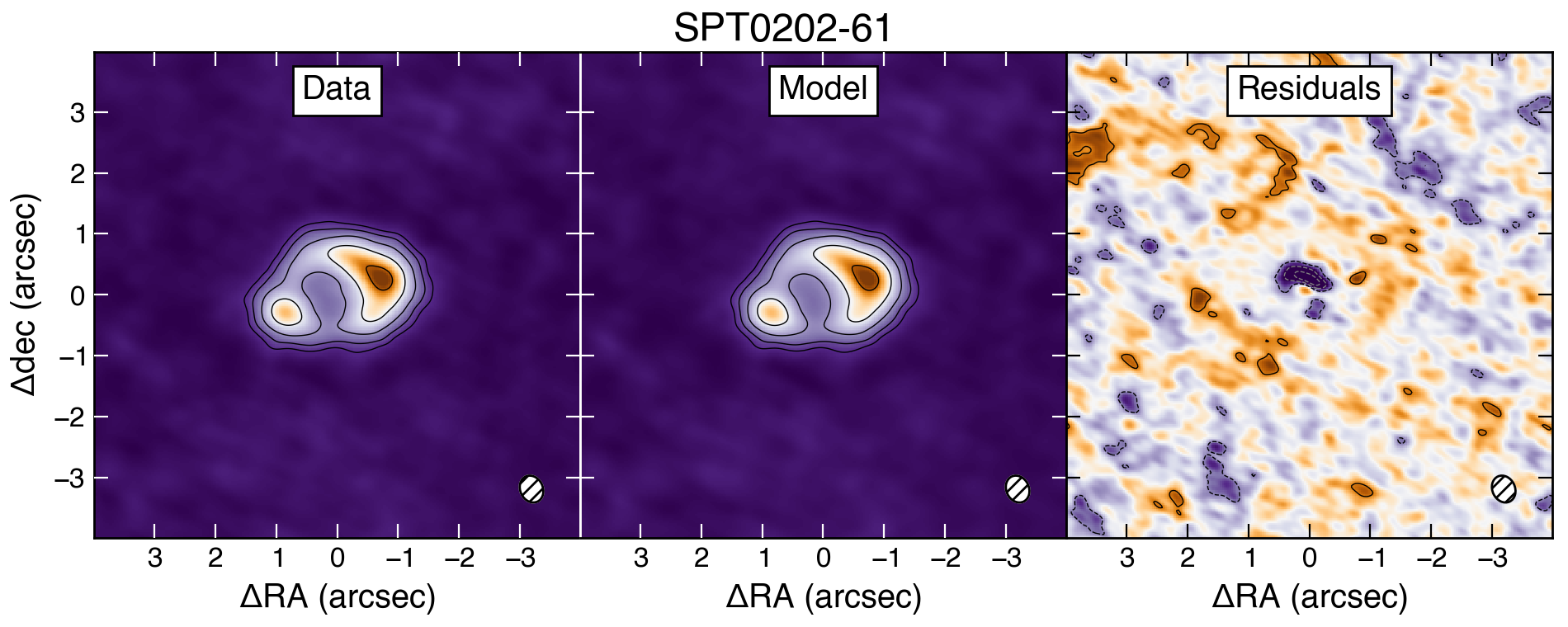}
\includegraphics[width=0.495\textwidth]{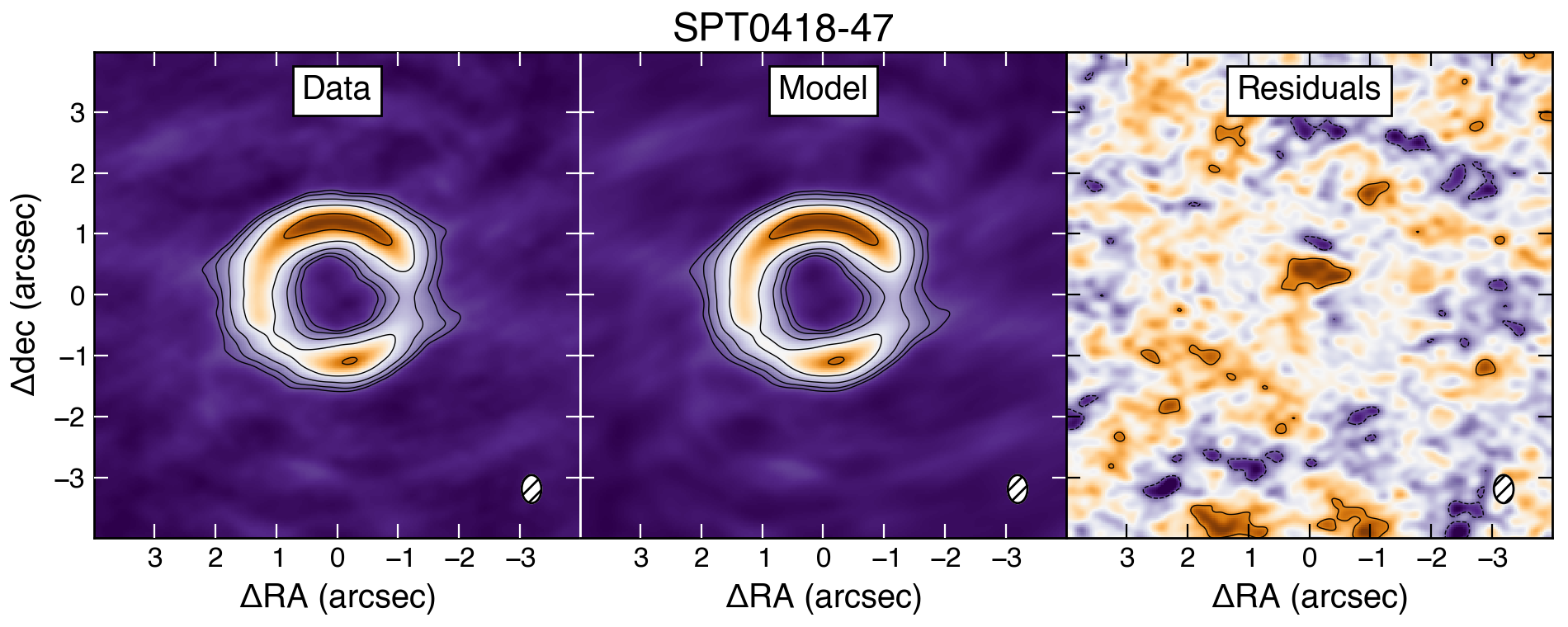}
\includegraphics[width=0.495\textwidth]{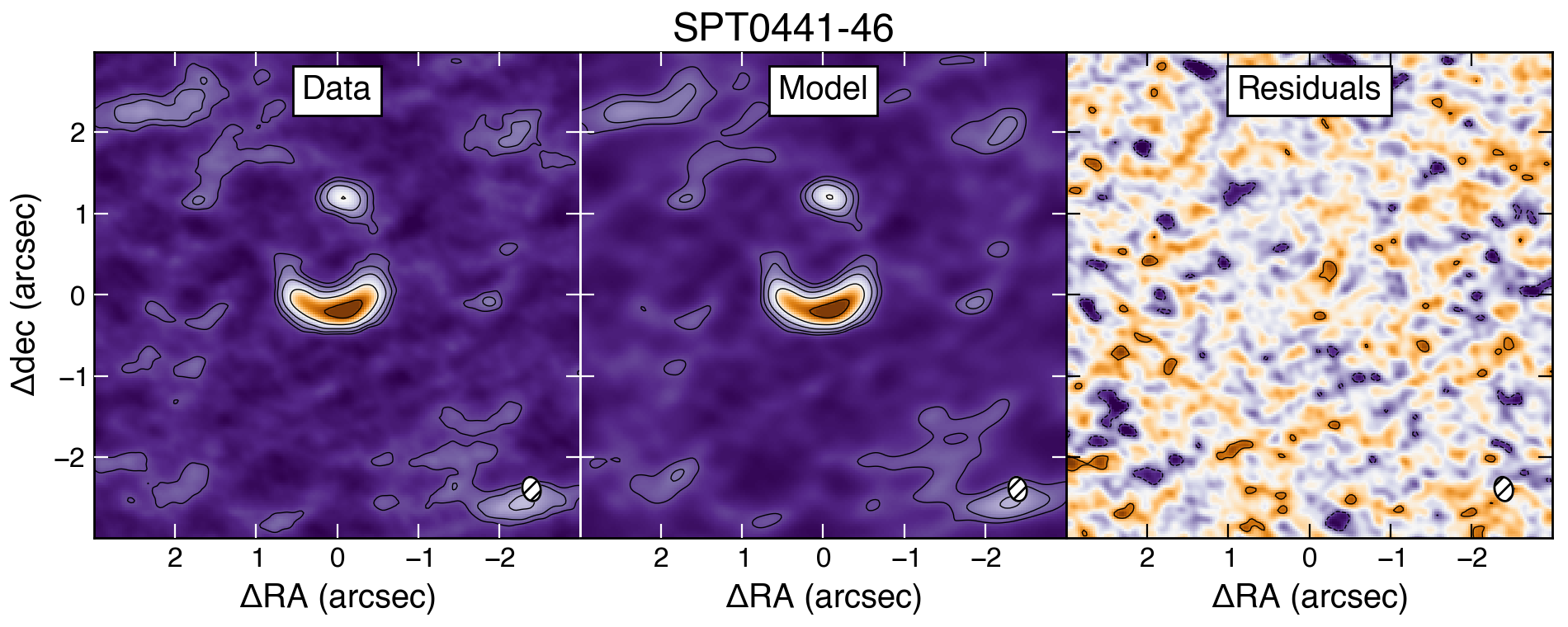}
\includegraphics[width=0.495\textwidth]{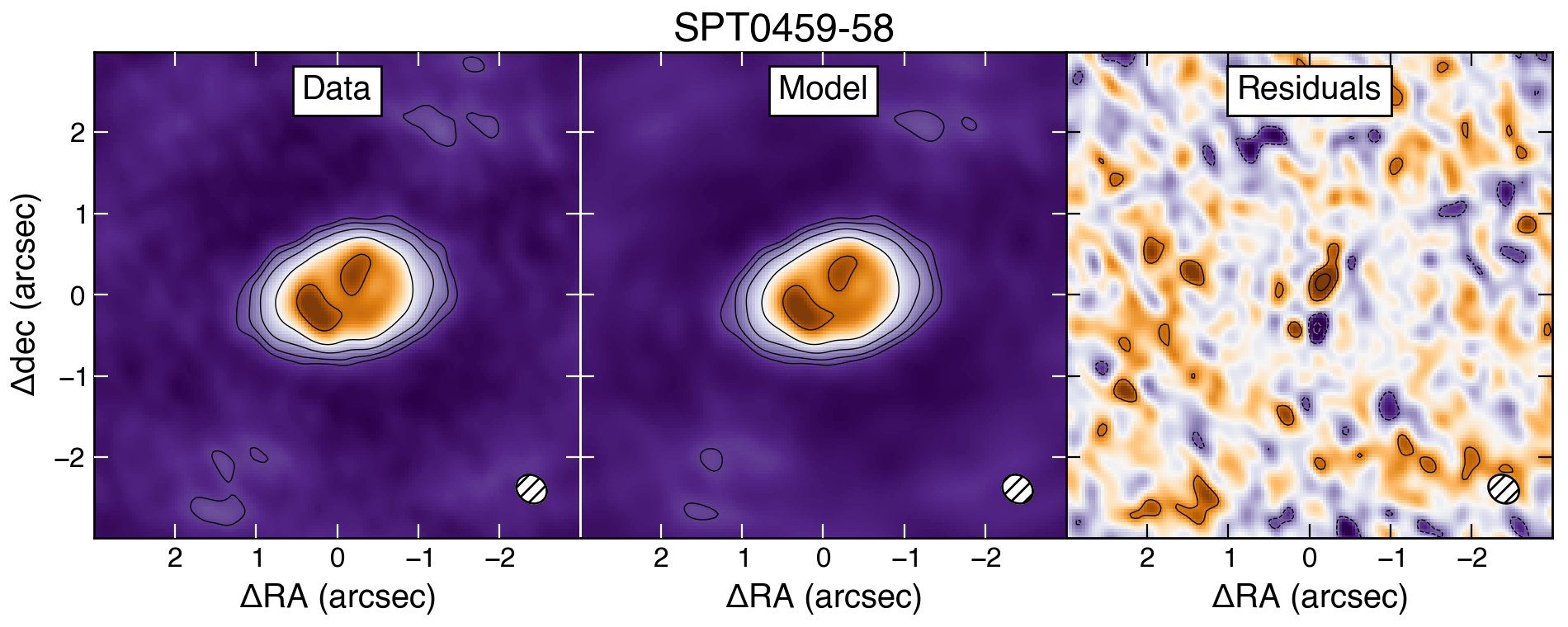}
\includegraphics[width=0.495\textwidth]{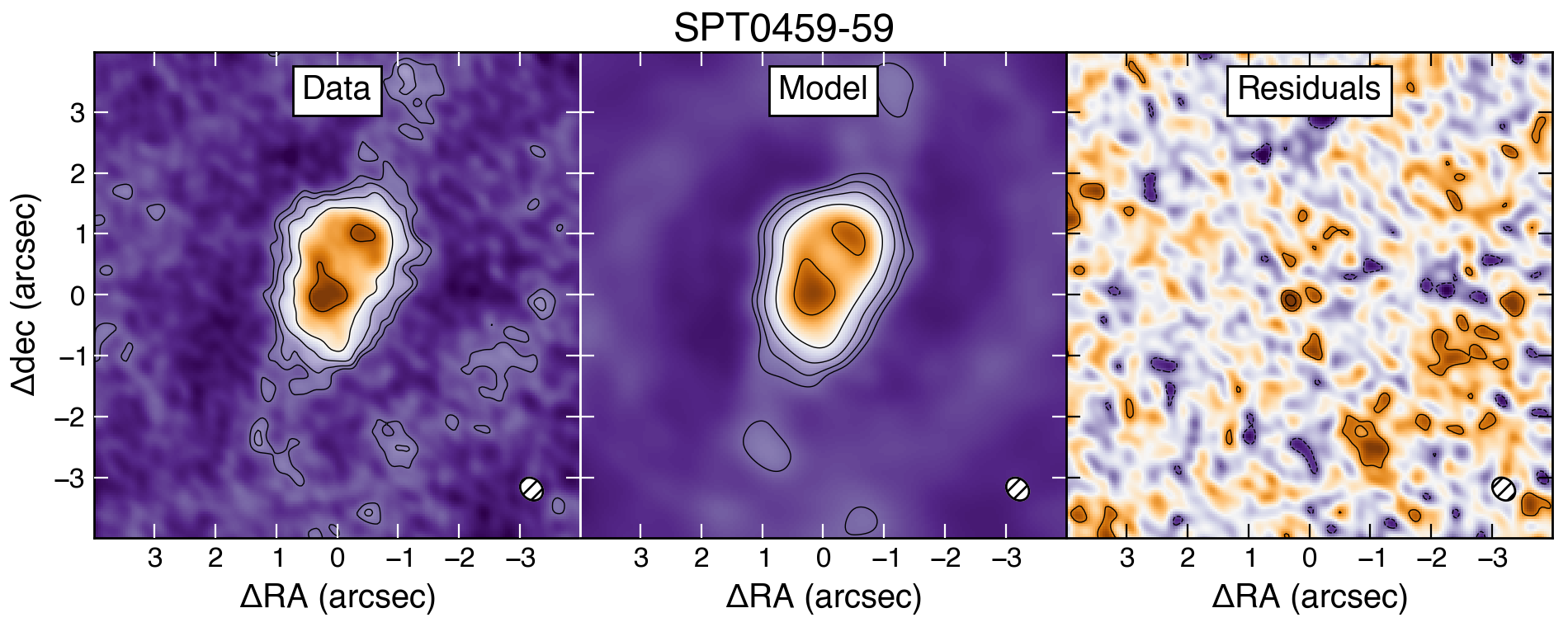}
\includegraphics[width=0.495\textwidth]{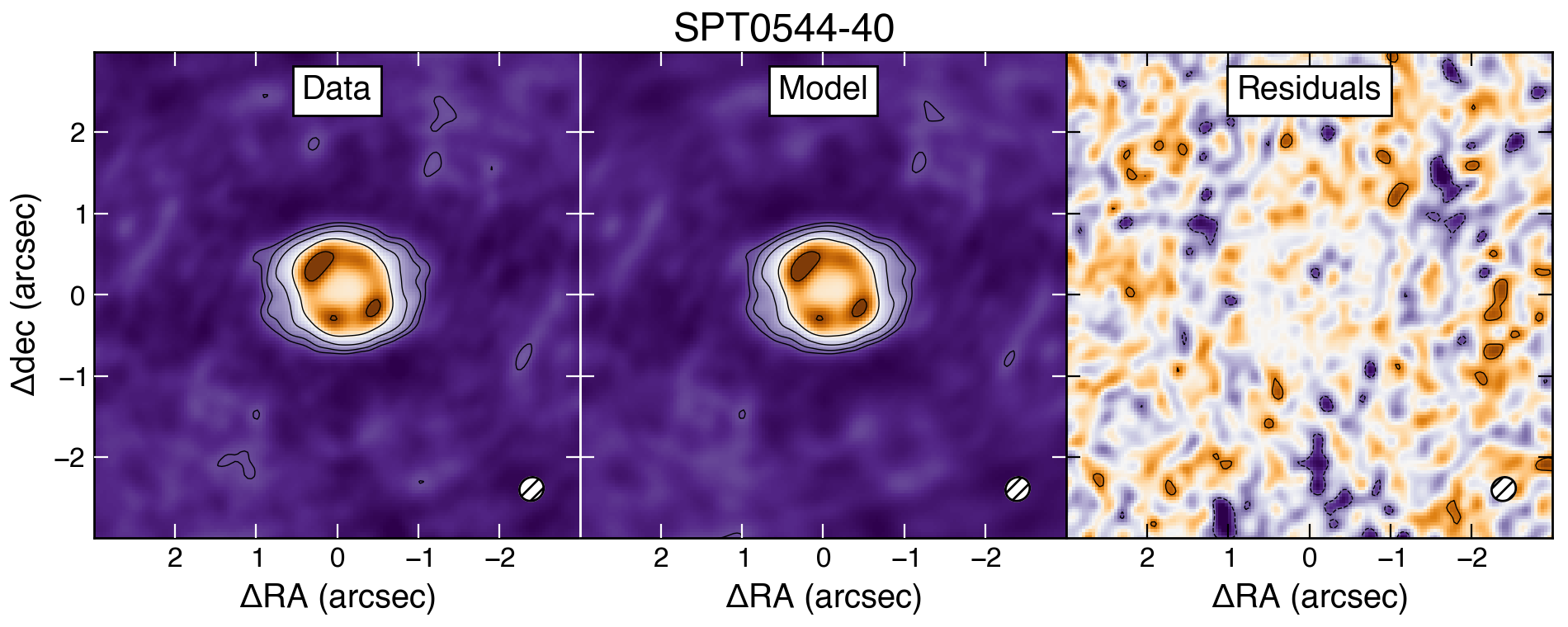}
\includegraphics[width=0.495\textwidth]{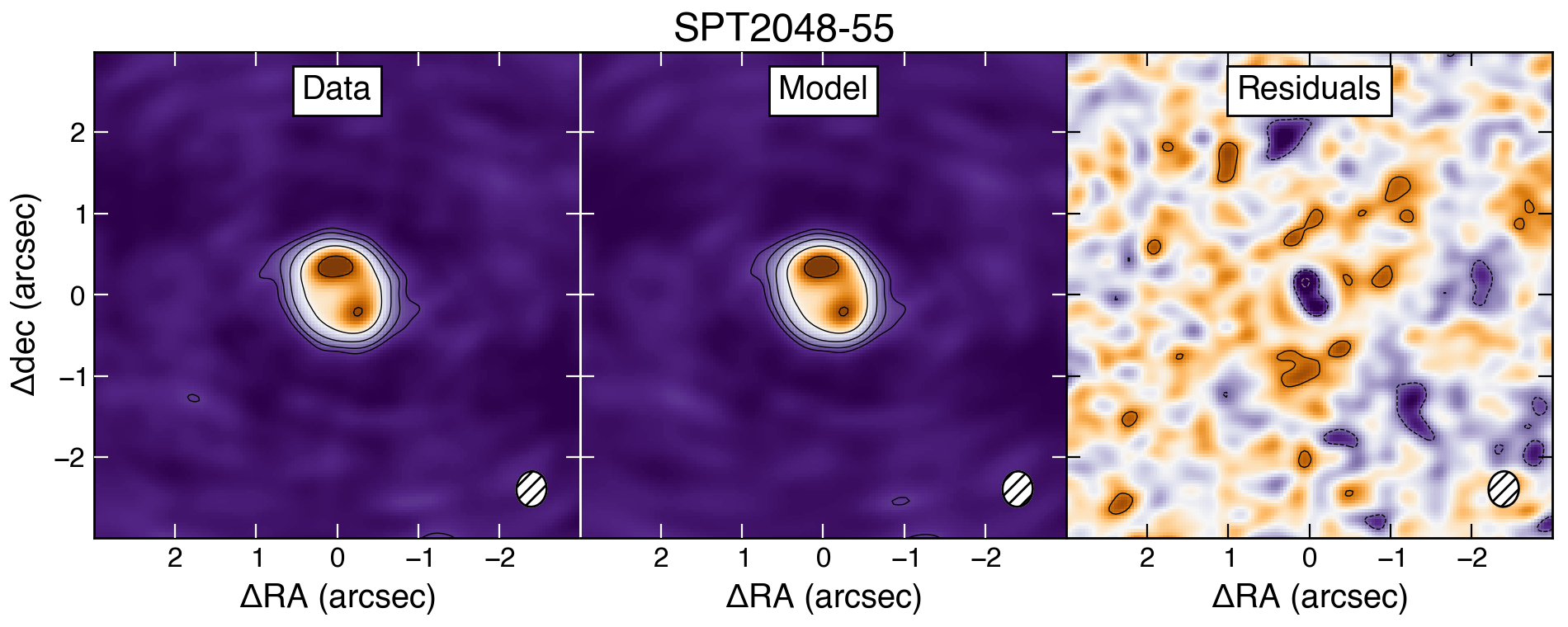}
\includegraphics[width=0.495\textwidth]{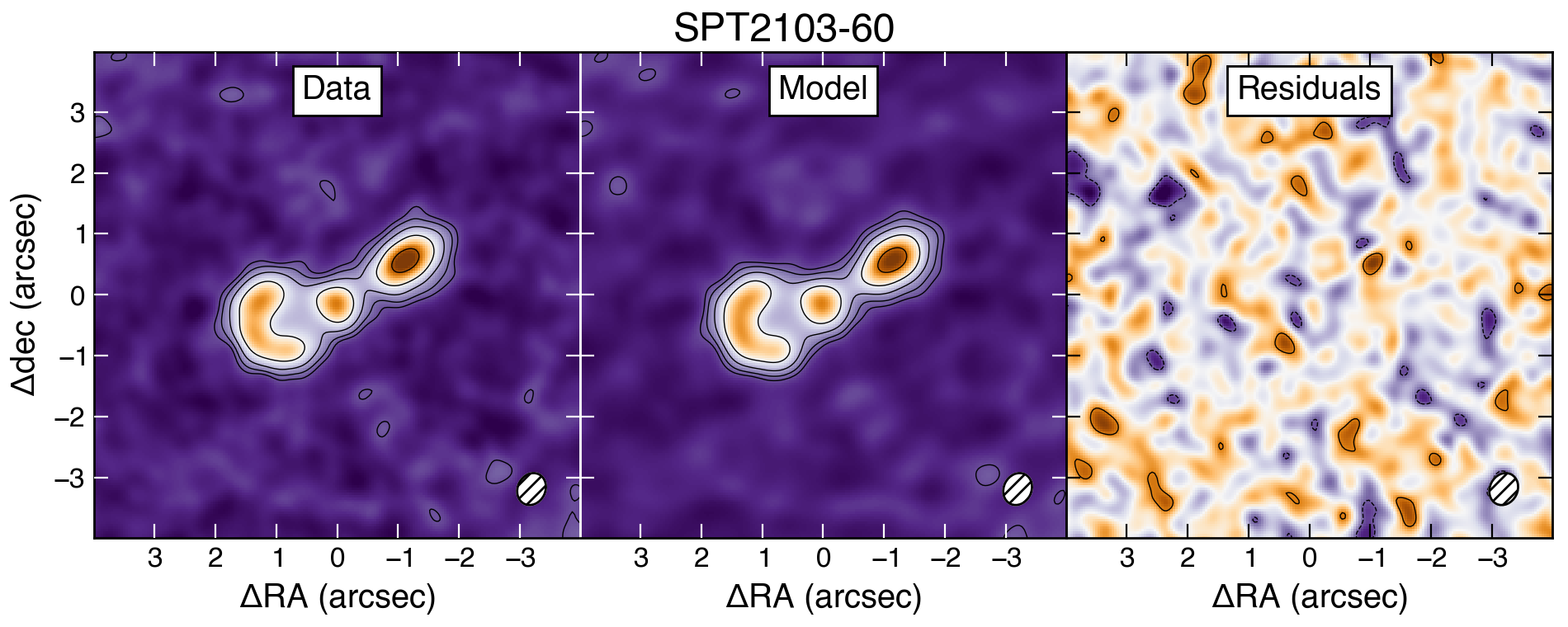}
\includegraphics[width=0.495\textwidth]{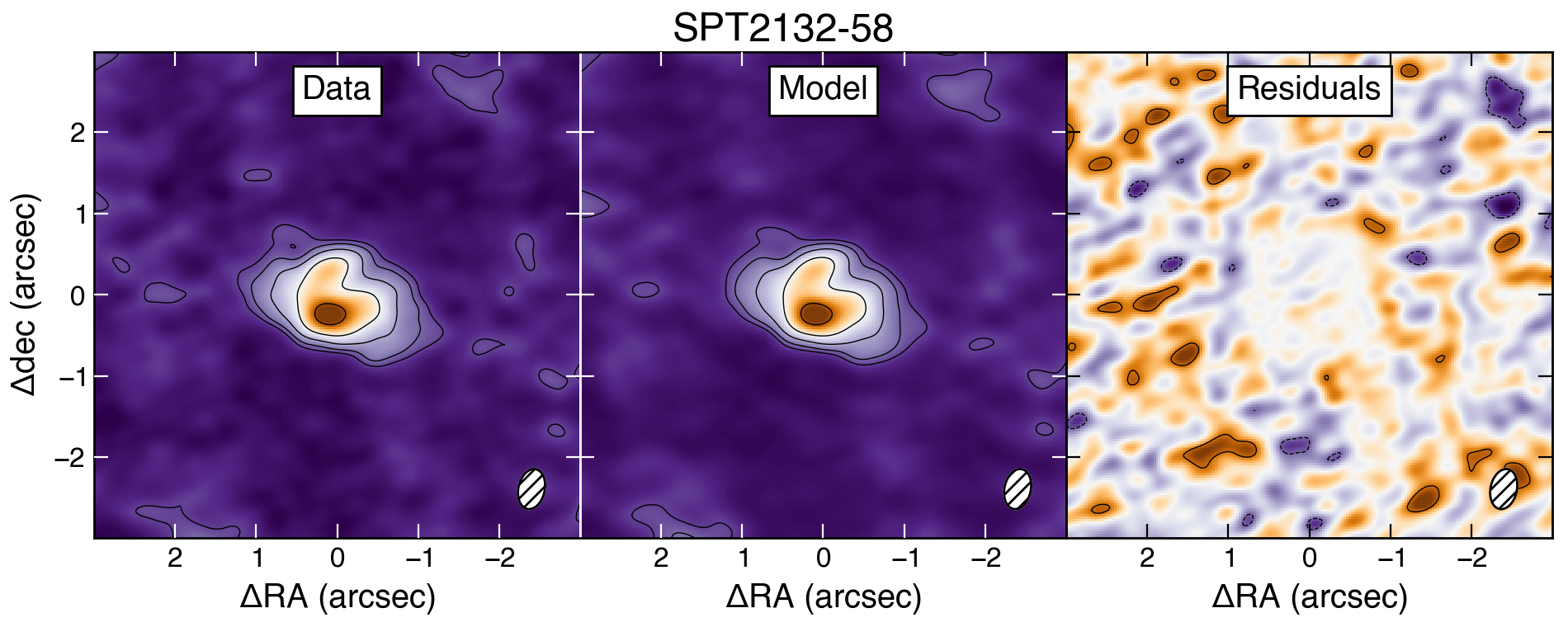}
\includegraphics[width=0.495\textwidth]{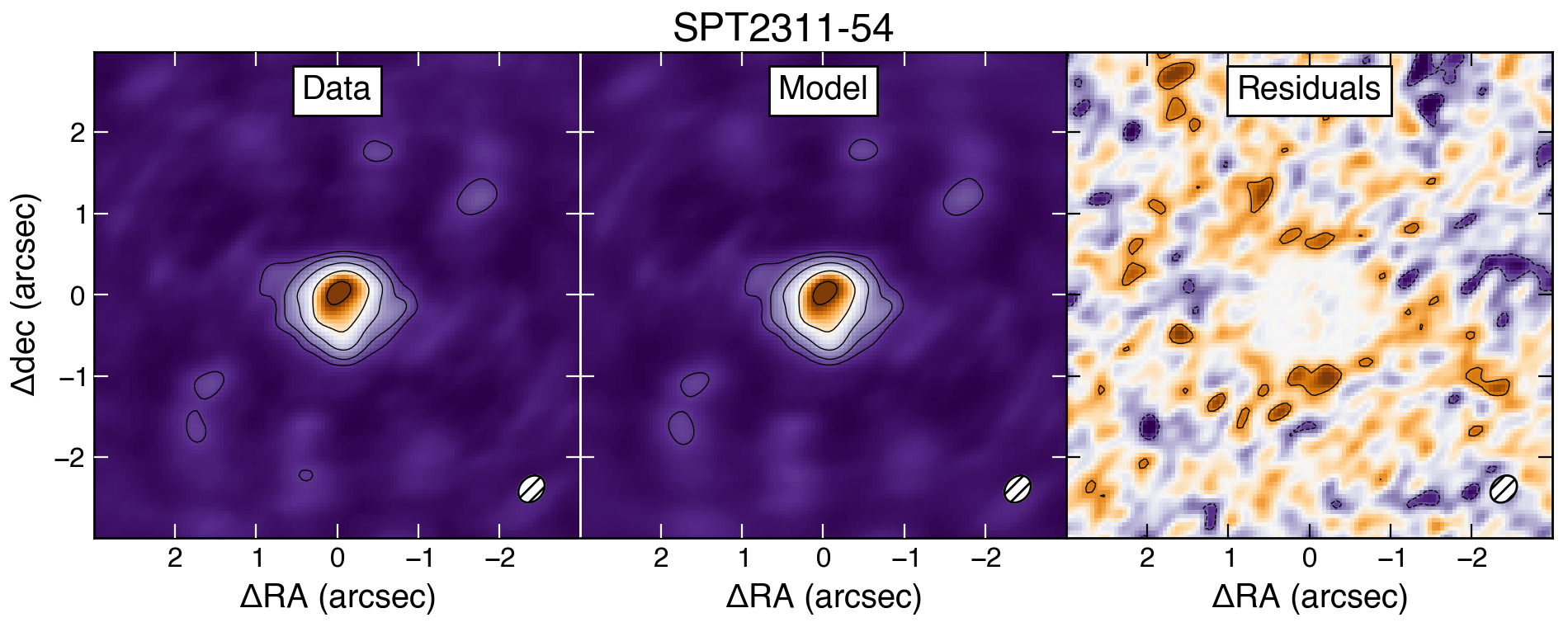}\\
\end{centering}
\caption{
Continuum lens modeling results showing the data, model, and residuals. Contours in the data and models are 5, 10, 20, ...\% of the peak, and contours in the residuals are in steps of $\pm$2$\sigma$. Axes are relative to the ALMA phase center. Data and model images are dirty (not deconvolved) because the fitting is performed in the Fourier domain; the sidelobe structure resulting from the $uv$ coverage of the observations in the data should be reproduced in the model. Emission from the lens in SPT0202-61 and southwestern source in SPT0459-59 has been subtracted in the visibilities prior to the lens modeling (visible in Figure~\ref{fig:contimages}); it is clear from the residual maps that this subtraction is imperfect.
}\label{fig:datamodresid}
\end{figure*}

Figure~\ref{fig:datamodresid} shows the data, model, and residuals from the lens modeling for the 119\,\um continuum. 

\begin{figure*}
\begin{centering}
\includegraphics[width=0.245\textwidth]{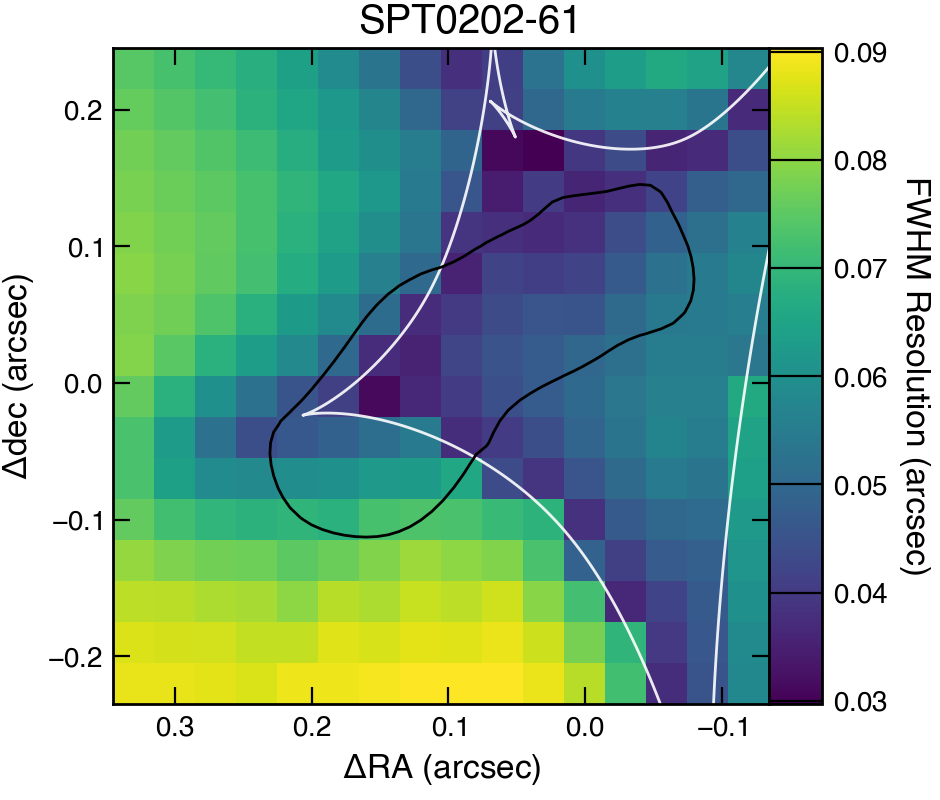}
\includegraphics[width=0.245\textwidth]{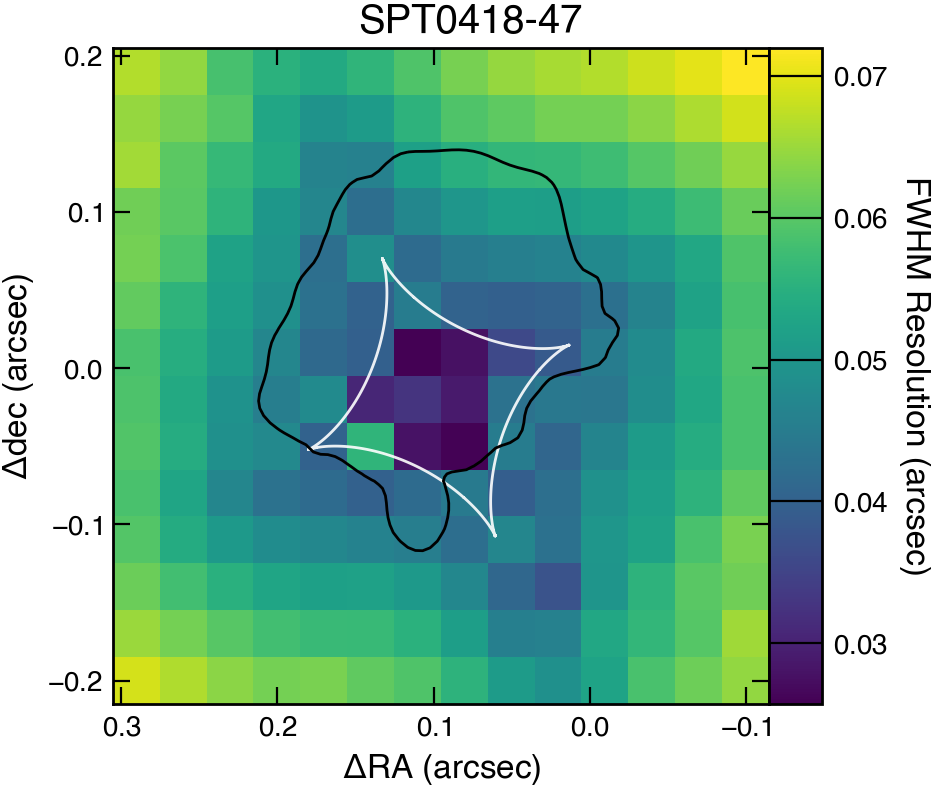}
\includegraphics[width=0.245\textwidth]{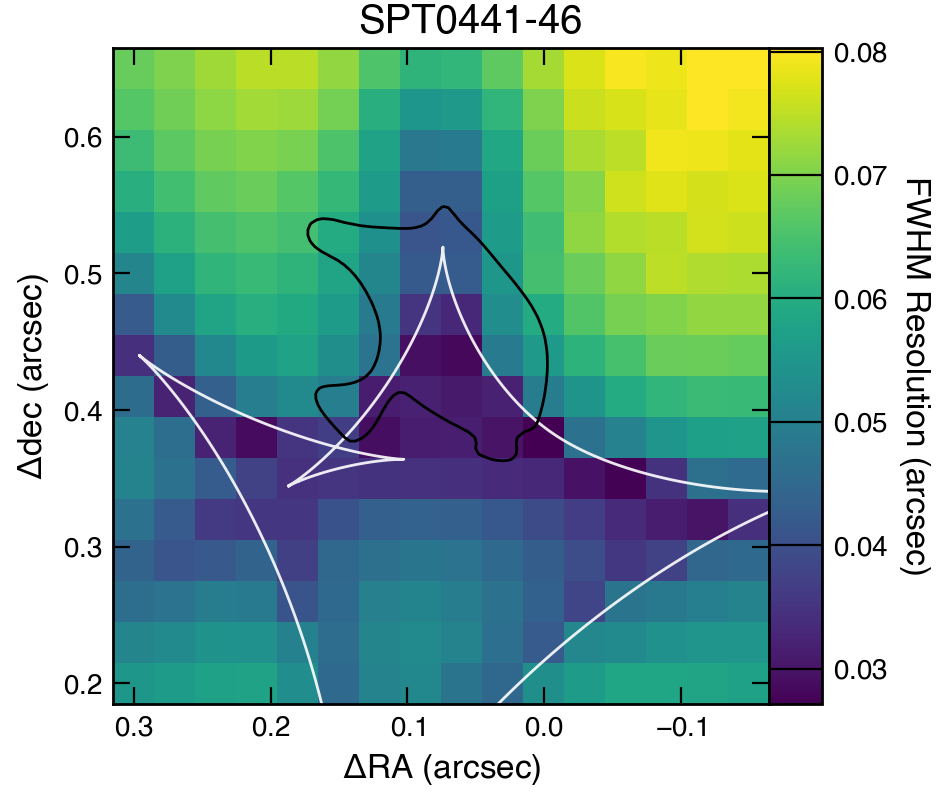}
\includegraphics[width=0.245\textwidth]{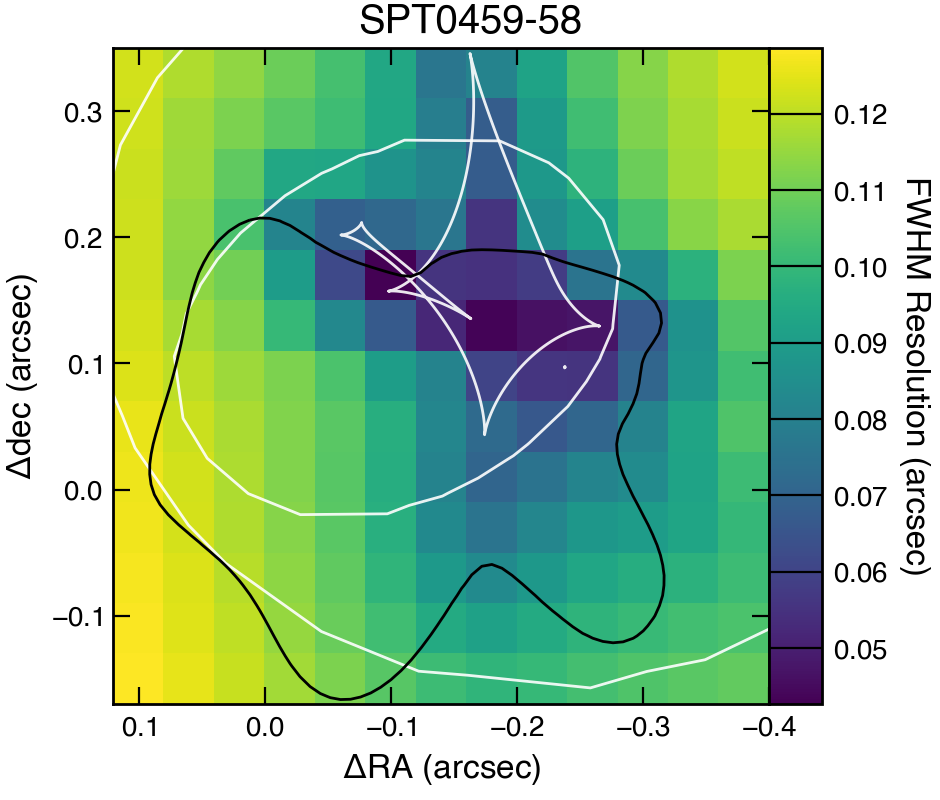}
\includegraphics[width=0.245\textwidth]{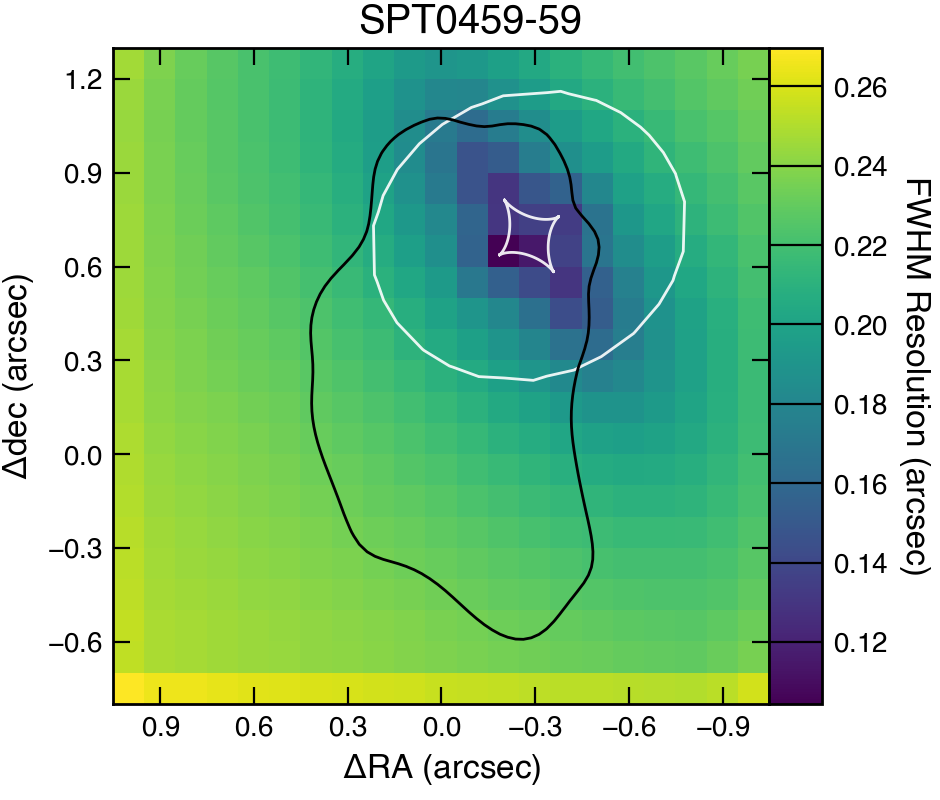}
\includegraphics[width=0.245\textwidth]{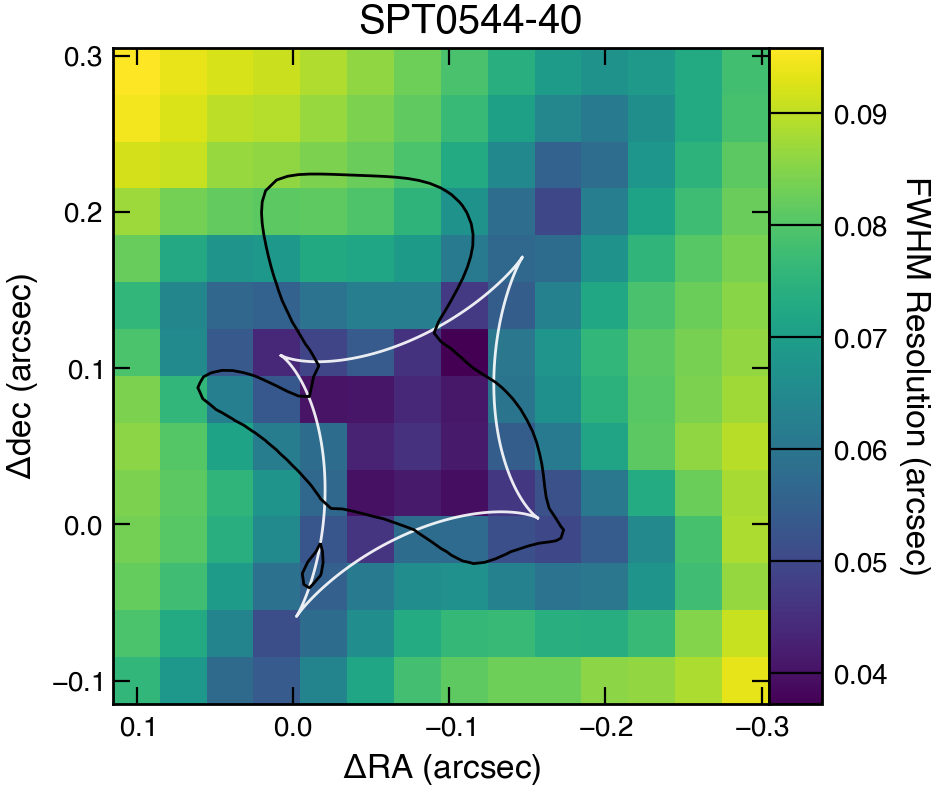}
\includegraphics[width=0.245\textwidth]{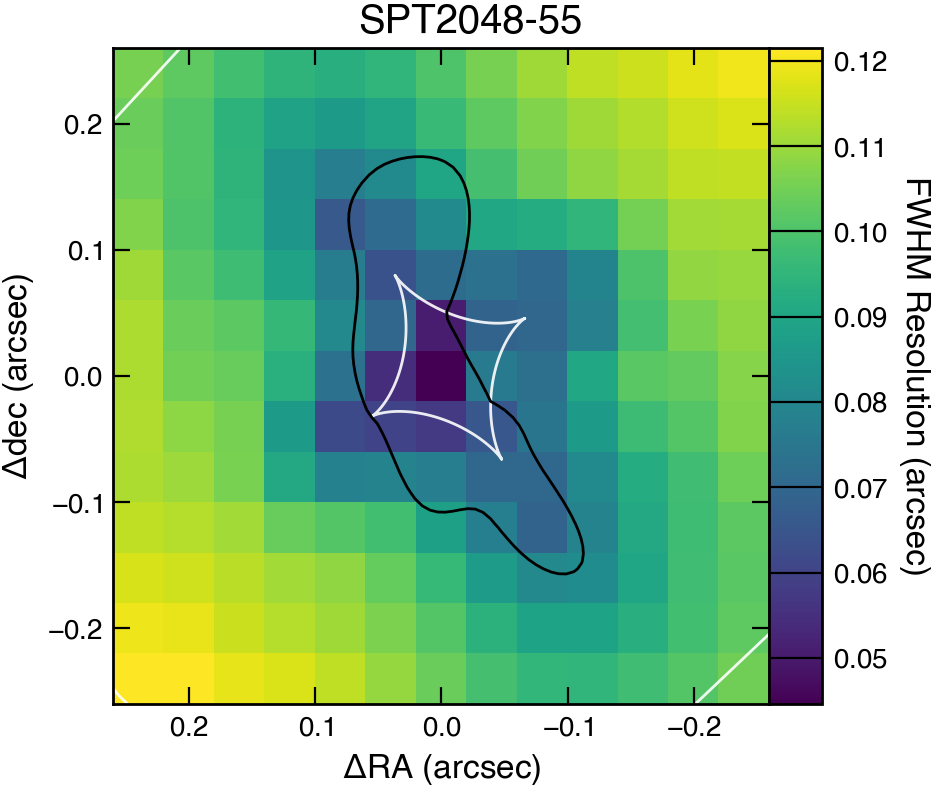}
\includegraphics[width=0.245\textwidth]{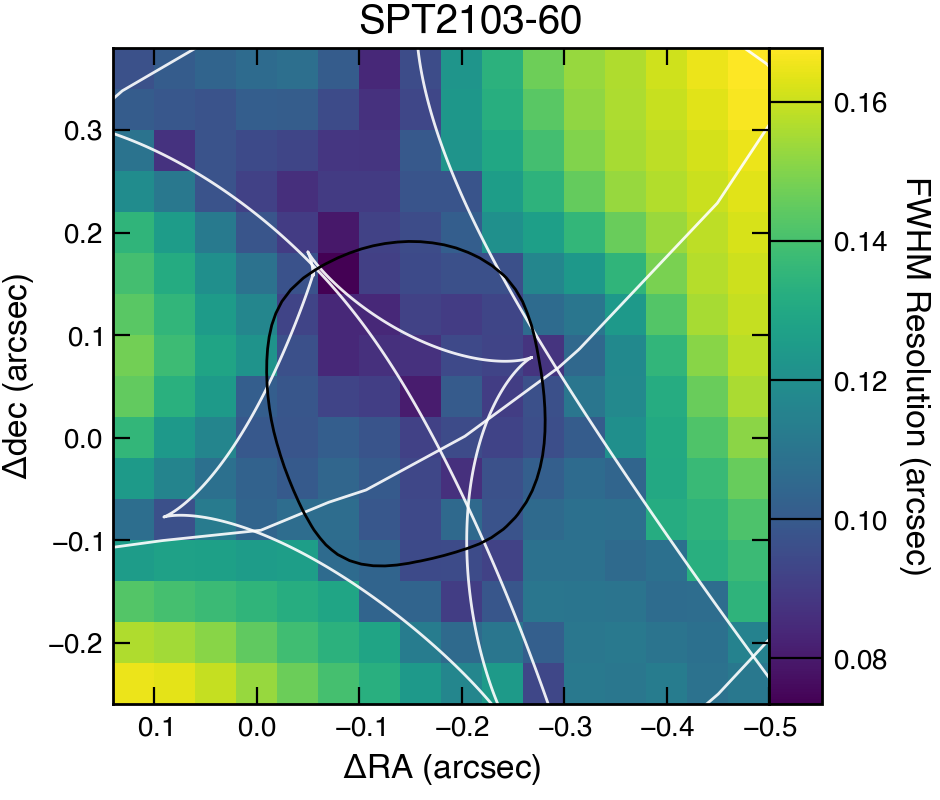}
\includegraphics[width=0.245\textwidth]{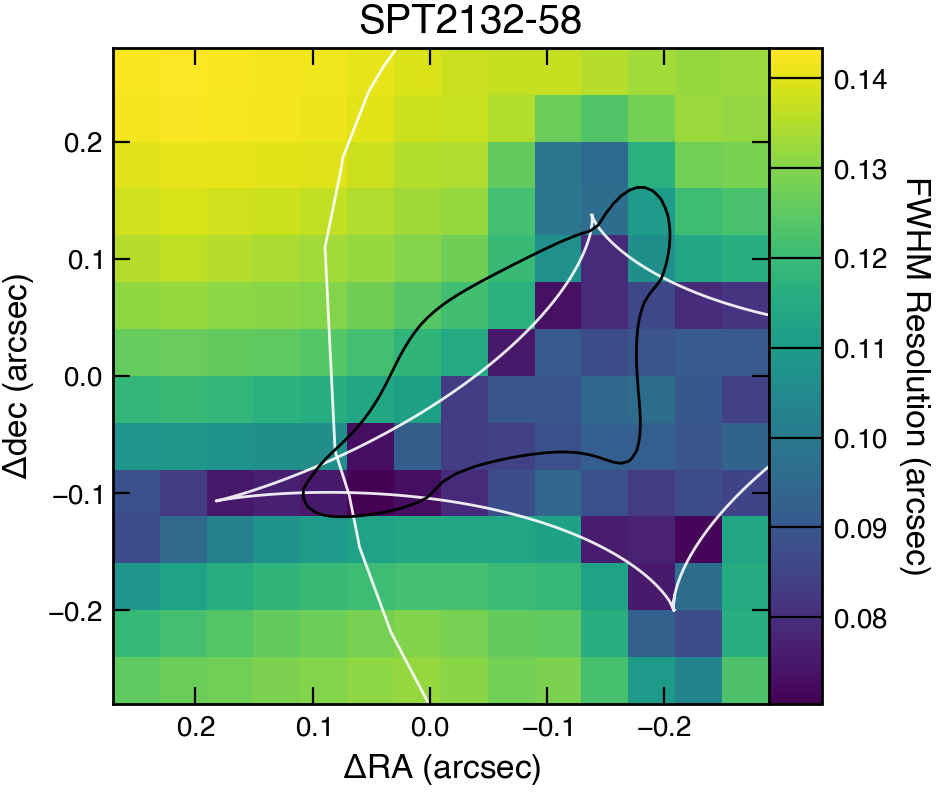}
\includegraphics[width=0.245\textwidth]{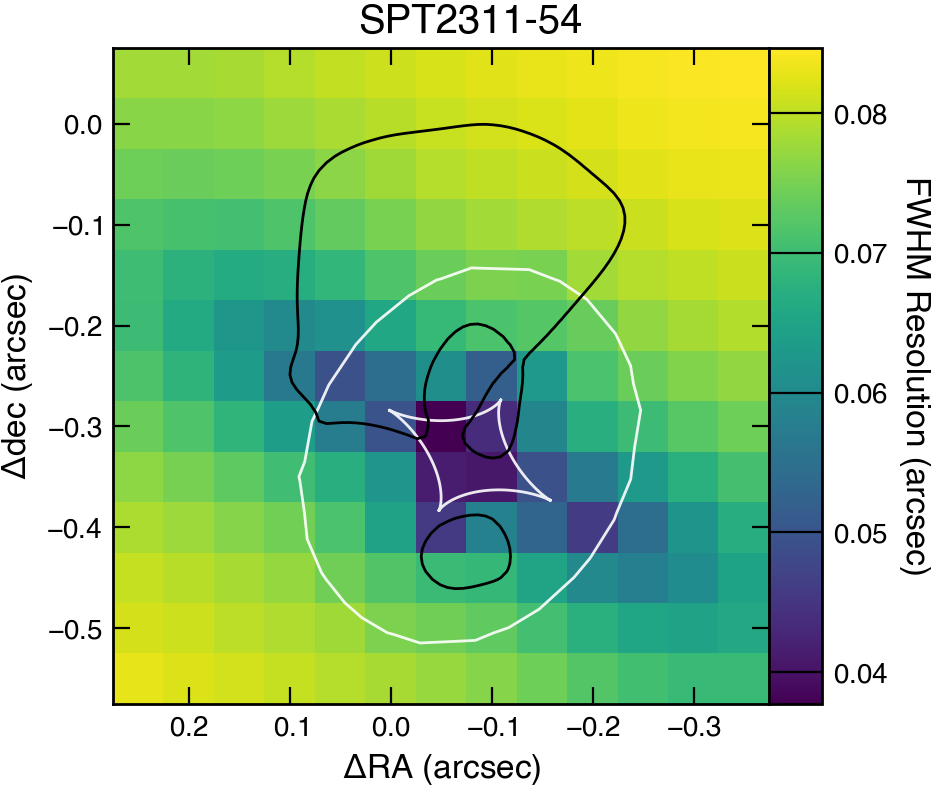}\\
\end{centering}
\caption{
Maps of the effective source-plane resolution for each source made from reconstructions of mock data; see Section~\ref{lenstests}. Each image shows the FWHM of a 2-D Gaussian fit to the reconstruction of a pointlike artificial source at the center of each map pixel, ``observed'' and analyzed identically to the real data. Axes are relative to the ALMA phase center. The black contour shows S/N$=$5 for the actual source as in Figure~\ref{fig:recon}. White contours show the lensing caustics.
}\label{fig:lensresmaps}
\end{figure*}

Figure~\ref{fig:lensresmaps} shows the effective FWHM resolution of the lensing reconstructions as a function of position in the source plane. These maps were made from reconstructions of mock data analyzed the same way as the real data; see Section~\ref{lenstests}. These images show the resolution for the 119\,\um continuum reconstructions. A similar exercise with the noise properties of the frequency ranges of the OH absorption components in each source gives nearly identical results. These maps make intuitive sense in that regions near the caustics with high magnification correspond to better source-plane resolution. We find good agreement with the usual rule-of-thumb that source-plane resolution is approximately the data resolution divided by the square root of the local magnification.

\begin{splitdeluxetable*}{lCCCCCBlCCCCCC}
\tabletypesize{\footnotesize}
\tablecaption{Best-fit lens model parameters\label{tab:lensparams}}
\tablehead{
\colhead{Source} &
\colhead{$\Delta x$} & 
\colhead{$\Delta y$} &
\colhead{$\log M(r<10\mathrm{kpc})/\Msol$} &
\colhead{$e_x$} &
\colhead{$e_y$} &
\colhead{Source} &
\colhead{$\gamma_x$} &
\colhead{$\gamma_y$} &
\colhead{$A_3$} &
\colhead{$B_3$} &
\colhead{$A_4$} &
\colhead{$B_4$}
}
\startdata
SPT0202-61 & +0.016\pm0.004 & +0.070\pm0.002 & 11.221\pm0.010 & +0.22\pm0.07 & -0.53\pm0.13 & 
SPT0202-61 & +0.281\pm0.007 & +0.074\pm0.004 & 0.087\pm0.005 & 0.026\pm0.003 & -0.042\pm0.006 & 0.071\pm0.006 \\
SPT0418-47 & +0.097\pm0.004 & -0.019\pm0.004 & 11.537\pm0.005 & +0.09\pm0.02 & -0.23\pm0.10 & 
SPT0418-47 &  0.000\pm0.005 & -0.011\pm0.005 & \nodata        & \nodata      & \nodata      & \nodata \\
SPT0441-46 & +0.079\pm0.010 & +0.338\pm0.007 & 11.385\pm0.013 & +0.03\pm0.04 & -0.19\pm0.06 & 
SPT0441-46 & -0.074\pm0.019 & -0.019\pm0.025 & -0.086\pm0.013 &0.044\pm0.012 & 0.063\pm0.017 & 0.057\pm0.015 \\
SPT0459-58 & -0.068\pm0.022 & +0.045\pm0.018 & 10.994\pm0.025 & +0.27\pm0.05 & +0.01\pm0.05 &
SPT0459-58 & \nodata & \nodata & \nodata & \nodata & \nodata & \nodata \\
           & -0.378\pm0.023 & +0.341\pm0.026 & 10.563\pm0.058 & +0.53\pm0.07 & -0.31\pm0.09 &
           &  &  &  &  &  & \\
SPT0459-59 & -0.280\pm0.016 & +0.699\pm0.016 & 11.246\pm0.034 & +0.27\pm0.09 & -0.12\pm0.07 & 
SPT0459-59 & -0.031\pm0.022 & -0.032\pm0.043 & \nodata & \nodata & \nodata & \nodata \\
SPT0544-40 & -0.075\pm0.002 & +0.056\pm0.003 & 11.055\pm0.010 & +0.40\pm0.02 & +0.26\pm0.02 & 
SPT0544-40 & -0.030\pm0.010 & -0.085\pm0.012 & \nodata & \nodata & \nodata & \nodata \\
SPT2048-55 & -0.006\pm0.017 & +0.007\pm0.016 & 11.020\pm0.007 & +0.17\pm0.21 & 0.21\pm0.15 & 
SPT2048-55 & +0.067\pm0.119 & +0.080\pm0.082 & \nodata & \nodata & \nodata & \nodata \\
SPT2103-60 & +0.851\pm0.005 & -0.485\pm0.006 & 11.151\pm0.001 & +0.42\pm0.02 & -0.24\pm0.02 &
SPT2103-60 & \nodata & \nodata & \nodata & \nodata & \nodata & \nodata \\
           & +0.117\pm0.020 & +1.113\pm0.021 & 11.161\pm0.002 & +0.43\pm0.02 & -0.45\pm0.02 &
           &  &  &  &  &  & \\
           & -1.423\pm0.043 & -1.839\pm0.149 & 11.022\pm0.003 & +0.84\pm0.02 & 0.11\pm0.02 & 
           &  &  &  &  &  & \\
SPT2132-58 & -0.174\pm0.023 & -0.031\pm0.014 & 10.874\pm0.003 & -0.15\pm0.04 & -0.74\pm0.05 &
SPT2132-58 & +0.101\pm0.050 & -0.035\pm0.036 & \nodata & \nodata & \nodata & \nodata \\
SPT2311-54 & -0.081\pm0.018 & -0.345\pm0.026 & 10.657\pm0.003 & +0.17\pm0.06 & -0.45\pm0.08 &
SPT2311-54 & \nodata & \nodata & \nodata & \nodata & \nodata & \nodata \\
\enddata
\tablecomments{Parameter descriptions are as follows. $\Delta x$, $\Delta y$: lens position in arcsec relative to the ALMA phase center. $\log M(r<10\mathrm{kpc})/\Msol$: mass contained within 10\,kpc of the lens center. $e_x$, $e_y$: lens ellipticity components. $\gamma_x$, $\gamma_y$: external tidal shear components. $A_3$, $B_3$: $m=3$ multipole components. $A_4$, $B_4$: $m=4$ multipole components. Entries with ellipses were fixed to 0 during fitting. The two sources with multiple lenses (SPT0459-58 and SPT2103-60) list each lens; no shear or multipole parameters were used for these sources.}
\end{splitdeluxetable*}

\end{CJK*}
\end{document}